\documentclass[journal, twocolumn]{IEEEtran}

\usepackage{graphicx}		
\usepackage{booktabs} 		
\usepackage{tabularx}		
\usepackage{array}			
\usepackage{epstopdf}   	
\usepackage{hyperref}		
\usepackage{xcolor}      	
\hypersetup{colorlinks,
	linkcolor={red!50!black},
	citecolor={red!50!black},
	urlcolor={blue!00!black}
}               			
\usepackage[all]{hypcap}    
	
\usepackage[cmex10]{amsmath}
\usepackage{dsfont}
\usepackage{amsthm}
\usepackage{amsfonts} 		
\usepackage{amssymb}		
\usepackage{cases}

\usepackage{array}
\usepackage{mdwmath}
\usepackage{mdwtab}

\usepackage[nocompress]{cite}

\usepackage{algorithm,algorithmic} 

\theoremstyle{definition}
\newtheorem{definition}{Definition}

\begin{document}
\title{An Overview on Resource Allocation Techniques for Multi-User MIMO Systems}

\author{Eduardo~Casta\~{n}eda,~\IEEEmembership{Member,~IEEE,}
	Ad\~{a}o~Silva,~\IEEEmembership{Member,~IEEE,}
	At\'{i}lio~Gameiro, 
	and~Marios~Kountouris,~\IEEEmembership{Senior Member,~IEEE}
	\thanks{Manuscript received February 23, 2016; revised August 31, 2016; accepted October 15, 2016. This work was supported by the Portuguese Funda\c{c}\~{a}o para a Ci\^{e}ncia e Tecnologia FCT/MEC through national funds, PURE-5GNET and VELOCE-MTC (UID/EEA/50008/2013) projects.}
	\thanks{E. Casta\~{n}eda, A. Silva, and A. Gameiro are with the Department
		of Electronics, Telecommunications and Informatics, and the Instituto de Telecomunica\c{c}\~{o}es (IT), Aveiro University, Aveiro, 3810-193, Portugal (e-mail: ecastaneda@av.it.pt; asilva@av.it.pt; amg@ua.pt). }
	\thanks{M. Kountouris is with the Mathematical and Algorithmic Sciences Lab, France Research Center, Huawei Technologies Co. Ltd. (e-mail: marios.kountouris@huawei.com).}
}


\maketitle

\begin{abstract}
Remarkable research activities and major advances have been occurred over the past decade in multiuser multiple-input multiple-output (MU-MIMO) systems. Several transmission technologies and precoding techniques have been developed in order to exploit the spatial dimension so that simultaneous transmission of independent data streams reuse the same radio resources. The achievable performance of such techniques heavily depends on the channel characteristics of the selected users, the amount of channel knowledge, and how efficiently interference is mitigated. 
In systems where the total number of receivers is larger than the number of total transmit antennas, user selection becomes a key approach to benefit from multiuser diversity and achieve full multiplexing gain. The overall performance of MU-MIMO systems is a complex joint multi-objective optimization problem since many variables and parameters have to be optimized, including the number of users, the number of antennas, spatial signaling, rate and power allocation, and transmission technique. The objective of this literature survey is to provide a comprehensive overview of the various methodologies used to approach the aforementioned joint optimization task in the downlink of MU-MIMO communication systems. 
\end{abstract}


\maketitle

\section{Introduction}
\label{section:introduction}

\begin{figure}[t!]
	\centering
	\includegraphics[width=0.9\linewidth]{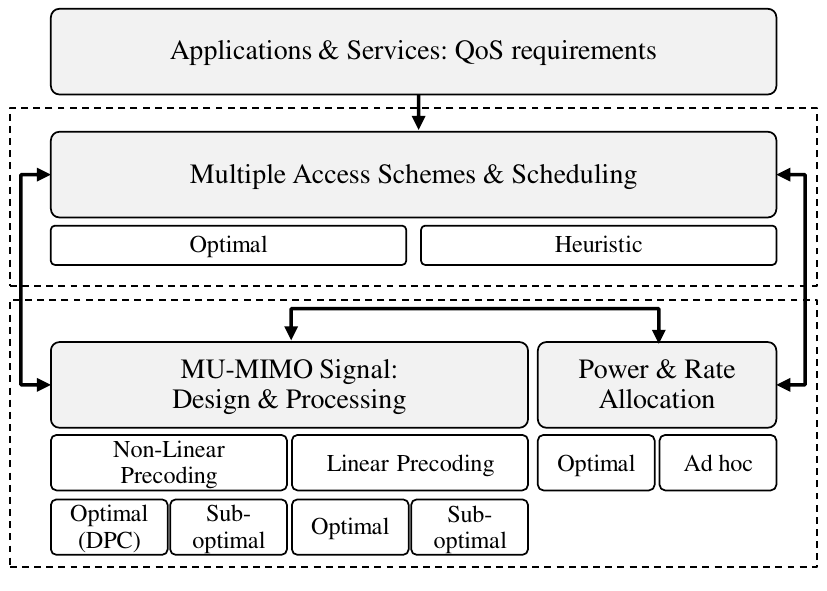}\\
	\caption{Components of the resource allocation policy in MU-MIMO systems} 
	\label{fig:mu-mimo-layered-tasks}
\end{figure}

\IEEEPARstart{F}{uture} wireless systems require fundamental and crisp understanding of design principles and control mechanisms to efficiently manage network resources. Resource allocation policies lie at the heart of wireless communication networks, since they aim at guaranteeing the required Quality of Service (QoS) at the user level, while ensuring efficient and optimized operation at the network level to maximize operators' revenue.
Resource allocation management in wireless communications may include a wide spectrum of network functionalities, such as scheduling, transmission rate control, power control, bandwidth reservation, call admission control, transmitter assignment, and handover \cite{Stanczak2009,Lee2014b,Ahmed2005}.  
In this survey, a resource allocation policy is defined by the following components: \textit{i}) a multiple access technique and a scheduling component that distributes resources among users subject to individual QoS requirements; \textit{ii}) a signaling strategy that allows simultaneous transmission of independent data streams to the scheduled users; and \textit{iii}) rate allocation and power control that guarantee QoS and harness potential interference. 
Fig.~\ref{fig:mu-mimo-layered-tasks} illustrates these components and the interconnection between them. The figure highlights the fact that each function of the resource allocation strategy can be performed in either optimal or suboptimal way, which is elaborated upon below.

The multiple access schemes can be classified as orthogonal or non-orthogonal. The former is a conventional scheme that assigns radio resources, e.g., code, sub-carrier, or time slot, to one user per transmission interval. The main characteristic of orthogonal multiple access schemes is their reliability, since there is no need to deal with co-resource interference. The resource allocation policy can optimize with reasonable complexity several performance metrics, such as throughput, fairness, and QoS \cite{Asadi2013}. The multiplexing gain, i.e., the number of scheduled users, is limited by the number of available radio resources in the system.  
In non-orthogonal multiple access,  a set of users concurrently superimpose their transmissions over the same radio resource, and potentially interfere with each other. 
In this scheme the co-resource interference can be mitigated by signal processing and transmission techniques implemented at the transmitter and/or receiver sides. Such techniques exploit different resource domains, e.g., power, code, or spatial domain, and a combination of them are envisaged to cope with the high data rate demands and system efficiency expected in the next generation of wireless networks \cite{Li2014a,Dai2015}.

Hereinafter, we focus on multiple access schemes based on multi-antenna transceivers operating at the spatial domain, i.e., multiple-input multiple-output (MIMO). MIMO communication, where a multi-antenna base station (BS) or access point (AP) transmits one or many data streams to one or multiple user equipments simultaneously, is a key technology to provide high throughput in broadband wireless communication systems. MIMO systems have evolved from a fundamental research concept to real-world deployment, and they have been integrated in state-of-the-art wireless network standards \cite{Li2010,Jones2015,Kim2015}, e.g., IEEE 802.11n, 802.11ac WLAN, 802.16e (Mobile WiMAX), 802.16m (WiMAX), 802.20 (MBWA), 802.22 (WRAN), 3GPP long-term evolution (LTE) and LTE-Advanced (E-UTRA). 
Resource allocation is particularly challenging in wireless communication systems mainly due to the wireless medium variability and channel randomness, which renders the overall performance location-dependent and time-varying \cite{Wang2007}. 
Nevertheless, high spectral efficiency and multiplexing gains can be attained in MIMO systems since multiple data streams can be conveyed to independent users. By exploiting the spatial degrees of freedom (DoF) offered by multiple antennas we can avoid system resource wastage \cite{Ajib2005}. Multiuser (MU) MIMO systems have been extensively investigated over the last years from both theoretical and practical perspective. In a recent evolution of MU-MIMO technology, known as massive MIMO or large-scale MIMO \cite{Lu2014,Zheng2015}, few hundreds of antennas are employed at the BS to send simultaneously different data streams to tens of users. Massive MIMO has been identified as one of the promising air interface technologies to address the massive capacity requirement required demanded by 5G networks \cite{Andrews2014,Boccardi2014}.

The downlink transmission is particularly challenging in MU-MIMO scenarios because the geographic location of the receivers is random and joint detection cannot be performed. The main goal is to convey independent data streams to a set of properly selected users, attaining spatial multiplexing gain offered by MU-MIMO. However, determining such a users set is a very challenging task, which depends on all elements of the resource allocation strategy, e.g., individual QoS requirements, signaling schemes, rate allocation and power control strategies implemented at the transmitter. 
MIMO systems allow for a plethora of mighty signal processing techniques that enhance the system performance by  exploiting a multi-dimensional pool of resources. This pool is composed of resources with different nature, e.g., signal spaces, transmission powers, time slots, sub-carriers, codes, and users. Efficient allocation designs over such a large set of resources implies that a tradeoff between optimality and feasibility. On the one hand, optimality can be reached by solving optimization problems over a set of integer and continuous variables, which may be a thoroughly complex task. Feasibility, on the other hand, implies that suboptimal resource allocation takes place by relaxing and reformulating optimization problems whose solutions can be found by practical and reliable algorithms.

{\color{black}{ 

\subsection{Contributions of the Survey} 
\label{section:introduction:contributions}

There exists a very rich literature on MIMO communications, and this paper complements it by providing a classification of different aspects of MU-MIMO systems and resource allocation schemes. Users with independent channels provide a new sort of diversity to enhance the overall performance. However, in contrast to systems where each user accesses a dedicated (orthogonal) resource \cite{Wang2007,Asadi2013,Capozzi2013,Li2014}, in MU-MIMO systems the additional diversity is realized when several users access the same resource simultaneously. 

Accounting for multiple antennas at both ends of the radio link allows spatial steering of independent signals using precoding schemes, which results in the coexistence of many data streams conveyed to the concurrent users. Some of the main contributions of this survey are the description and classification of linear and non-linear precoding schemes, considering the amount of channel information available at the transmitter, the network scenario (e.g. single-cell or multi-cell), and the antenna settings. Each precoding scheme relies on different characteristics of the MU-MIMO channels to fully exploit the spatial domain. The paper provides a comprehensive classification of metrics that quantify the spatial compatibility, which can be used to select users and improve the precoding performance.

The spectral efficiency, error rates, fairness, and QoS are common criteria to assess performance in the MU-MIMO literature. Optimizing each one of these metrics requires specific problem and constrains formulations. The type of precoding, antenna configuration, and upper layer demands can be taken into account to design robust resource allocation algorithms, i.e. cross layer designs. Other contributions of this paper are the description and classification of different optimization criteria and general constraints used to characterize MU-MIMO system. The proposed classification  incorporates the antenna configuration, the amount of channel information available at the transmitter, and upper layer requirements. 

Early surveys on MU-MIMO have pointed out that resource allocation can be opportunistically enhanced by tracking the instantaneous channel fluctuations for scenarios with a single transmitter \cite{Ajib2005,Gesbert2007a}. However, in recent years, a large number of techniques have been developed for very diverse and heterogeneous MIMO scenarios. The paper presents a classification of state-of-the-art scheduling algorithms for MU-MIMO scenarios for single and multiple transmitter scenarios. We consider the channel state information, the objective functions optimized by the scheduler, the degree of cooperation/coordination between transmitters, and the power allocation techniques. 

The goal of this survey is not to describe in detail the theory behind precoding design, rate allocation, power control or user scheduling, but rather to use their fundamental principles to get insight on the interplay among them. Our aim is to describe state-of-the-art processing techniques for MU-MIMO, point out practical challenges, and present general guidelines to design efficient resource allocation algorithms. The material favors broad intuition over detailed mathematical formulations, which are left to the references. Although the list of references is certainly not intended to be exhaustive, the cited works and the references therein may serve as a starting point for readers aiming to go beyond a tutorial.

}}

\subsection{Organization} 
\label{section:introduction:organization}

The paper is organized as follows.
In Section~\ref{section:preliminaries} we present the basic ideas behind MIMO wireless communications, introduce MU-MIMO systems, and discuss the main challenges.
In Section~\ref{section:system-set-ups} we introduce the most commonly studied MU-MIMO channel and system models, their characteristics and conventional assumptions.
Section~\ref{section:precoding} is devoted to signal design and precoding schemes under different conditions of channel information. 
In Section~\ref{section:metrics-spatial-compatibility} we introduce the most common metrics of spatial compatibility, which are used to categorize users and reduce the scheduling complexity.
Section~\ref{section:performance-uf}  presents a classification of optimization criteria  and describes the usual constraints considered in MU-MIMO systems. 
In Section~\ref{section:algorithms} we propose a classification of the several techniques to address the user scheduling problem. The specific characteristics, limitations and use cases for each technique are discussed.    
Section~\ref{section:algorithms:selection-partial-csit} is focused on scheduling algorithms with partial channel information at the transmitter. We categorize the existing approaches and present guidelines to minimize complexity and improve efficiency.
In Section~\ref{section:power-allocation} we present the most common power allocation schemes and discuss their role in MU-MIMO systems.
Finally, we conclude the paper in Section~\ref{section:remarks}.
The reader can find a list of technical terms and abbreviations summarized in Table~\ref{table:abbreviations}.

We adopt the following notation: matrices and vectors are set in upper and lower boldface, respectively. $(\cdot)^{T}$, $(\cdot)^{H}$, $|\cdot|$, $\|\cdot\|_{p}$ denote the transpose, the Hermitian transpose, the absolute value, and the $p$-norm, respectively. $rank(\mathbf{A})$, $null(\mathbf{A})$ denote the rank and null space of matrix $\mathbf{A}$. $Span(\mathbf{A})$ and $Span(\mathbf{A})^{\perp}$ denote the subspace and orthogonal subspace spanned by the columns of matrix $\mathbf{A}$. Calligraphic letters, e.g. $\mathcal{G}$, denote sets, and $|\mathcal{G}|$ denotes cardinality.  $\mathbb{R}_{+}$ is the set of nonnegative real numbers and $\mathbb{C}^{N \times M}$ is the space of $N \times M$ matrices. $\mathcal{CN}(\mathbf{a},\mathbf{A})$ is the complex Gaussian distribution with mean $\mathbf{a}$ and covariance matrix $\mathbf{A}$. $\mathbb{E}[\cdot]$ denotes expectation.

\section{Preliminaries}
\label{section:preliminaries}

\subsection{Multiple Antenna Systems}
\label{section:introduction:mimo-systems}

A MIMO system employs multiple antennas at the transmitter ($M$) and receiver ($N$) sides to improve communication performance.
The seminal works \cite{Foschini1998,Telatar1999} provide a mathematical motivation behind multiple antenna processing and communications. Theoretical analysis has shown that the spectral efficiency, i.e., the amount of error-free bits per second per Hertz (bps/Hz), follows the scaling low $\min(M,N)$, without increasing the power or bandwidth requirements.
The signal processing techniques in multi-antenna systems can be classified as \textit{spatial diversity techniques} and \textit{spatial multiplexing techniques} \cite{Mietzner2009}.

Spatial diversity techniques (see \cite{Mietzner2009} and references therein), provide transmission reliability and minimize error rates. This is attained by transforming a fading wireless channel into an additive white Gaussian noise (AWGN)-like channel, i.e., one can mitigate signal degradation due to fading \cite{Ajib2005}. The probability that multiple statistically independent channels experience simultaneously deep fading gets very low as the number of independent paths increases. The spatial diversity  techniques can be applied at both transmission and reception sides of the link. Transmit diversity schemes include space diversity, polarization diversity, time diversity, frequency diversity, and angle diversity. Examples of receive diversity schemes are selection combining, maximum ratio combining (MRC), and equal gain combining \cite{Tse2005,Goldsmith2005}.

\subsection{Multiuser MIMO}
\label{section:introduction:mu-mimo}
This paper focuses on spatial multiplexing techniques, which exploit the DoF provided by MIMO. Spatial multiplexing is tightly related to multiuser communications and smart antennas processing \cite{Tse2005}. In multiuser systems, spatial multiplexing gains can be attained by steering signals toward specific receivers, such that the power to intended users is boosted. Simultaneously, co-channel interference to unintended users can be partially or completely suppressed.

In MU-MIMO systems, the available resources (power, bandwidth, antennas, codes, or time slots) must be assigned among $K$ active users.
There are two kinds of multiuser channels: the downlink channel, also known as \textit{broadcast channel} ({BC}), where a single transmitter sends different messages to many receivers; and the uplink channel, also called \textit{multiple access channel} ({MAC}), where many transmitters communicate with a single receiver.
There are several \textit{explicit} differences between {BC} and {MAC}. In the former, the transmitted signal is a combination of the signals intended for all co-scheduled users, subject to total transmit power, $P$, constraints. In contrast, in the {MAC} channel, the signal from the $k$-th user is affected by other co-scheduled users, subject to individual power constraints, i.e. $P_k$, \cite{Tse2005}. 
There exists an \textit{implicit} connection between BC and MAC, known as \textit{duality}, which establishes the relationship between the capacity regions of both access channels \cite{Tse2005,Goldsmith2005}. The BC-MAC duality has been fundamental to define optimal policies for power allocation, signaling, and QoS guaranteeing in MU-MIMO systems, see \cite{Schubert2004,Schubert2006,Yu2006a,Chiang2008}.
The capacity regions include  operative point where transmission to multiple users do not interfere with each other. Every transmission is performed over orthogonal signaling dimensions, which is a signal separation called duplexing \cite{Goldsmith2005}. This operation is performed by allocating communications across different time slots, known as time-division-duplex ({TDD}), or across separated frequency bands, known as frequency-division-duplex ({FDD}).

\begin{figure}[t!]
	\centering
	\includegraphics[width=0.98\linewidth]{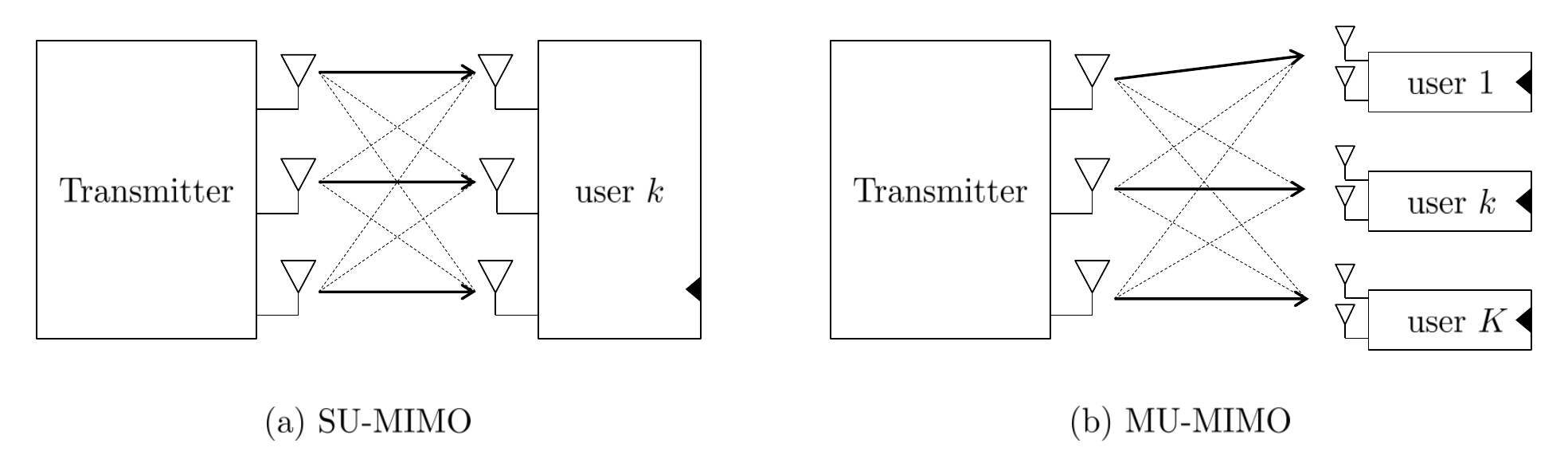}\\
	\caption{(a) Single-user (SU) and (b) Multi-user (MU) MIMO modes}
	\label{fig:su-mu-mimo}
\end{figure}

In the literature of {MU-MIMO}, two types of diversity are studied: \textit{spatial multiplexing diversity} and \textit{multiuser diversity} ({MUDiv}). The former is a consequence of the independent fading across MIMO links of different users. This means that independent data streams can be transmitted over parallel spatial channels, increasing the system capacity \cite{Zheng2003}. The latter arises when users that are geographically far apart have channels that fade independently at any point in time. Such independent fading processes can be exploited so that users with specific channel conditions are simultaneously scheduled \cite{Viswanath2002}. 
There are two modes of transmission in MIMO systems, see Fig.~\ref{fig:su-mu-mimo}: single user (SU) and multiuser (MU) mode. The SU-MIMO mode improves the performance of a single user, allocating one or many data streams in the same radio resource. In the MU-MIMO mode, different data streams are sent to different users such that a  performance metric is optimized, e.g., the average sum rate.

Selecting between SU- or MU-MIMO transmission modes depends on the accuracy of the channel state information at the transmitter (CSIT), the amount of allowed interference, the target rate per user, the number of user, the signal-to-noise ratio (SNR) regime, and the achievable capacity in each mode \cite{Tang2010}.
Nonetheless, by assuming sufficient CSIT knowledge, MU-MIMO processing techniques provide several performance gains \cite{Gesbert2007a}: multiple antennas attain diversity gain, which improves bit error rates ({BER}); directivity gains realized by {MUDiv}, since the spatial signatures of the users are uncorrelated, which mitigates inter-user interference (IUI); immunity to propagation limitations in SU-MIMO, such as rank loss or antenna correlation; and multiplexing gains that scale, at most, with the minimum number of deployed antennas.

\begin{figure*}[th!]
	\centering
	\includegraphics[width=0.98\linewidth]{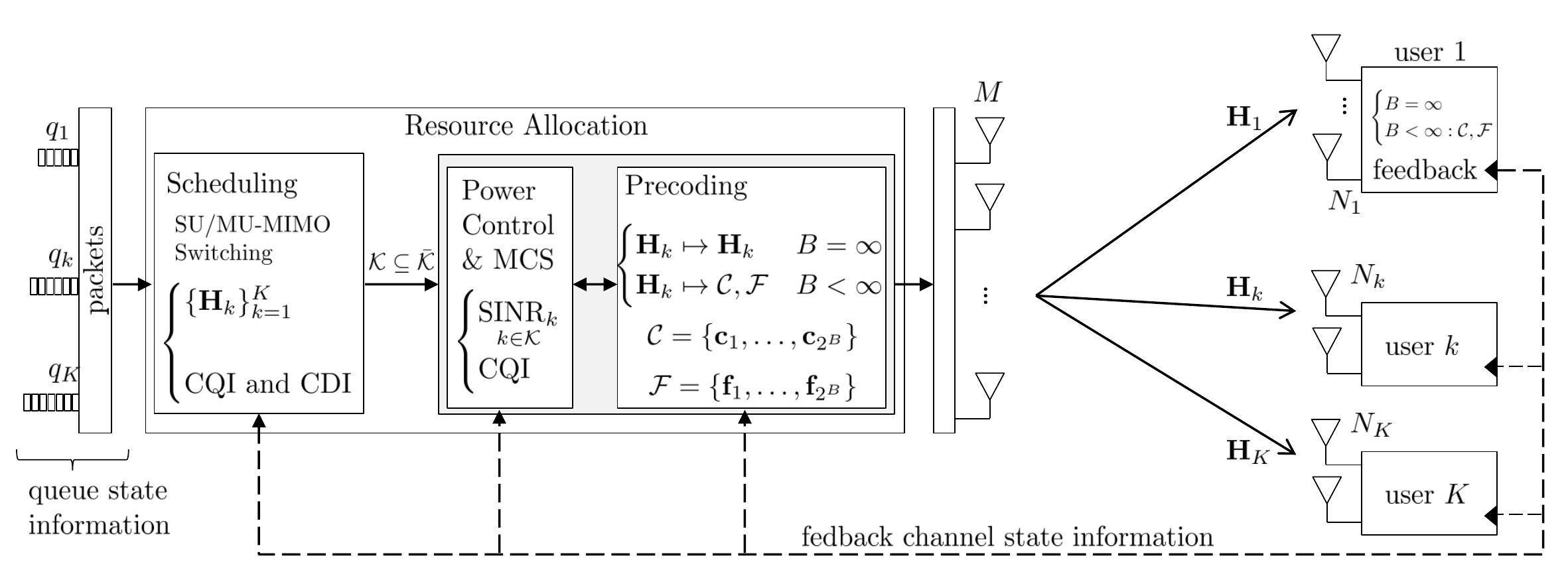}\\
	\caption{Processing blocks and input signals in a MU-MIMO scenario. A transmitter with $M$ antennas serves $K$ multiple antenna receivers.} 
	\label{fig:system-processing-blocks}
\end{figure*} 

\subsection{The need of User Scheduling}
\label{section:introduction:user-selection}

In {MU-MIMO} {BC} systems, the overall performance depends on how efficiently the resource allocation algorithms manage the hyper-dimensional pool of resources (carriers, time slots, codes, power, antennas, users, etc.). 
Consider a system with a transmitter equipped with $M$ antennas, and let $\bar{\mathcal{K}}=\{1,2,3, \ldots, K\}$ be the set of all active users, illustrated in Fig.~\ref{fig:system-processing-blocks}. To qualitatively determine the objectives of scheduling, we provide the following definitions:     

\begin{definition}\label{defn:quality-of-service}
	\textit{Quality of Service}. We say that QoS defines a set of prescribed network-/user-based performance targets (e.g., peak rates, error rates, average delays, or queue stability), that can be measured, improved, and guaranteed for a specific upper layer application. 
\end{definition}

\begin{definition}\label{defn:user_scheduling}
	\textit{User scheduling}. We say that a set of radio resources (e.g. time slot, codes, sub-channels, powers, etc.), has been assigned to a group of scheduled users, $\mathcal{K} \subseteq \bar{\mathcal{K}}$, so that a global performance metric is optimized subject to power and QoS constraints. Moreover, each user $k \in \mathcal{K}$, achieves non-zero rate with successful information reception.
\end{definition}

Consider that each user $k \in \bar{\mathcal{K}}$, is equipped with $N_k$ antennas. By having more receive than transmit antennas ($M< \sum_{k} N_k$), one can solve a selection problem to achieve MUDiv gains in fading fluctuating channels \cite{Ajib2005,Sharif2005,Sharif2007,Gesbert2007a,Yang2011}. A fundamental task in resource management is to select a subset of users $\mathcal{K} \subseteq \bar{\mathcal{K}}$, and assign resource to it, so that a given performance metric is optimized.
For the sake of illustration, consider a single-transmitter scenario, and let us formulate a general user scheduling problem for a single resource (sub-carrier, time slot, or code) as follows:

\begin{IEEEeqnarray}{rrCl}
	& {\text{maximize}} & \quad &  \sum_{k =1 }^{K} \xi_{\pi(k)} U_{\pi(k)}  \label{eq:general-combinatorial-problem:cost-function}   \\
	& \text{subject to} && \sum_{k =1}^{K} \xi_{k} p_{k} \leq P \IEEEyessubnumber \label{eq:general-combinatorial-problem:total-power}\\
	&&& 0 \leq \xi_{k} p_{k} \leq P_{k} \ \ \ \forall k \in \bar{\mathcal{K}} \IEEEyessubnumber \label{eq:general-combinatorial-problem:individual-power} \\
	&  & & \sum_{k =1}^{K} \xi_{k} \leq c_{p} \IEEEyessubnumber \label{eq:general-combinatorial-problem:set-cardinality} \\
	&&& \xi_{k} \in \{0, 1\} \ \ \ \forall k \in \bar{\mathcal{K}} \IEEEyessubnumber \label{eq:general-combinatorial-problem:binary-vars}
\end{IEEEeqnarray}

Our goal is to maximize the sum of the utility functions, $U_{\pi(k)}$, $\forall k$, which depends on several parameters: the multiuser MIMO channel, $\mathbf{H}_{k}$, the allocated power, $p_{k}$, the individual data queues, $q_{k}$, and the encoding order, $\pi(\cdot)$. The QoS requirements can be included in the definition of $U_{\pi(k)}$, as individual weights [cf. Section~\ref{section:performance-uf}]. Equations (\ref{eq:general-combinatorial-problem:total-power}) and (\ref{eq:general-combinatorial-problem:individual-power}) define total and individual power constraints, which are set according to the scenario [cf. Section~\ref{section:system-set-ups}]. The term $\xi_{\pi(k)}$ is a binary variable with value equal to 1 if the $\pi(k)$-th user is scheduled and 0 otherwise. The set of selected user is given by $\mathcal{K} = \{ k \in \bar{\mathcal{K}}: \xi_{k} = 1 \}$. 
The system operates in MU-MIMO mode if $1 < |\mathcal{K}| \leq c_{p}$, where $c_{p}$ in (\ref{eq:general-combinatorial-problem:set-cardinality}) denotes the maximum number of users or transmitted data streams, that can be sent over $M$ antennas. If the solution of (\ref{eq:general-combinatorial-problem:cost-function}) only exists for $|\mathcal{K}|=1$, the system operates in SU-MIMO mode. In such a case, the optimization problem can be formulated to attain MUDiv, multiplexing (high rate), and diversity (high reliability) gains, see \cite{Tse2005,Heath2005,Chen2006}.
Depending on the CSIT, the type of signaling design and coding applied to the data, theoretical analysis show that the number of users with optimal nonzero power is upper bounded\footnote{The upper bound is tight for small values of $M$ and it becomes loose as the number of transmit antennas grow large. Numerical results comparing the upper bound of $|\mathcal{K}|$ for several coding techniques can be found in \cite{Boccardi2006}. } as $|\mathcal{K}| \leq c_p \leq M^2$, \cite{Yu2006a}. 
In practical systems, multiplexing gain can be scaled up to $|\mathcal{K}| \leq M$, by means of linear signal processing   [cf. Section~\ref{section:precoding:linear}].

The mathematical formulation in (\ref{eq:general-combinatorial-problem:cost-function}) resembles a knapsack or subset-sum problem \cite{Rondeau2009,Williamson2010}, which is known to be non-polynomial time complete (NP-C). Although the users are fixed items that must be chosen to construct $\mathcal{K}$, their associated utility functions change according to the channel conditions and the resource allocation of the co-selected users. This implies that the optimization variables are, in general, globally coupled.
Finding the optimal set $\mathcal{K}$, is a combinatorial problem due to the binary variables $\xi_{k}$, and the encoding order $\pi(\cdot)$. Moreover, depending on $U_{k}$, problem (\ref{eq:general-combinatorial-problem:cost-function}) might deal with non-convex functions on the multiple parameters, e.g. $K$, $M$, $N_k$, etc. The feasibility of (\ref{eq:general-combinatorial-problem:cost-function}) relies on the constraints and processing, e.g. the precoding schemes, the power allocation, the CSIT accuracy, \cite{Gesbert2007a}, [cf. Section~\ref{section:performance-uf}]. The scheduling problem can be solved optimally by exhaustively searching (ExS) over all possible set sizes and user permutations. However, the computational complexity of ExS is prohibitively high, even for small values of $K$ \cite{Ajib2005}. Furthermore, problem (\ref{eq:general-combinatorial-problem:cost-function}) can be modified to include additional dimensions, such as multiple carriers for OFDMA systems (e.g. \cite{Chan2007,Moretti2013}) or codes for CDMA systems (e.g. \cite{Driouch2012}).

\subsection{The need of CSI availability} 
\label{section:introduction:needof-csit}

Channel knowledge at the transmitter can be modeled taking into account instantaneous or statistical information, e.g. variance, covariance, angles of arrival/departure, and dominant path in line-of-sight \cite{Vu2007}. 
There are two main strategies used to obtain CSI, reciprocity and feedback, which provide different feedback requirements and robustness to CSI errors. The former, known as open-loop feedback, uses uplink channel information to define the downlink channel in the next transmission interval. It is suitable for TDD since transmit directions are in identical frequencies, and the channel can be reversed. The latter, known as closed-loop feedback, requires sending the downlink channel to the transmitter using dedicated pilots, and is commonly used in FDD.  
The majority of the papers reviewed in Section~\ref{section:algorithms} assume closed-loop with full or limited channel feedback, and we refer the reader to \cite{Love2008,Vu2007,Kobayashi2011} and references therein for further discussion on CSI acquisition and its impact on system performance.

To perform multiple antenna processing, interference mitigation, user scheduling, efficient power allocation, and to profit from MUDiv, knowledge of CSIT is compulsory.  The complete lack of CSIT reduces the multiplexing gain to one, and cannot use MUDiv for boosting the achievable capacity \cite{Jafar2005, Sharif2005,Hassibi2007}. In such scenarios, the optimum resource allocation and transmission schemes are performed over orthogonal dimensions \cite{Koutsopoulos2008}.    
In the literature of MU-MIMO, a large number of works assume full CSI (error-free),  at both the receiver (CSIR) and transmitter (CSIT) sides. 
In practical systems, a strong downlink pilot channel, provided by the transmitter, is available to the users, hence the CSIR estimation error is negligible relative to that of the CSIT \cite{Lau2009}. For simplicity, it is widely assumed that CSIR is perfectly known at the mobile terminals. 
{\color{black}{
In cellular MU-MIMO systems, channel estimation relies on having orthogonal pilots allocated to different users. The orthogonality is guaranteed for the users within the same cell, but not for those scattered across different cells. The number of BS antennas and bandwidth constraints may not allow orthogonal pilots for each user in the system, resulting in pilot contamination \cite{Marzetta2010}. Under universal frequency reuse, the pilots can be drastically polluted by users at adjacent cells, when the serving BS performs channel estimation \cite{Elijah2016}.
}}

Achieving full CSIT (ideal noiseless and delay-free feedback) is highly challenging in practice. Feeding back the CSI requires rates that grow rapidly with the transmit power and the number of antennas \cite{Jindal2006}. However, by assuming full CSIT, it is possible to derive upper bounds on the performance of different signal processing techniques and scheduling algorithms. The information-theoretic and numerical results using full CSIT provide useful insights regarding the system performance bounds (e.g., \cite{Caire2003,Weingarten2006,Sharif2007}). Resource allocation strategies that optimize spectral efficiency, fairness, power consumption, and error probability can be designed to characterize optimal operating points \cite{Schubert2006,Bjornson2013,Chiang2008}. Analytical results for full CSIT reveal the role of each parameter in the system, e.g., number of deployed transmit and receive antennas, number of active users, SNR regime, etc. 

If channel knowledge is obtained via partial (rate-limited) feedback, the information available at the transmitter has finite resolution, resulting in quantization errors. \textit{Partial CSIT} is comprised of two quantities: channel quality information (CQI) and channel direction information (CDI) \cite{Gesbert2007a}. The CQI measures the achievable SINR, the channel magnitude, or any other function of the link quality. The CDI is the quantized version of the original channel direction, which is determined using codebooks [cf. Section~\ref{section:precoding:partial-csit}]. The transmitter uses both indicators for scheduling [cf. Section~\ref{section:algorithms:selection-partial-csit}], and the CQI is particularly used for power control, link adaptation, and interference management \cite{Huang2012a}. The CSI feedback interval highly depends on the users' mobility,\footnote{The authors in \cite{Kobayashi2007a} showed that mobility defines the best reliable transmission strategy for capacity maximization, i.e., space-time coding or space division multiplexing. The results in \cite{Zhang2011} defined acceptable mobility ranges for MU-MIMO scenarios.} and even for short-range communications (e.g., WiFi), immediate feedback is needed to achieve and maintain good performance \cite{Jones2015}.
By considering high mobility and limited feedback rates, one cannot rely on instantaneous or full CSI. In such cases, the transmitters perform resource allocation based on \textit{statistical} CSI, which vary over larger time scales than the instantaneous CSI. The statistics for the downlink and uplink are reciprocal in both FDD and TDD, which can be used to perform resource allocation, see \cite{Gao2009} and references therein.
Table~\ref{tab:channel-information} summarizes the different types of CSIT in MU-MIMO systems, and the antenna configuration at the receivers. Partial CSIT refers to quantized channel information, which will be elaborated upon Sections \ref{section:precoding:partial-csit} and \ref{section:algorithms:selection-partial-csit}.

\begin{table}[t!] \scriptsize
	\renewcommand{\arraystretch}{1.3}	
	\caption{Summary of the Type of CSI at the Transmitter for MISO ($N=1$) and MIMO ($N>1$) Configurations}
	\label{tab:channel-information}%
	\centering
	\begin{tabularx}{0.48\textwidth}{l *{2}{>{\centering\arraybackslash}X} } 
		\toprule
		&  \textbf{MISO}  &  \textbf{MIMO} \\
		\midrule
		\textbf{Full CSIT} & \cite{Dimic2005, Huang2013, Tu2003, Wang2008a, Yoo2006,  Jiang2006, Dai2009, Lee2014, Huang2012b, Razi2010, Mao2012, Bjornson2014, Shi2012, Maciel2010, Chung2010, Driouch2012, Yoo2005a, Christopoulos2013, Xia2009, Matskani2008, Lau2005a, Cottatellucci2006, Wang2015, Yang2011, Koutsopoulos2008, Kobayashi2007, Boccardi2006, Destounis2015, Zhang2007b, Swannack2004, Tsai2008, Shirani-Mehr2011, Hammarwall2007, Seifi2011, Castaneda2015, Huh2012, Jang2011, Park2014, Ku2015, Song2008, Tu2003, Maciel2010,Bogale2015,Liu2015a} &  \cite{Bayesteh2008, Shen2006, Wang2010, Bayesteh2008, Ko2012, Elliott2009, Chan2007, Tran2010, Wang2010a, Nam2014a, Torabzadeh2010, Souihli2010, Moretti2013, Zhang2009, Cui2011a, Yu2013, Park2010, Lossow2013,  Jagannathan2007, Chen2007a, Fuchs2007, Tejera2006, Zhang2005a, Wang2006, Chen2008, Sigdel2009a, Yi2011, Sun2010, Tran2012, Elliott2012, Hei2009, Lim2009, Cheng2014, Aniba2007} \\
		\midrule
		\textbf{Statistical} &  \cite{Liu2014a, Nam2014, Lee2014, Adhikary2013, Adhikary2013a, Stridh2006a,Kountouris2006,Kountouris2006b, Hammarwall2008,Hammarwall2008a, Dartmann2013,YiXu2014,Lee2014c,Liu2015a}  & \cite{Raghavan2007,Fuchs2007, Wang2006, Rico-Alvarino2014, Fan2014,Gao2009} \\
		\midrule
		\textbf{Outdated} & \cite{Liu2014a,Driouch2012, Lau2009, Kobayashi2007,Kobayashi2007a, Tang2010, Shirani-Mehr2010, Zhang2011,Liu2015a} & \cite{Wang2006} \\
		\midrule
		\textbf{Correlated} & \cite{Kountouris2006b,Xu2009a,Weich2006,Gao2009, Raghavan2015,Liu2014a,Liu2015a} & \cite{Schellmann2010,Chen2007a,Raghavan2007,Weich2006,Tran2012,Wang2006,Chen2008,Xu2009a,Bjornson2009a,Bjornson2010a,Godana2013a,Rao2013} \\
		\midrule
		\textbf{Partial CSIT} & \cite{ Moon2013, Xia2009, Wagner2008,Yang2011,Jindal2006, Zorba2008, Conte2010, Tang2010, Huang2009a, Vicario2008, Yoo2007, Kountouris2007, Kountouris2008, Dai2008, Kountouris2006, Zhang2011, Ravindran2012, Kountouris2006a, Wang2007a, Choi2007c, Sohn2010, Kountouris2008a, Kountouris2005a, Kountouris2005, Xu2010, Khoshnevis2013, Simon2011, Zakhour2007, Huang2012a, Huang2007} & \cite{Raghavan2007,Chae2008, Chen2013, Trivellato2008, Jindal2008, Nam2014a, VanRensburg2009, Hosein2009, Schellmann2010,Min2013, Sohn2012,  Zhang2007a, Fan2014, Wang2012d} \\
		\bottomrule
	\end{tabularx}%
\end{table}%

{\color{black}{
\section{MU-MIMO Channel and System Models}\label{section:system-set-ups}
}}

\begin{figure}[b!]
	\centering
	\includegraphics[width=0.98\linewidth]{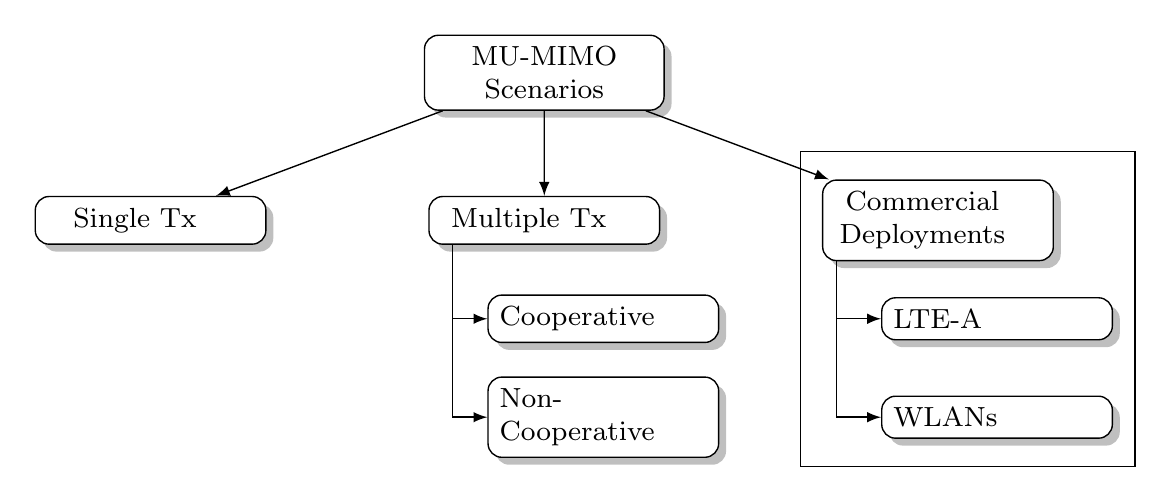}\\
	\caption{Classification of MU-MIMO Scenarios and two examples of commercial technologies} \label{fig:generic-deployments}
\end{figure}

The signal processing and scheduling algorithms described in the following sections have been developed and studied for single-hop MU-MIMO scenarios. We have classified the scenarios in two groups, see Fig.~\ref{fig:generic-deployments}: single transmitter and multiple transmitters scenarios.

The implemented resource allocation strategies, user scheduling, and signal processing techniques depend on the number of coordinated transmitters, the number of antennas ($M$ and $N$), the number of users ($K$), the SNR regime, and the CSIT accuracy. The system optimization relies on close-loop (e.g. \cite{Vu2007, Mietzner2009}) or open-loop (e.g. \cite{Destounis2015}) feedback,  to achieve spatial multiplexing gains, multiuser diversity gains, and to combat interference.
In cellular systems, there are two main sources of interference \cite{Andrews2005}: other active devices in the same co-channel and same cell, i.e., intra-cell or IUI; and from transmissions in other cells, i.e., inter-cell interference (ICI). The techniques to mitigate IUI and ICI depend on the type of scenario and the optimization criterion. There exist a number of scenarios where the interference cannot be reduced, see \cite{Jindal2006,Lozano2013}, whose characteristics are described in the following definition:

\begin{definition}\label{defn:interference-limited-system}
	\textit{Interference-limited system}. An MU-MIMO system is said to be interference limited if the performance metric saturates (ceiling effect) with the transmit SNR. This might occur due to CSIT inaccuracy, highly correlated multiuser channels (IUI), and irreducible ICI.
\end{definition}

\subsection{Scenarios with a single transmitter}
\label{section:system-set-ups:single-cell-scenarios} 

The objective of MU-MIMO processing is to accommodate many users per resource. Therefore, resource allocation strategies are commonly analyzed at the basic resource unit, e.g., code, single-carrier, time-slot, or frequency-time resource block. This can be done regardless the global system model (single-carrier, OFDM, or CDMA), since the same resource allocation strategy is applied over all resources, e.g., \cite{Lau2006,Fuchs2007,Koutsopoulos2008,Vicario2008,Tsai2008,Conte2010, Driouch2012, Huang2013, Cheng2014, Fan2014,Bogale2016a}. We adopt a signal model using the most general approach in the reviewed references.  
Consider a scenario where the transmitter is equipped with $M$ antennas, and $K$ active users are equipped with $N$ antennas. Let $\mathbf{H}_{k} \in \mathbb{C}^{N \times M}$, be the discrete-time complex baseband MIMO channel  of the $k$-th user for a given carrier. The received signal can be expressed as:

\begin{equation}\label{eq:received_signal}
\mathbf{y}_{k} = \mathbf{H}_{k}\mathbf{x} + \mathbf{n}_{k}  
\end{equation}
where $\mathbf{x} \in \mathbb{C}^{M \times 1}$, is the joint transmitted signal for all users.
The MIMO channel is usually assumed to be ergodic, i.e., it evolves over time and frequency in an independent and identically distributed (i.i.d.) manner. The channel is commonly modeled as \textit{Rayleigh fading}, which is suitable for non line-of-sight communications. The complete spatial statistics can be described by the second-order moments of the channel \cite{Raghavan2015}. Define the channel covariance matrix as $\mathbf{\Sigma}_{k} = \mathbb{E}[\mathbf{H}_{k}^{H} \mathbf{H}_{k}]$, which depends on the antenna configuration, propagation environment, scattering conditions and mobility. The channel can be decomposed as $\mathbf{H}_{k} = \sqrt{\mathbf{\Sigma}_{k}} \mathbf{H}_{iid}$, where $\mathbf{H}_{iid}$ has i.i.d. entries with distribution $\mathcal{CN}(0,1)$.

Assuming spatially uncorrelated Rayleigh fading channels, i.e.,  $\mathbf{\Sigma}_{k} = \gamma \mathbf{I}$, $\forall k$ and some $\gamma > 0$, is the most common practice in the literature \cite{Biglieri1998}. Physically, this implies rich scattering environments with sufficient antenna spacing at both ends of the radio link \cite{Foschini1998}. Under these conditions, the fading paths between the multi-antenna transmitter and receiver become independent. The MU-MIMO channels are modeled as narrow band, experiencing frequency-flat (constant) fading, where there is no inter-symbol interference \cite{Mietzner2009}. A common simplification used in OFDM broadband systems, is to assume multiple flat-fading sub-channels \cite{Lau2006,Wang2007}. The received signal per sub-channel can be modeled as  (\ref{eq:received_signal}), avoiding frequency selectivity \cite{Ho2009}. More realistic channel models for broadband MIMO systems and their performance analysis were proposed in \cite{Couillet2011}.
The Rayleigh model can be thought as a particular case of the asymmetric \textit{Ricean fading} channel model. In this case, the entries of $\mathbf{H}_{iid}$ have non-zero mean and there exists a dominant line-of-sight (LoS) component that increases the average SNR \cite{Goldsmith2005}. 
{\color{black}{Such a model is less common in MU-MIMO systems operating at microwave (i.e., sub-6 GHz bands), since the channels become more static and the benefits of MUDiv vanish with the magnitude of the LoS paths \cite{Hammarwall2008a}.
In recent years, wireless communication over the millimeter wave (mmWave) frequency range (30-300 GHz) has proved to be a feasible and reliable technology with a central role to play in 5G \cite{Rappaport2014,Heath2016,Bogale2016}. The mmWave technology rely on directional antennas to overcome propagation loss, penetration loss, and rain fading. High directivity implies that Ricean fading channels can be used to characterize both LoS and non-LoS components present in the mmWave channels \cite{Bai2015,Shokri-Ghadikolaei2015,Kutty2016,Bogale2016}. 		
}}

Several authors model the MIMO channel such that the  correlation at transmit and receive antennas is distinguishable, and $\mathbf{\Sigma}_{k} \neq \gamma \mathbf{I}$, $\forall k$, see references in Table~\ref{tab:channel-information}. There are two approaches to model and analyze performance under MIMO correlation, the \textit{jointly correlated model} \cite{Weich2006,Gao2009} and the simplified \textit{Kronecker model} \cite{Jorswieck2006,Hanlen2012}. The former assumes separability between transmit and receive eigen-directions, and characterizes their mutual dependence. The latter assumes complete correlation separability between the transmitter and receiver arrays, see \cite{Shiu2000,Raghavan2010,Bjornson2010a,Hanlen2012} and references therein.\footnote{Experimental results in \cite{Czink2009} and theoretical analysis in \cite{Raghavan2010} have shown that the conventional Kronecker model may not be well suited for MU-MIMO scenarios resulting in misleading estimates for the capacity of realistic scattering environments. This occurs due to the sparsity of correlated channel matrices, and the fact that the parameters of the Kronecker model change with time and position. The model can be used in scenarios with particular conditions on the local scattering at the transmitter and receiver sides \cite{Jorswieck2006}.}  We refer to \cite{Couillet2011} for a comprehensive analysis of Rayleigh and Rician correlated MU-MIMO channels. Note that in some MIMO propagation scenarios with uncorrelated antennas, the MIMO capacity can be low as compared to the SISO one due to the \textit{keyhole} or \textit{pinhole} effect. This is related to environments where rich scattering around the transmitter and receiver leads to low correlation of the signals, while other
propagation effects, like diffraction or waveguiding, lead to a rank reduction of the transfer function matrix \cite{Keyhole2002,Keyhole2006}.

The MIMO channels $\mathbf{H}_{k} \ \forall k$, may also include large-scale fading effects due to shadowing and path loss \cite{Goldsmith2005,Tse2005}. Depending on the type of access technique (OFDMA, CDMA, or TDMA), the channel model can take into account multi-path components, correlation, Doppler spread, and angular properties \cite{Koutsopoulos2008}.  Another common assumption to avoid frequency dependency (particularly in low mobility scenarios), is to account for \textit{block-fading channels} \cite{Biglieri1998}. This means that the CSI is constant (within a coherence time duration) for a block of consecutive channel uses before changing independently for the next block.

The noise $\mathbf{n}_{k} \sim \mathcal{CN}(\mathbf{0}, \mathbf{I}_{N})$, is usually modeled by i.i.d. normalized entries according to a circular normal (complex) Gaussian distribution with zero mean and unit variance \cite{Love2008}. 
The transmitted signal, $\mathbf{x}$, can be defined according to the encoding applied over the user data, the number of spatial streams per user, and the power allocation. If linear precoding is used [cf. Section~\ref{section:precoding}], the transmitted signal is defined as

\begin{equation}\label{eq:transmitted_signal}
\mathbf{x} = \sum_{k=1}^{K} \mathbf{W}_{k}\mathbf{d}_{k} 
\end{equation}
where $\mathbf{W}_{k} \in \mathbb{C}^{M \times d_k}$, is the precoding matrix, $\mathbf{d}_{k} \sim \mathcal{CN}(\mathbf{0}, \mathbf{I}_{d_k})$ is the data signal, and $d_k$ is the number of multiplexed data streams of user $k$.
In single-transmitter  MU-MIMO scenarios with signal models defined by (\ref{eq:received_signal}) and (\ref{eq:transmitted_signal}), the ICI is negligible or assumed to be part of the additive background noise. Therefore, IUI is the main performance limiting factor, which can be addressed by precoding the user data, [cf. Section~\ref{section:precoding}].

The active users might experience similar average long-term channel gains (large-scale fading or path loss) and SNR regimes. For some practical cellular systems, assuming homogeneous users is valid if open-loop power control is used to compensate for cell-interior and cell-edge path losses. Therefore, the resultant effective multiuser channels have quasi-identical variances  \cite{Huang2012a,Destounis2015}. The user distribution affects the fading statistics of $\mathbf{H}_k$ $\forall k$, and the general design of resource allocation algorithms \cite{Jorswieck2006}.

A possible single-transmitter MU-MIMO scenario arises in satellite communications. However, due to the characteristics of the satellite channels, marginal MIMO gains can be realized. The absence of scatterers in the satellite vicinity yields a Rician-type channel with a strong line-of-sight component, turning off the capabilities of MIMO processing. Due to the large coverage area in satellite communications, the users have heterogeneous long-term channel gains, which directly affects the resource allocation decisions. Regardless of the limited literature about MU-MIMO satellite communications, recent works show promising results and discussions about how intensive frequency reuse, user scheduling, and multibeam signal processing will be implemented in next generation broadband satellite systems \cite{Arapoglou2011,Zorba2008,Cottatellucci2006}.

\subsection{Scenarios with multiple transmitters}
\label{section:system-set-ups:multi-cell-scenarios}

In this scenarios, the channel models and assumptions aforementioned are applied. Yet, some additional considerations are made so that signaling and connections between network entities can be modeled.   
Deploying several transmitters across a geographic area can provide reliable communication for heterogeneous mobile terminals (different path losses or SNR regimes relative to each transmitter). This kind of infra-structure based systems include cellular and wireless local area networks (WLAN). 
The resource allocation and access control can be performed based on CSI and knowledge of the interference structure \cite{Osseiran2011}. If the transmitters are allowed to cooperate, e.g., through a central processing unit (CU), IUI can be mitigated using CSIT for signal design [cf. Sections~\ref{section:introduction:needof-csit} and \ref{section:precoding}]. Global knowledge or estimation of the interference can be used to avoid poor spectral efficiency or inaccurate assignment of radio resources.

This paper focuses on scenarios where roaming (mobility) and  reuse of resources are central management tasks. The premise behind cellular communications, is to exploit the power falloff with distance of signal propagation to reuse the same channel at spatially-separated locations. This means that the serving area is divided in non overlapping cells. Any cell site within a neighborhood cannot use the same frequency channel, which makes the same reused frequency channels sufficiently far apart \cite{Gesbert2010,Nguyen2014}. 
In traditional cellular systems, a given user belongs to only one cell at a time and resource allocation is performed unilaterally by its serving {BS} (non-cooperative approach in Fig.~\ref{fig:generic-deployments}). 
Each transmitter serves its own set of users, transmission parameters are adjusted in a selfish manner by measuring ICI (simple interference-awareness), and there is no information exchange between BSs \cite{Nguyen2014}.
If frequency reuse is employed, the {BS} can make autonomous resource allocation decisions and be sure that no uncoordinated {ICI} appears within the cell \cite{Bjornson2013}. However, in many practical systems, universal frequency reuse is applied, which means that neighboring cells can access the same frequencies and time-slots simultaneously. This might increase the ICI and potentially degrade performance \cite{Dahlman2013}. 
The mitigation of {ICI} is a fundamental problem since the transmit strategy chosen by one {BS} will affect the reception quality of the users served by adjacent {BS}s.

A cluster of BSs can coordinate the resource allocation, scheduling decisions, and ICI mitigation techniques  (cooperative approach in Fig.~\ref{fig:generic-deployments}). Dynamic clustering is an ongoing research topic (see \cite{Gesbert2010,Bjornson2013,Nguyen2014,Li2014} and references therein), which promises to meet the requirements established in the third generation partnership project ({3GPP}) standards \cite{Dahlman2013}.
Different forms of {ICI} control have been proposed over the last years. Extensions of space multiple access techniques for multi-cell systems have received several names, \textit{coordinated multi-point} ({CoMP}) \cite{Marsch2011,Lee2012a,Dahlman2013}, \textit{network {MIMO}} \cite{Zhang2009}, or  \textit{joint signal transmission/processing} ({JT}) \cite{Nguyen2014}. These techniques exploit the spatial dimensions, serving multiple users (specially cell edge users \cite{Bjornson2013}), while mitigating {ICI} of clustered {BS}s.\footnote{Theoretical analysis in \cite{Lozano2013} shows that in the high {SNR} regime, the achievable system capacity is fundamentally interference-limited due to the out-of-cluster interference. This occurs regardless the level of coordination and cooperation between clustered BSs. However, coordinated scheduling and user clustering can provide means to improve spectral efficiency and mitigate interference at high SNR.} 
For these approaches, a cluster can be treated as a super cell, for which mathematical models from the single-cell scenario can be applied straightforwardly, e.g. \cite{Liu2011,Khoshnevis2013,Marsch2011}.

If user data is shared among {BS}s, the use of \textit{proactive} interference mitigation within a cluster can take place. This implies that coordinated BSs do not separately design their physical (PHY) and media access control layer parameters. Instead, the BSs coordinate their coding and decoding, exploiting knowledge of global data and {CSI} \cite{Gesbert2010}. However, to guarantee large performance gains for these systems, several conditions must be met \cite{Bjornson2013,Nguyen2014}: global {CSI} and data sharing availability, which scales up requirements for  channel estimation, backhaul capacity, and cooperation; coherent joint transmission and accurate synchronization; and  centralized  resource allocation algorithms, which may be infeasible in terms of computation load and scalability.

There is another approach of multi-cell cooperation, coined as \textit{coordinated scheduling} (CS) with \textit{coordinated beamforming} ({CBF}), which is a form of coordinated transmission for interference mitigation \cite{Dahlman2013,Nguyen2014}.  CS/CBF refers to the partial or total sharing of {CSI} between {BS}s to estimate spatial signaling, power allocation, and scheduling without sharing data or performing signal-level synchronization \cite{Gesbert2010}.  {CBF} implies that each {BS} has a disjoint set of users to serve, but selects transmit strategies jointly with all other {BS}s to reduce {ICI}. In this approach, exchange of user data is not necessary, but control information and {CSI} can be exchanged to simultaneously transmit to a particular set of users \cite{Lee2012a}.
CBF is more suitable than JT for practical implementations, since it requires less information exchange. Nevertheless, CSI acquisition, control signaling, and coordinated scheduling are challenging tasks due to limited feedback bandwidth and finite capacity backhaul \cite{Li2014}.

\subsection{Commercial Deployments}
\label{section:system-set-ups:practical-deployments}

This paper covers two wireless technologies, whose specifications already support MU-MIMO communications:  LTE-Advanced for cellular networks and IEEE 802.11ac for wireless local area networks (WLAN).

$\bullet$ \textit{LTE-Advanced}: This is the 3GPP cellular system standard for 4G and beyond communications \cite{Dahlman2013}. Several capabilities have been added to the LTE standard to increase capacity demands and integrated a large number of features in the access network. Among these attributes, the ones related to MIMO processing are the most relevant in the context of this paper: enhanced downlink MIMO, multi-point and coordinated transmission schemes, and multi-antenna enhancements. Due to the fact that LTE is a cellular technology, most of the studied deployments in the literature lie in the category of multiple transmitters scenarios, see Fig.~\ref{fig:generic-deployments}.

MU-MIMO communication has been incorporated in  LTE  with the following maximum values: four users in MU-MIMO configuration, two layers (spatial streams) per user, four simultaneous layers, and robust CSI tracking. Practical antenna deployments at the transmitter use dual-polarized arrays, and the expected number of co-scheduled users for such configurations will be two for most cases \cite{Lim2013,Fan2014}. LTE provides a mechanism to improve performance by switching between MU or SU-MIMO mode on a per sub-frame basis, based on CSI, traffic type, and data loads. The goal of dynamic mode switching is to balance the spectral efficiency of cell edge and average cell users. This can be achieved, for instance, by using transmit diversity for users at the cell edge, or implementing spatial multiplexing for cell center users \cite{Liu2012}.

$\bullet$ \textit{IEEE 802.11ac}: This WLAN standard supports multiuser downlink transmission, and the number of simultaneous data streams is limited by the number of antennas at the transmitter. In MU-MIMO mode, it is possible to simultaneously transmit up to 8 independent data streams and up to 4 users \cite{Kim2015,Jones2015}. Spatial multiplexing is achieved using different modulation and coding schemes per stream. MU-MIMO transmission prevents the user equipment with less antennas to limit the achievable capacity of other multiple antenna users, which generates rate gains for all receivers. A unique \textit{compressed explicit feedback} protocol, based on channel sounding sequences, guarantees interoperability and is used to estimate CSI and define the steering matrices (beamforming). Other methods for channel estimation are described in \cite{Liao2016} and references therein. 
Compared to cellular networks, WLANs usually have fewer users moving at lower speeds, the APs are less powerful that the BSs and the network topology is ad hoc. Although multiple APs could be deployed, most of the works in the literature focus on MU-MIMO systems with a single transmitter.
{\color{black}{Nonetheless, joint transmission from several APs to different mobile users is feasible in WLANs, which requires coordinated power control, distributed CSI tracking, as well as synchronization in time, frequency, and phase. The authors in \cite{Hamed2016} have shown that distributed MIMO can be achieved by enhancing the physical layer for coordinated transmission, and by implementing time-critical functions for the media access control layer.
It is likely that in the next generation of 802.11 standard, coordination schemes between APs will be adopted to enable MU-MIMO communications \cite{Liao2016}.
}}

\section{MU-MIMO Precoding}
\label{section:precoding}

The spatial dimension provided by the multi-antenna transceivers can be used to create independent channelization schemes. In this way, the transmitter serves different users simultaneously over the same time slot and frequency band, which is known as space-division multiple access (SDMA) \cite{Godara2002Ch18, Tse2005, Gesbert2007a, Mietzner2009}.
The spatial steering of independent signals consists of manipulating their amplitude and phases (the concept of \textit{beamforming} in classic array signal processing), in order to add them up constructively in desired directions and destructively in the undesired ones \cite{Mietzner2009,Bjornson2013}. By jointly encoding all (co-resource) signals using channel information, it is possible to increase the signal-to-interference-plus-noise ratio (SINR) at the intended receiver and mitigate interference for non-intended receivers.

In the literature of MU-MIMO systems, the term beamforming refers to the signal steering by means of beams to achieve SDMA. The term \textit{precoding} is used to denote the scaling and rotation of the set of beams, so that, their power and spatial properties are modified according to a specific goal. Hereinafter, we use the term precoding\footnote{Some authors denote as precoding all processing techniques over the transmitted signals, which achieve multiplexing or diversity gains, i.e., both space-time coding and beamforming \cite{Shirani-Mehr2010}.} to describe the signal processing (i.e., beam vector/matrix computation, scaling, rotation, and projection), applied to the independent signals prior to transmission. 
In this section we describe the most common precoding techniques used in MU-MIMO scenarios, and their characteristics according to CSIT. Table~\ref{tab:precoding} summarizes the precoding schemes used in the surveyed literature, as well as their associated methods for user selection, which will be elaborated upon in the following sections. 
An important performance metric determined by the precoders $\mathbf{W}_{k}$, $\forall k$, related to the delivered energy through the MU-MIMO channels, is provided in the following definition.

\begin{definition}\label{defn:effective_channel_gains}
	\textit{Effective channel gain}. It is the magnitude of the channel projected onto its associated precoding weight. Let $\mathbf{H}_{k}^{(eff)} = \mathbf{H}_{k} \mathbf{W}_{k}$, be the \textit{effective channel} after spatial steering, thus, the effective channel gain is given by: \textit{i}) $|\mathbf{H}_{k}^{(eff)}|^2$ for MISO scenarios; \textit{ii}) for MIMO scenarios, it is given by a function of the eigenvalues of $\mathbf{H}_{k}^{(eff)}$:  $\det \left(  \mathbf{H}_{k}^{(eff)}( \mathbf{H}_{k}^{(eff)} )^ {H} \right)$ or $\| \mathbf{H}_{k}^{(eff)}\|_{F}^{2}$.
\end{definition}

\begin{table*}[t] \scriptsize
	\renewcommand{\arraystretch}{1.3}	
	\caption{Precoding Schemes and their associated Scheduling Methods}
	\label{tab:precoding}%
	\centering		
	\begin{tabularx}{\textwidth}{ *{6}{>{\centering\arraybackslash}X} }  
		\toprule	
		\textbf{Method} &   \textbf{Utility-based}    &   \textbf{CSI-Mapping}    & \textbf{Metaheuristic}\newline \textbf{(stochastic)} & \textbf{Classic}\newline \textbf{Optimization}& \textbf{Exhaustive} \textbf{Search} \\
		\midrule
		\textbf{DPC}   & \cite{Boccardi2006}  &   -    & \cite{Elliott2009}  &   -    & - \\		
		\midrule
		\textbf{THP}   & \cite{Boccardi2006}  &   -    &   -    &   -    &  - \\
		\midrule
		\textbf{VP} &  \cite{Boccardi2006}  & \cite{Razi2010}  &    -   &   -    & - \\
		\midrule
		\textbf{MRT}   & \cite{Moon2013, Zhang2011}  & \cite{Wang2015}  & \cite{Bjornson2014}  &   -    & - \\
		\midrule
		\textbf{ZFBF}  & \cite{Dimic2005, Moon2013, Huang2013, Wang2008a, Huang2012b, Zhang2005a, Lau2005a, Chan2007, Conte2010, Chae2008, Kountouris2007, Shirani-Mehr2010, Zhang2011, Ravindran2012, Jindal2008, Boccardi2006, Tsai2008, Huh2012, Park2010, Lossow2013,Hammarwall2008,YiXu2014,Bogale2015}  & \cite{Wang2008a, Yoo2006, Nam2014, Bayesteh2008, Adhikary2013, Adhikary2013a, Wang2010, Mao2012, Shi2012, Maciel2010, Zhang2005a, Chung2010, Driouch2012, Yoo2005a, Wang2006, Sun2010, Wang2015, Yang2011, Conte2010, Huang2012a, Chae2008, Yoo2007, Kountouris2007, Dai2008, Kountouris2006a, Wang2007a, Trivellato2008, Min2013, Sohn2010, Souihli2010, Swannack2004, Shirani-Mehr2011, Xu2010, Sohn2012, Khoshnevis2013, Castaneda2015, Jang2011,Hammarwall2008,Lee2014c,Liu2015a}  & \cite{Bjornson2014, Lau2005a, Cottatellucci2006}   & \cite{Chan2007}   & \cite{Chan2007, Zakhour2007, Destounis2015, Swannack2004,Bogale2015}   \\
		\midrule
		\textbf{ZFDP} & \cite{Dimic2005, Tran2010}   & \cite{Tu2003, Jiang2006, Dai2009, Tejera2006, Sigdel2009a, Sun2010}   &    -   &   -    &  - \\
		\midrule
		\textbf{SZF}   &    -   & \cite{Zorba2008}  & \cite{Elliott2012}   &   -    &  -  \\
		\midrule
		\textbf{CIZF}  &   -    & \cite{Christopoulos2013}  &    -   &   -    &  - \\
		\midrule
		\textbf{BD}    & \cite{Shen2006, Chen2007a, Chen2008, Ko2012, Tran2012, Lim2009, Cheng2014, Rico-Alvarino2014, Zhang2009,YiXu2014}   &  \cite{Nam2014, Shen2006, Lee2014, Adhikary2013, Adhikary2013a, Chen2007a, Fuchs2007, Wang2006, Sigdel2009a, Yi2011, Lim2009, Wang2010a, Nam2014a}  & \cite{Hei2009}   & \cite{Chan2007, Moretti2013}  & - \\
		\midrule
		\textbf{MMSE/SLNR}  &  \cite{Zhang2011, Torabzadeh2010, Cui2011a, Seifi2011,Liu2014a}  & \cite{Zorba2008, Lau2009, Kountouris2005a,Kountouris2006b}, \cite{Xia2009, Castaneda2015}, \cite{Adhikary2013}  & \cite{Cottatellucci2006}   &   -    &  - \\
		\midrule
		\textbf{Adaptive}  & \cite{Koutsopoulos2008, Kobayashi2007, Boccardi2006, Hammarwall2007, Dartmann2013, Yu2013}   & \cite{Zhang2007b}  &   -    &  \cite{Stridh2006a, Matskani2008, Ku2015, Song2008}  &  - \\
		\midrule
		\textbf{Codebook-based} & \cite{Fan2014, Wang2012d, Tang2010, Vicario2008, Kountouris2006, Ravindran2012}  & \cite{Kountouris2006b,Lee2014, Yang2011, Huang2009a, Kountouris2008, Chen2013, Zhang2007a, Choi2007c, Kountouris2008a, Kountouris2005a, Kountouris2005, Huang2007, VanRensburg2009, Hosein2009, Schellmann2010, Simon2011,Xu2009a}&  -     &   -    &  - \\
		\bottomrule
	\end{tabularx}%
\end{table*}%

\subsection{Non-linear Precoding with full CSIT}
\label{section:precoding:non-linear}

From an information-theoretic perspective, the optimal transmit strategy for the MU-MIMO BC is \textit{dirty paper coding} ({DPC}) \cite{Costa1983}, and theoretical results showed that such strategy  achieves the entire BC capacity region \cite{Caire2003,Weingarten2006}. The principle behind this optimum coding technique is that the transmitter knows the interference for each user. Therefore, interference can be pre-subtracted (from the information theoretical standpoint) before transmission, which yields the capacity of an interference free channel. DPC is a non-linear process that requires successive encoding and decoding, whose performance depends on the particular sequential order $\pi(\cdot)$ assigned to the co-scheduled users \cite{Caire2006}. Although the implementation complexity of DPC for practical systems is prohibitively high, it establishes the fundamental capacity limits for MU-MIMO broadcast channels \cite{Weingarten2006,Hassibi2007}. 

Suboptimal yet more practical non-linear precoding schemes have been proposed as an alternative to DPC \cite{Boccardi2006}. The error rate and interference can be minimized at the symbol level by the Tomlinson-Harashima precoding (THP),\footnote{The application of THP in MU-MIMO has been of particular interest in recent research on multibeam satellite communications \cite{Arapoglou2011}.} which is not limited by the number of transmit or receive antennas \cite{Tse2005}.  By modifying or perturbing the characteristics of the transmitted signal, the power consumption can be minimized (compared to traditional channel inversion filtering), using the non-linear vector perturbation (VP) scheme \cite{Hochwald2005, Kurve2009}. This technique requires a multidimensional integer-lattice least squares optimization, whose solution can be found by several approaches.

\subsection{Linear Precoding with full CSIT}
\label{section:precoding:linear}

Linear precoding is a generalization of traditional SDMA \cite{Gesbert2007a}, which matches the signal on both ends of the radio link. This is attained by decoupling the input data into orthogonal spatial beams and allocating power according to CSIT \cite{Vu2007}. The precoders weights are  vectors or matrices jointly designed at the transmitter, according to several parameters: the type of CSIT, the coding order, the performance metric (e.g., mean-square error (MSE), error probability, power consumption, or achievable SNR), and the system constraints (e.g., power and QoS). 
The optimal precoder technique (linear filtering in the spatial domain), would be able to balance between signal power maximization and interference power minimization \cite{Larsson2008,Jorswieck2008}. Precoding design subject to general constraints can be performed using standard optimization techniques (see \cite{Hong2012} for a comprehensive survey) or by heuristic approaches, e.g., maximizing the signal-to-leakage-plus-noise ration (SLNR\footnote{SLNR is also known in the literature as the transmit Wiener filter, transmit MMSE beamforming, or virtual SINR beamforming \cite{Bjornson2014b}. }) \cite{Sadek2007, Bjornson2013}.
However, determining the optimal precoders is an NP-hard problem for many performance metrics \cite{Liu2011,Weeraddana2012}, whose evaluation is performed by computationally demanding algorithms \cite{Schubert2004,Christensen2008,Godara2002Ch18}. Therefore, many works in the literature focus on more suboptimal, yet practical, schemes that can achieve spatial multiplexing gains with low computational complexity.  
A large number of linear precoding techniques have been developed, for which having more transmit than receive antennas, i.e., $M\geq N$, is a condition required in most cases. There are some particular precoding schemes (e.g. SLNR) that can be implemented if $M < N$, but the system becomes interference limited in the moderate and high signal-to-noise (SNR) regimes, [cf. Definition~\ref{defn:interference-limited-system}]. Therefore, power control and recursive adaptation of the precoders are mandatory to operate in those SNR regimes \cite{Sadek2007}.

In a MISO antenna configuration, the matched filter or maximum ratio transmission (MRT) precoding maximizes the signal power at the intended users. This is performed by projecting the data symbol onto the beamforming vector given by the spatial direction of the intended channel \cite{Bjornson2013}. A similar precoding scheme for MIMO scenarios is the singular value decomposition (SVD) beamforming \cite{Tse2005}, which uses the eigenvectors of the channel as beamforming weights.
The zero-forcing beamforming\footnote{Precoding based on ZFBF obtains a multiplexing gain of $M$ and can asymptotically achieve the DPC performance when $K \rightarrow \infty$ \cite{Yoo2006}.} (ZFBF) \cite{Caire2003}, also known as channel inversion precoding \cite{Peel2005}, completely suppresses IUI in MISO scenarios. This technique is based on prefiltering the transmit signal vector by means of the Moore-Penrose inverse \cite{Gentle2007}. An extension of ZFBF for MIMO scenarios is the block diagonalization (BD) precoding, where multiple data streams can be transmitted per user \cite{Spencer2004a}.

In MISO scenarios, the regularized channel inversion (often coined as MMSE precoding) enhances ZF, taking into account the noise variance to improve performance in the low SNR regime \cite{Peel2005}. An extension of MMSE for the general MIMO scenario is the regularized BD \cite{Stankovic2008}.
Zero-forcing dirty paper (ZFDP) coding \cite{Caire2003}, is a technique designed for MISO settings. For a given user $k$, ZFDP suppresses the interference coming from the next encoded users $\{k+1,\ldots,K\}$, combining QR decomposition \cite{Gentle2007} and DPC. Extensions of ZFDP for the MIMO scenarios were defined in \cite{Dabbagh2007,Spencer2004a} (an iterative SVD method), and  \cite{Sun2010} (combining ZF, DPC, and eigen-beamforming). Successive zero-forcing (SZF) was proposed in \cite{Dabbagh2007}, for MIMO settings. SZF partially suppresses IUI, by encoding users similar to ZFDP, but DPC is not applied in the encoding process.
The generalization of precoding schemes based on ZF for multiple antenna receivers is not trivial. This is because applying MISO decompositions methods to the MIMO channel is equivalent to treating each receive antenna as an independent user. This process does not completely exploit the multiplexing and diversity gains of MIMO systems \cite{Dabbagh2007}.

The aforementioned precoding schemes can be classified as user-level precoders, i.e., independent codewords intended to different users are transmitted simultaneously. There is another class of precoding schemes, where simultaneous transmitted symbols are addressed to different users \cite{Alodeh2015}. This class of symbol-level precoding, e.g. constructive interference zero forcing (CIZF) precoding \cite{Masouros2009,Alodeh2015}, has been developed for MISO settings. CIZF constructively correlates the interference among the spatial streams, rather than decorrelating them completely as in the case of user-level precoding schemes.

\subsection{Precoding with partial CSIT}
\label{section:precoding:partial-csit}

In the literature of limited feedback systems, CSIT acquisition relies on {\color{black}{collections of predefined codewords (vector or matrix weights) or codebooks,}} that are known a priori at the transmitter and receiver sides. The codewords can be deterministic or randomly constructed, which defines the type of signal processing applied to achieve spatial multiplexing and interference mitigation \cite{Love2008}:

$\bullet$ \textit{Channel quantization and precoding}: The codebook $\mathcal{C} = \{ \mathbf{c}_{1}, \ldots, \mathbf{c}_{b}, \ldots, \mathbf{c}_{2^B} \}$, is used by user $k$ to quantize its channel direction with $B$ bits. This means that each user feeds back the index $b$, of the most co-linear codeword to its channel, [cf. Section~\ref{section:metrics-spatial-compatibility:spatial-clustering}]. 
This is illustrated in Fig.~\ref{fig:hyperslab_cones}, where the user $k$ would report the index $b$, related to the cone where $\mathbf{H}_{k}$ has been clustered. The precoding weights are usually computed using ZFBF over the quantized channels. Due to quantization errors, the signals cannot be perfectly orthogonalized, and the sum-rate reaches a ceiling as the SNR regime increases \cite{Jindal2006}. In other words, resource allocation is performed over non-orthogonal spatial directions.

The optimal codebook design has not been fully solved in the literature. However, if the channel is assumed to have spatially i.i.d. entries, ($\mathbf{\Sigma}_{k} =  \mathbf{I}$, and homogeneous long-term channel gains, $\forall k$), off-line designs of $\mathcal{C}$ can be realized using different approaches, e.g., the Grassmannian design \cite{Love2003}, random vector quantization (VQ) \cite{Jindal2006}, quantization cell approximation (QCA) \cite{Mukkavilli2003},  and other techniques described in \cite{Love2008,Trivellato2008}. This sort of isotropically distributed codebooks achieve acceptable performance, since they mirror the statistical properties of the eigen-directions of $\mathbf{H}_{k}$, $\forall k$. For correlated channels ($\mathbf{\Sigma}_{k} \neq  \mathbf{I}$), non-uniform or skewed codebooks must be constructed, taking into account the statistical characteristics of the dominant eigen-directions of $\mathbf{\Sigma}_{k}$, $\forall k$, see \cite{Raghavan2007,Rao2013,Godana2013a,Raghavan2015}.

$\bullet$ \textit{Random beamforming (RBF)} \cite{Sharif2005}: The available precoding vectors,  $\mathcal{F} = \{ \mathbf{f}_{1}, \ldots, \mathbf{f}_{b}, \ldots, \mathbf{f}_{2^B} \}$, are constructed at the transmitter according to a known distribution,\footnote{The goal is to generate $2^B$ codewords that are i.i.d., according to the stationary distribution of the unquantized beamforming vector \cite{Love2008}.}  or by methods that yield random orthonormal basis in $\mathbb{C}^{M \times 1}$, \cite{Choi2007c}.  The transmitter sends pilot symbols to different user through the beams in $\mathcal{F}$, and each user feeds back the index $b$, of its best beam [cf. Section~\ref{section:algorithms:selection-partial-csit}]. RBF is an extension of the opportunistic beamforming scheme in \cite{Viswanath2002}, and attempts to sustain multiuser diversity over fading channels with partial CSIT. Random changes of amplitude and phase at each transmit antenna result in channels experiencing accelerated fluctuations, which enhances the MIMO processing gains. This approach is only effective when the number of users is large,\footnote{RBF asymptotically achieves the performance of downlink MIMO systems with full CSIT as $K \rightarrow \infty$. However, it does not achieve the full multiplexing gain at high SNR \cite{Hassibi2007}, and suffers from large quantization error as $M$ grows \cite{Kim2005a}.} $K \gg M$, and the number of antennas at the transmitter $M$ is moderate \cite{Choi2007c}. In some scenarios the performance can be enhanced by power control, or employing statistical channel information to compensate poor MUDiv.

We hasten to say that defining the optimum $B$, either for $\mathcal{C}$ or $\mathcal{F}$, is not a trivial optimization problem. In practice, only a tradeoff between feed back load and performance should be sought. Under mild conditions, $B$ can be minimized for certain codebook designs while guaranteeing diversity gains \cite{Castanheira2014a}. Large codebooks provide accurate CSIT at a price of factors, large feedback signaling overhead and memory requirements at the receiver, which increase exponentially with $B$ \cite{Wang2007a}.

\subsection{Precoding in Multiple Transmitters Scenarios}
\label{section:precoding:multiple-transmitters}

Recent research work have focused on transmission cooperation and coordination in multi-cellular and heterogeneous networks. The signal processing techniques performed at the transmitters depend on several aspects  \cite{Gesbert2010,Nguyen2014}: shared or non-shared user data; global or local CSIT; full or partial CSI; level of synchronization; per-transmitter power, backhaul, and delay  constraints; full or limited coordination; heterogeneous SNR regime across the receivers; and the number of clustered transmitters.

In multi-cell scenarios, assuming full CSIT and shared user data across the transmitter, the precoders can be computed by techniques described in Section~\ref{section:precoding:linear}. However, the power should be constrained on a per-transmitter basis, instead of global power allocation as in single-transmitter scenarios \cite{Silva2013}.
The precoding design can be oriented to optimize different objective functions, e.g. power minimization or sum-rate maximization. Optimizing such functions deals, in general, with non-convex problems, whose solutions are non-linear precoders \cite{Nguyen2014}. Precoding weights designed for common utility functions, e.g. weighted-sum-rate maximization or mean-square-error (MSE) minimization, can be realized via standard optimization techniques and sophisticated high-performance algorithms, see \cite{Hong2012,Kaviani2012} and references therein. 

However, a more pragmatic approach for MU-MIMO, is given by linear precoding schemes (e.g. ZFBF, SLNR, or BD), which can be extended for multi-cell systems and providing reliable and low-complexity solutions \cite{Zhang2009,Bjornson2010}. In \textit{network MIMO} scenarios, multi-cell BD precoding can remove ICI for the clustered cells using centralized processing at a central controller (full coordination) \cite{Zhang2009}.
In MISO multi-cell scenarios, linear precoding can be performed in a distributed fashion or with partial coordination. There are two fundamental schemes \cite{Bjornson2013}: MRT (a competitive or egoistic scheme) and inter-cell interference cancellation (ICIC is a cooperative or altruistic scheme). For the MRT scheme, the transmitters ignore the ICI that they cause to unintended receivers in their vicinity. The goal is to maximize the received signal power of local users (effective in non-cooperative systems). The ICIC is a ZF-based scheme, which means that the transmitters design their precoders so that no interference is cause to non-intended receivers (effective in cooperative systems). The design of ICIC/ZF precoding is subject to constraints in the number of transmit and receive antennas, and its effectiveness highly depends on the CSIT across the transmitters. A close-to-optimal\footnote{The optimal precoding design maximizes the received signal powers at low SNRs, minimizes the interference leakage at high SNRs, and balances between these conflicting goals at moderate SNRs.} distributed precoding scheme (for an arbitrary SNR), is attained by balancing MRT and ICIC, whose mathematical formulation is discussed in \cite{Bjornson2013,Bjornson2014b}.
If the system operates under a limited feedback constraint, the CSIT is acquired using the principles described in Section~\ref{section:precoding:partial-csit}. Precoding design extensions from single to multi-cell scenarios are not straightforward. For instance, in multi-cell cooperative systems, the codebooks sizes may significantly increase \cite{Gesbert2010}, and depending on the delay tolerance and codebook granularity, the clustered transmitters should switch between MRT and ICIC schemes \cite{Godana2013}.

Several multi-cell CBF designs attempt to suppress IUI and ICI simultaneously with minimum coordination or control signaling between transmitters. These schemes rely on instantaneous or statistical CSI, and can be classified as \cite{Zheng2015}: \textit{hierarchical} and \textit{coupled} precoders. The hierarchical approach is implemented in systems where each transmitter suppresses ICI using ZF. This is attained through the sequential construction of outer and inner precoders that reduce ICI and IUI, respectively. The ICI cancellation is achieved by aligning interference subspaces at the receivers, whilst the IUI is suppressed at the transmitters using linear precoding and local CSI, e.g., \cite{Adhikary2013,Suh2011,Liu2014a,YiXu2014,Ferrand2014,Chen2014}.
The coupled or nested-structure approach, is implemented in interference limited system, [cf. Definition~\ref{defn:interference-limited-system}]. The precoding design (beamforming weights and power allocation) optimizes an objective function subject to a set of QoS constraints, [cf. Section~\ref{section:performance-uf}]. This kind of optimization problems has been extensively studied in the literature, covering power allocation \cite{Chiang2008} and beamforming design \cite{Hong2012}. 
Under certain conditions, two specific problem formulations admit global optimal solutions via standard  optimization in multi-cellular scenarios \cite{Boyd2004,Stanczak2009,Tan2012}: 
\textit{i) power minimization subject to individual SINR constraints}. The CBF approach in \cite{Zakhour2013} mitigates ICI by iteratively adjusting beamforming weights and transmit powers according to the experienced interference per user.
\textit{ii) the maximization of minimum SINR subject to power constraints}. The authors in \cite{Huang2013b} tackled this problem using the Perron-Frobenius theory \cite{Stanczak2009,Chiang2008}. The joint beamforming and power allocation design is reformulated as an eigenvalue-eigenvector optimization problem, whose optimal solution can be found by geometrically-fast convergent algorithms.

\subsection{Precoding in LTE-Advanced}
\label{section:precoding:lte-advanced}

In the LTE specification, CSI acquisition is called \textit{implicit feedback} and relies on the following parameters \cite{Li2010,Liu2012}: rank indicator (RI), which defines the number of data streams recommended for SU-MIMO transmission; precoding matrix index (PMI), which is the index of the best precoding matrix in the codebook; and channel quality indicator (CQI), which contains information of the channel quality corresponding to the reported RI and PMI \cite{Dahlman2013}. The RI and PMI indexes provide the CDI of the MIMO channels, while the CQI indicates the strength of the corresponding spatial direction. 
The precoding matrices are defined at the BSs using different approaches \cite{Li2014,Lim2013}: codebook-based precoding or non-codebook-based precoding (arbitrary precoder selection based on RI). The standard also supports a dual codebook structure in MU-MIMO mode. This increases codebook granularity, suppresses IUI more efficiently, and enhances the overall performance. The idea is that one codebook tracks the wideband and long-term channel characteristics, while the other tracks the frequency-selective and short-term channel variations \cite{Lim2013}. In this way, the transmitter has more flexibility and accuracy when designing the precoding matrix. 
Another supported scheme is the \textit{Per Unitary basis stream User and Rate Control} (PU2RC), which allows multiple orthonormal bases per codebook, increasing quantization granularity \cite{Huang2009a, Tang2010}. Although PU2RC uses deterministic codebooks, random codebooks can be used to simplify theoretical analysis, e.g. \cite{Sharif2005,Jindal2006,Huang2009a,Tang2010}. The optimum number of bases in the codebook depends on the number of active users $K$, and should be optimized to maximize MUDiv and multiplexing gains \cite{Huang2009a}.

LTE standard supports interference mitigation based on linear precoding schemes, such as SLNR and ZFBF with quantized channels \cite{Lee2012a}. The precoders can be dynamically recalculated after each CSI update, i.e., tracking channel variations. Linear precoding based on quantized channels outperforms codebook-based precoding (RBF or PU2RC), when then number of active users is small, i.e., \textit{sparse networks}, $K \approx  M$ \cite{Gesbert2007a}. The reverse holds as MUDiv increases, $K \gg  M$, \cite{Huang2009a}.

\subsection{Precoding in 802.11ac}
\label{section:precoding:208_11ac} 

{\color{black}{
Sounding frames (null data packet) for MU-MIMO beamforming were introduced in 802.11n for channel estimation purposes. The transmitter sends known symbol patters from each antenna, allowing the receiver to sequentially construct the channel matrix, which is compressed and sent back to the transmitter \cite{Rico-Alvarino2014}. The method employed to calculate the steering matrix is implementation and vendor specific relying on explicit CSI feedback, and is not defined by the 802.11ac standard \cite{Bejarano2013,Arubanetworks2014,Liao2016}.}} 
One popular technique to construct the steering or precoding matrix is through SVD precoding \cite{Rico-Alvarino2014,Kim2015}. Since knowledge of CSIT is mandatory and feedback rates are limited, the receivers represent their estimated precoding matrix with orthogonal columns using Givens rotations \cite{Gentle2007}. Then, the set of calculated parameters (angles) at the receivers are adjusted, quantized, and fed back to the transmitter. In general, the final precoding matrix calculated by the transmitter will be different from the weights reported by the users due to the orthogonalization process \cite{Kim2015}.

Another practical approach is to compute the steering matrix using linear precoding schemes, e.g. ZFBF or MMSE  \cite{Jones2015}. In \cite{Cheng2014}, the authors used BD and regularized BD with geometric-mean decomposition for a MU-MIMO scenario. The designs' goal is to balance the achievable SNR across the users, meeting the requirements of the standard. If the number of transmit antennas is larger than the total number of receive antennas, the additional DoF can be used to efficiently nullify inter-stream interference \cite{Jones2015}.

{\color{black}{
		
\subsection{Discussion and Future Directions}
\label{section:precoding:massive-mimo} 

The objective of precoding is to achieve spatial multiplexing, enhance link reliability and improve coverage in MIMO systems. Every physically realizable precoding design depends on the objective function, the CSIT accuracy, the number of transmitters involved, and more recently, the hardware characteristics \cite{Bjornson2014c,Kutty2016,Heath2016}. The schemes described in previous subsections can be classified as \textit{full digital precoders}, where the signal processing happens in the baseband at sub-6 GHz bands. However, these schemes cannot be directly implemented in state-of-the-art transceiver architectures with many antenna elements, neither upon higher frequency bands, i.e., mmWave \cite{Heath2016}.

Massive MIMO \cite{Lu2014} differentiates itself from classical MU-MIMO by the fact that the number of antennas at the transmitter is larger compared to the number of served users. In conventional digital precoding, [cf. Section~\ref{section:precoding:linear}], each antenna element requires a radio frequency (RF) chain, i.e., signal mixer and analog-to-digital converter (ADC). In massive MIMO, although the number of antennas and RF chains is much larger than in conventional MU-MIMO, hundreds of low-cost amplifiers with low output power are used to replace the high power amplifier used in the latter. In order to keep the hardware cost and the circuit power consumption low, cost-effective and power-efficient hardware components are employed, e.g. low resolution ADC. And yet, this results in hardware impairments, especially low resolution quantization, that may affect the system performance. However,  recent results and implementations show that the effect of hardware impairments, similarly to the effect of noise and interference, is averaged out due to the excess number of antennas \cite{Bjornson2014c}. Furthermore, the effect of quantization and AD conversion with very low number of bits can be taken into account in the precoding and signal design, providing schemes with promising spectral efficiency performance. In TDD massive MIMO, the major limiting factors were considered to be pilot contamination and channel reciprocity/calibration. However, both issues are now well studied and understood, and efficient transceiver designs compensating for pilot contamination and imperfect calibration are available. The major bottleneck for practical implementations of TDD-based massive MIMO remains the amount of training required. High peak rates have been demonstrated in downlink massive MIMO (in TDD and sub-6 GHz bands) with linear and non-linear precoding by several companies. Nevertheless, the amount of uplink training required can reduce the net throughput by at least half. A major challenge for massive MIMO is its successful deployment in FDD systems, in which CSI should be obtained by feedback. Efficient approaches for channel representation and CSI quantization and compression are necessary for viable implementations. Codebook-based approaches would require a relatively high number of bits for channel quantization and feedback, which not only will reduce the net throughput but also is not supported by current standards. Various new approaches for precoding and feedback in FDD massive MIMO are expected in the near future. 

The current trend of using frequencies above 6 GHz for broadband wireless communications put an extra stress to massive MIMO systems. Current MIMO transceiver architectures may not be cost-effective and realistic in mmWave frequencies due to extremely high cost and power consumption. Different transceiver architectures have been recently proposed to address the hardware limitations. These new schemes require the joint optimization of precoding weights in the digital and analog domains, the so called \textit{hybrid precoding} \cite{Ayach2014,Gao2015,Kutty2016,Heath2016}. In the digital domain, the low-dimensional precoding weights are computed using microprocessors. In the analog domain the RF precoders are implemented by phase shifters and variable gain amplifiers \cite{Bogale2015}. 
The main goal of hybrid precoding schemes is to achieve the performance of full digital precoders, but with a reduced number of RF chains \cite{Sohrabi2016}. The performance gap between the full digital and hybrid precoders depends on the spatial load at the transmitter, i.e., the ratio between the number of active data streams over the number of RF chains, $M_{RF}$, \cite{Bogale2016a}. Notice that the number of co-scheduled users per resource is limited by $M_{RF}$ in hybrid transceiver architectures, [cf. Section~\ref{section:algorithms:aggregated-utility:massive-mimo}]. 
The optimal design of hybrid precoders has not been fully understood, and due to the power and amplitude constraints the sum-rate optimization problem becomes non-convex \cite{Gao2016}. Therefore, ongoing research is focused on designing sub-optimal, yet efficient and practical, architectures that improve the joint performance of digital and analog precoders \cite{Sohrabi2016,Yu2016,Kutty2016,Bogale2016a,Heath2016}. Although hybrid precoding provides a compromise between system performance and hardware complexity, it still remains challenging to implement reliable and cost-effect analog beamforming schemes at mmWave. Despite the complexity reduction using hybrid precoding, the performance gains using fully digital beamforming remain attractive. Hardware complexity may not be the major issue with digital beamforming at mmWave; significant challenges will arise in signal compression and CPRI protocols \cite{DelaOliva2016}, which will require innovative solutions.
}}

\section{Spatial Compatibility Metrics}
\label{section:metrics-spatial-compatibility}

Let $\mathcal{K} \subseteq \bar{\mathcal{K}}$ and $\mathcal{K}' \subseteq \bar{\mathcal{K}}$ be two sets of user with at least one non-common user, where $\mathbf{H}(\mathcal{K})= \{\mathbf{H}_{i}\}_{i \in \mathcal{K}}$ and $\mathbf{H}(\mathcal{K}')=\{\mathbf{H}_{j}\}_{j \in \mathcal{K}'}$ are their associated channels. A metric for spatial compatibility is a function of the CSIT that maps the spatial properties of the multiuser MIMO channels to a positive scalar value quantifying how efficiently such channels can be separated in space \cite{Maciel2010}, i.e., $f(\mathbf{H}(\mathcal{K})):\mathbb{C}^{ |\mathcal{K}|N \times M} \mapsto \mathbb{R}_{+}$. 
Consider the two subsets $\mathcal{K}$ and $\mathcal{K}'$, a metric of spatial compatibility can be used to estimate the achievable performance of their associated multiuser channels, e.g., having $f(\mathbf{H}(\mathcal{K})) > f(\mathbf{H}(\mathcal{K}'))$ may imply that $\mathcal{K}$ is the set that achieves the maximum capacity. The mapping function $f(\cdot)$ can be used for scheduling purposes, depending on its definition and other system parameters (e.g. SNR regime and $M$), which will be discussed in Section~\ref{section:algorithms:aggregated-utility:csi-mapping}. 
Below we provide two definitions related to the design of user scheduling policies.

\begin{definition}\label{defn:compatible_set_of_users}
	\textit{Spatially compatible users}. A feasible set of users $\mathcal{K}$, is spatially compatible if the multiuser MIMO channels, $\mathbf{H}(\mathcal{K})$, can be separated in the spatial domain by means of beamforming/precoding.
\end{definition}

\begin{definition}\label{defn:user_grouping}
	\textit{User grouping}. It is the task of forming a subset of users $\mathcal{K}$, according a compatibility criterion, e.g. spatial separability in Definition~\ref{defn:compatible_set_of_users}, to maximize the resource allocation and scheduling efficiency.  
\end{definition}

{\color{black}{
		
User grouping can be the initial step in a MU-MIMO scheduling algorithm since the characteristics of the joint channels dictate the transmission reliability and the resource allocation feasibility [cf. Definition~\ref{defn:resource_feasibility}].  
In MU-MIMO scenarios, there exists a correspondence between the precoding capability to reduce IUI and the user grouping technique. The performance achieved by a precoder scheme is determined by the characteristics of the selected multiuser channels, i.e., providing \textit{spatially compatible users} to the precoding processing block (see Fig.~\ref{fig:system-processing-blocks}), is fundamental to guarantee high attainable SINRs at the receivers. 
It is worth mentioning that the vast majority of scheduling algorithms in the literature focuses on constructing sets of users with orthogonal or semi-orthogonal MIMO channels. Yet, for some particular signal designs, the best scheduling strategy is to group users whose channels are parallel or semi-parallel, see \cite{Jang2011,Christopoulos2013}. This can also be the case in scheduling for non-orthogonal multiple access (NOMA)-MIMO schemes \cite{Liu2015,Dai2015}, where the notion of spatial compatibility may be revised.
The spatial compatibility metrics are used in the MU-MIMO literature to pair users and optimize the performance of a particular utility function. They are also used to quantize the channels in systems with limited feedback rates. 

}}

\subsection{Null Space Projection}
\label{section:metrics-spatial-compatibility:projection-spaces}

One of the objectives of MU-MIMO technology is to multiplex independent data streams to different users, which implies that only a subset of the transmitted data symbols are useful for each co-scheduled user. In such a scenario, a fundamental problem is to mitigate IUI, i.e., suppress the information intended to other receivers. There exist several signal processing techniques that can achieve such a goal, e.g., linear precoding or interference alignment. The effectiveness of such techniques rely upon the characteristics of the subspaces spanned by the MIMO channels, i.e, IUI is a function of the overlapped interference subspaces \cite{Jafar2011}.

\begin{figure}[t!]
	\centering
	\includegraphics[width=0.9\linewidth]{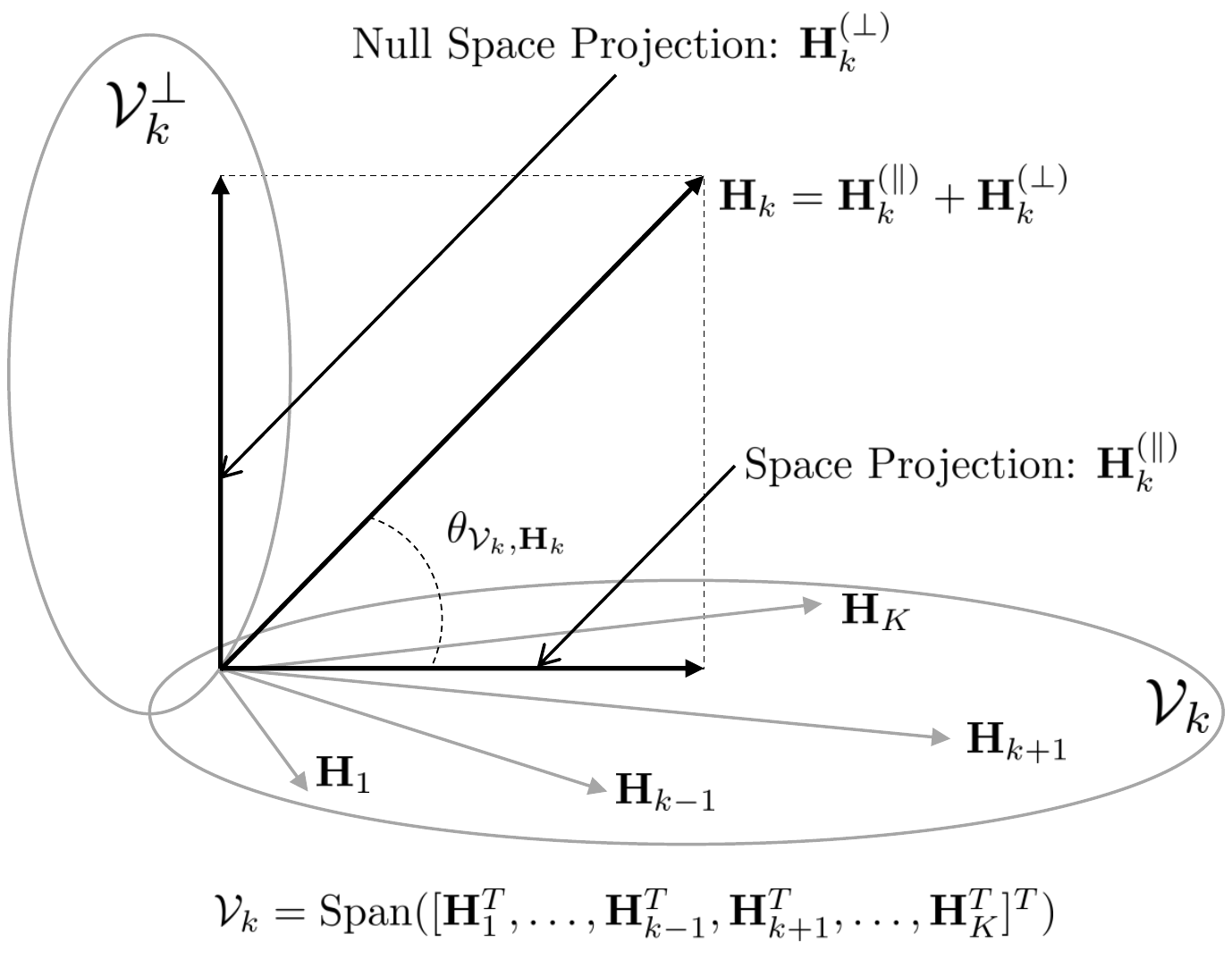}\\
	\caption{Decomposition of the channel $\mathbf{H}_{k}$ given by its projection onto the subspaces $\mathcal{V}_{k}$ and $\mathcal{V}_{k}^{\perp}$.}
	\label{fig:space-projections}
\end{figure}

Consider a set of user $\mathcal{K}$ with $K$ users, let $\tilde{\mathbf{H}}_{k} = [\mathbf{H}_{1}^{T}, \ldots, \mathbf{H}_{k-1}^{T}, \mathbf{H}_{k+1}^{T},\ldots, \mathbf{H}_{K}^{T}]^{T}$, be the aggregated interference matrix of the user $k$ such that $M>\max_{k} \ \text{rank}(\tilde{\mathbf{H}}_{k})$, which is a necessary condition to suppress IUI \cite{Wang2006}. Define $\mathcal{V}_{k} = \text{Span}(\tilde{\mathbf{H}}_{k})$, as the subspace spanned by the channels of the subset of users $\mathcal{K} \setminus \{k\}$, and let $\mathcal{V}_{k}^{\perp}  = \text{Span}(\tilde{\mathbf{H}}_{k})^{\perp}$, be its orthogonal complement subspace. 
In other words, $\mathcal{V}_{k}^{\perp}$ spans the null space of $\tilde{\mathbf{H}}_{k}$, i.e., $null(\tilde{\mathbf{H}}_{k}) = \{\mathbf{x} \in \mathbb{C}^{M \times 1} : \tilde{\mathbf{H}}_{k} \mathbf{x} = \mathbf{0} \}$.
The channel of the $k$-th user can be expressed as the sum of two vectors $\mathbf{H}_{k} = \mathbf{H}_{k}^{(\parallel)} + \mathbf{H}_{k}^{(\perp)}$, each one representing the projection of $\mathbf{H}_{k}$ onto the subspaces $\mathcal{V}_{k}$ and $\mathcal{V}_{k}^{\perp}$ respectively, as illustrated in Fig.~\ref{fig:space-projections}.

For all user-level ZF-based precoding schemes described in Section~\ref{section:precoding:linear}, $\mathcal{V}_{k}^{\perp} = \bigcup_{i=1, i \neq k}^{K}  \text{Span}(\mathbf{H}_{i})^{\perp}$, i.e., $\mathcal{V}_{k}^{\perp}$ contains all overlapped interference subspaces of channel $\mathbf{H}_{k}$. The component $\mathbf{H}_{k}^{(\parallel)}$ is related to the signal degradation of the user $k$ due to channel correlation, whereas $\mathbf{H}_{k}^{(\perp)}$ defines the \textit{zero-forcing direction}, i.e. the spatial direction that is free of IUI. The squared magnitude of $\mathbf{H}_{k}^{(\perp)}$ is known as the null space projection (NSP), and directly computes the effective channel gain obtained by ZF precoding \cite{Caire2003}. 

A common approach in the literature is to define the mapping $f(\mathbf{H}(\mathcal{K}))$ as a function of the exact or approximated \textit{effective channel gains} [cf. Definition~\ref{defn:effective_channel_gains}]. In other words, the spatial compatibility metric has to consider the precoding scheme, the SNR regime, and the available spatial DoF. Analytical results in \cite{Lau2006,Chan2007,Shi2012} have shown that the set of users $\mathcal{K}$ that maximizes the product of their effective channel gains\footnote{Some works in the literature (e.g. \cite{Swannack2004,Zakhour2007,Castaneda2014b}) optimize the sum of effective channel gains instead of their product, which yields similar performance for the high SNR and large number of users, i.e. $K \gg M$.} is also the set that maximizes the sum-rate in the high SNR regime. The NSP has been extensively used for user grouping and scheduling purposes whenever ZF-based precoding is implemented [cf. Section~\ref{section:algorithms:aggregated-utility}]. Note that a set of spatially compatible users would have large NSP components, which means that their associated channels are semi-orthogonal. 

The computation of NSP for the general MU-MIMO scenario where $\mathbf{H}_{k} \in \mathbb{C}^{N_k \times M}$ $\forall k$, can be performed by several approaches, namely SVD \cite{Golub1996,Gentle2007,Yanai2011,Wang2010,Tran2010}, orthogonal projection matrix \cite{Golub1996,Yanai2011,Fuchs2007,Caire2003}, Gram-Schmidt orthogonalization (GSO) \cite{Golub1996,Gentle2007,Yoo2006,Sigdel2009a,Shen2006}, QR decomposition \cite{Tran2012}, products of the partial correlation coefficients \cite{Yanai2011,Castaneda2014b}, and ratio of determinants \cite{Castaneda2014b,Razi2010}. 
Several works find sets of spatially compatible users by computing an approximation of the NSP (e.g. \cite{Fuchs2007,Bjornson2013,Cheng2014,Castaneda2015,Xu2010,Bayesteh2008,Hammarwall2008}), which yields efficient scheduling designs that require low computational complexity.
Moreover, recent works (\cite{Manolakos2015} and references therein), have proposed efficient methods to compute and track null spaces for a set of co-scheduled users, which is fundamental to mitigate IUI for ZF-based precoding in single and multiple transmitter scenarios.

\subsection{Spatial Clustering}
\label{section:metrics-spatial-compatibility:spatial-clustering}

{\color{black}{
Spatial clustering refers to the association of a channel matrix to a spatial subspace. This technique has been used in the literature to perform user/channel grouping or scheduling in compliance with the precoding scheme and CSIT availability. In feedback limited scenarios, spatial clustering is also used to quantize the channel or precoder weights [cf. Section~\ref{section:precoding:partial-csit}].  
Let $\measuredangle(\mathbf{H}_{i},\mathbf{H}_{j})$ denote the angle between i.i.d. channels $\mathbf{H}_{i},\mathbf{H}_{j} \in \mathbb{C}^{1 \times M}$, and define the normalized inner product, also known as \textit{coefficient of correlation}, as \cite{Gentle2007}:
}}

\begin{equation}\label{eq:coefficient_correlation}
\cos (\measuredangle(\mathbf{H}_{i},\mathbf{H}_{j})) =  \frac{|\mathbf{H}_{i}\mathbf{H}_{j}^{H}|}{ \| \mathbf{H}_{i} \| \| \mathbf{H}_{j} \|},
\end{equation}
where $\cos(\measuredangle(\mathbf{H}_{i},\mathbf{H}_{j})) \in [0, 1]$, and $\measuredangle(\mathbf{H}_{i},\mathbf{H}_{j}) = \frac{\pi}{2}$, implies that the channels are spatially uncorrelated or orthogonal.
The coefficient of correlation indicates how efficiently the transmitter can serve user $i$ without affecting user $j$, and vice versa. For particular MU-MISO scenarios, in which the set of scheduled users is subject to the cardinality constraint $|\mathcal{K}|=2$, the NSP is a function of $\sin^2 (\measuredangle(\mathbf{H}_{i},\mathbf{H}_{j}))$, which can be evaluated from  (\ref{eq:coefficient_correlation})  simplifying the user pairing and scheduling design \cite{Yang2011}. Several works focused on sum-rate analysis (e.g., \cite{Lu2009,Yang2011,Wang2015}) limit the cardinality to $|\mathcal{K}|=2$ for simplicity and tractability. Other works analyzed scenarios with more practical constraints and considered two or four users, which are common values of $K$ in standardized systems, see \cite{Kountouris2008,Lim2013,Jones2015}. 
The metric (\ref{eq:coefficient_correlation}) has been extensively used to define policies for user grouping. In MU-MIMO systems a set of user $\mathcal{K}$ is called $\epsilon$-orthogonal if $\cos (\measuredangle(\mathbf{H}_{i},\mathbf{H}_{j})) < \epsilon$ for every $i \neq j$ with $i,j \in \mathcal{K}$ \cite{Swannack2004,Yoo2005a,Bayesteh2008}. The users can be grouped into disjoint sets according to a desired threshold $\epsilon$, and the semi-orthogonal sets can be scheduled over independent carriers (e.g. \cite{Zhang2005a,Chung2010,Driouch2012}), codes (e.g. \cite{Driouch2012}), or time slots (e.g. \cite{Yoo2006}). 
The optimum value of $\epsilon$ depends on the deployment parameters ($K$ and $M$) and is usually calculated through simulations, since for $M>2$ it is very hard or impossible to find the optimal $\epsilon$ in closed-form. Nevertheless, there exist closed-form expressions to compute the ergodic capacity as a function of $\epsilon$ for MU-MISO scenarios where the transmitter has two antennas $M=2$ and the users have homogeneous \cite[Ch. 7]{Yang2011} or heterogeneous \cite{Wang2015} large-scale fading gains.

\begin{figure}[t!]
	\centering
	\includegraphics[width=0.8\linewidth]{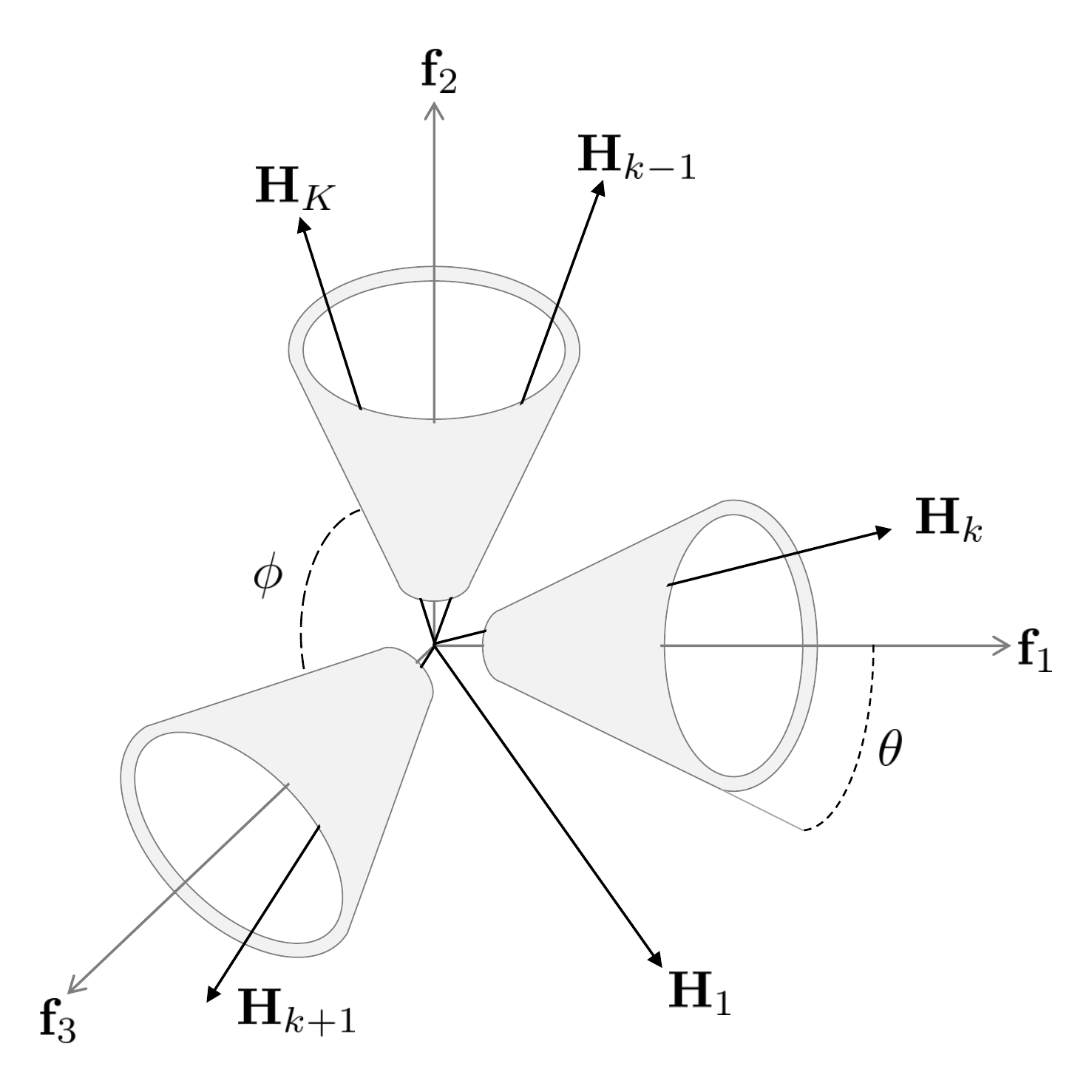}\\
	\caption{The channels $\{\mathbf{H}_{j}\}_{j=1}^{K}$ can be clustered within cones given a set $\mathcal{F} = \{\mathbf{f}_{1},\mathbf{f}_{2},\mathbf{f}_{3}\}$ formed either by orthonormal basis or codebook words. }
	\label{fig:hyperslab_cones}
\end{figure}

{\color{black}{
Consider that a codebook, $\mathcal{F}$, is used to define the CDI of the MIMO channels, i.e., assume partial CSIT, [cf. Section~\ref{section:precoding:partial-csit}]. In such a scenario, spatial clustering can be performed for a given  parameter $\theta$ and the codeword $\mathbf{f}_{i} \in \mathbb{C}^{1 \times M}$. Define the hyperslab $\mathfrak{F}_{i}(\theta)$ as \cite{Lee2014}:
\begin{equation}\nonumber %
	\mathfrak{F}_{i}(\theta) =\left\{ \mathbf{H}_{k} \in \mathbb{C}^{1 \times M}, k \in \mathcal{K}: \cos (\measuredangle(\mathbf{H}_{k},\mathbf{f}_{i})) \leq \cos(\theta) \right\}.	\label{eq:hyperslab}
\end{equation}		
The hyperslab defines a vector subspace whose elements attain a spatial correlation not greater than $\cos(\theta)$, w.r.t. the codeword $\mathbf{f}_{i}$, as illustrated in Fig.~\ref{fig:hyperslab_cones}. The parameter $\cos(\theta)$ is set to guarantee a target $\epsilon$-orthogonality. 
}}
The generalization of the hyperslab clustering\footnote{Some authors (e.g., \cite{Swannack2004,Lau2009}) define $\mathfrak{F}_{i}(\theta)$ as a function of two parameters, $\theta$ and the minimum acceptable channel magnitude. In this way only sets of spatially compatible and strong channels can be constructed.} for MIMO settings is straightforward, i.e. $\mathbf{H}_{k} \in \mathbb{C}^{N_{k} \times M}$, $\mathbf{f}_{i} \in \mathbb{C}^{M \times M}$, and the coefficient of correlation between matrices can be defined as in \cite{Sigdel2009a,Tran2012}. 
The parameter $\phi$ shown in Fig.~\ref{fig:hyperslab_cones} can be adjusted to fix the maximum co-linearity between cones \cite{Wang2012d}. 

Several scheduling algorithms based on spatial clustering (e.g., \cite{Wang2006,Razi2010,Yoo2006,Kountouris2007,Wang2008a,Lau2009,Lu2009,Sun2010,Sohn2010,Mao2012,Tran2012,Min2013,Lee2014}), can achieve MUDiv gains and improve the overall performance by adjusting the parameter $\theta$ (or threshold $\epsilon$) according to the number of competing users \cite{Yoo2006}, the SNR regime, the large-scale fading gain, and the precoding scheme \cite{Wang2015}. 
In \cite{Lau2009}, it was shown that the optimal cardinality of the user set grouped based on spatial clustering is a function of $K$, $\theta$, and $M$.
The statistical properties of (\ref{eq:coefficient_correlation}) have been extensively studied in the literature cf. \cite{Mukkavilli2003,Swannack2004,Yoo2006,Jindal2006,Au-Yeung2007,Zhang2007b,Jagannathan2007}, and such properties depend on the type of CSIT (full or partial), the MIMO channel distribution, and the system parameters (e.g. $M$ and $B$). 

\subsection{Compatibility between Subspaces}
\label{section:metrics-spatial-compatibility:other-metrics}

{\color{black}{
In heterogeneous MU-MIMO scenarios where each user is equipped with $N_{k}$ antennas $\forall k$, the signal spaces span several dimension and mutual interference between user can be measured either in the angular or in the subspace domains \cite{Wang2006,Yi2011}. Moreover, metric (\ref{eq:coefficient_correlation}) cannot be used directly in heterogeneous MU-MIMO scenarios to measure spatial correlation between channels of different dimensions, thus, several alternative metrics have been proposed in the literature.}}
Consider a set of users $\mathcal{K}$ and its MU-MIMO channel $\mathbf{H}(\mathcal{K})$; the spatial compatibility can be measured as a function of the corresponding eigenvalues:
\begin{enumerate}
	
	\item  The \textit{orthogonality defect} \cite{Bayesteh2008} is a metric derived from Hadamard's inequality \cite{Cover2006} that measures how close a basis is to orthogonal. It quantifies the energy degradation of the channel matrix due to the correlation between all its column vectors. User grouping algorithms based on this metric have been developed for homogeneous MIMO systems, e.g. \cite{Tran2010,Castaneda2014b}. 
	
	\item The determinant geometrically represents the volume of the parallelepiped defined by the column vectors of the channel matrix. Larger determinant values implies that the column vectors of a matrix are more orthogonal \cite{Gentle2007}. The \textit{matrix volume ratio} used in \cite{Wang2010a,Castaneda2015} measures the volume reduction of $\mathbf{H}(\mathcal{K})$ w.r.t. $\mathbf{H}(\mathcal{K} + \{k'\})$, where $k' \notin \mathcal{K}$ is a candidate user attempting to be grouped. Different approximations of such a metric can be done using the arithmetic-geometric mean inequality over the squared singular values of $\mathbf{H}(\mathcal{K})$, e.g., \cite{Xu2009a,Castaneda2014b}.
	
	\item The \textit{geometrical angle} (see \cite{Yi2011,Nam2014a} and references therein), is a metric similar to the matrix volume ratio that measures the spatial compatibility as a function of the all possible correlation coefficients [cf. (\ref{eq:coefficient_correlation})] between the basis of two subspaces. It can be computed from the eigenvalues of two MIMO matrices  $\mathbf{H}_{i}  \in \mathbb{C}^{N_{i} \times M}$, $\mathbf{H}_{j}  \in \mathbb{C}^{N_{j} \times M}$, whose subspaces have different dimensions, i.e., $N_{i} \neq N_{j}$.
	
	\item If the elements of a multiuser channel matrix are highly correlated, the matrix is said to be ill-conditioned, i.e., it is close to singular and cannot be inverted. In numerical analysis, the \textit{condition number} \cite{Gentle2007} quantifies whether a matrix is well- or ill-conditioned, and is computed as the ratio between the maximum and minimum eigenvalues. This metric is used to measure how the eigenvalues of $\mathbf{H}(\mathcal{K})$ spread out due to spatial correlation. The ratio between the condition numbers of two MIMO channels can be used to quantify their spatial distance and compatibility \cite{Czink2009}. 
\end{enumerate}	

For a given user $k \in \mathcal{K}$, the spatial compatibility between $\mathbf{H}_{k}$  and $\tilde{\mathbf{H}}_{k}$, can be measured by a function of $\theta_{\mathcal{V}_{k}\mathbf{H}_{k}}$, which is the angle between $\mathcal{V}_{k}$ and $\text{Span}(\mathbf{H}_{k})$, see Fig.~\ref{fig:space-projections}. The following metrics assess the spatial compatibility based on geometrical properties of the multiuser channels:
\begin{enumerate}
	\item The \textit{principal angle between subspaces} \cite{Wang2006,Yi2011,Nam2014a}  measures the relative orientation of the basis of $\mathcal{V}_{\mathbf{H}_{k}} = \text{Span}(\mathbf{H}_{k})$ regarding the basis of $\mathcal{V}_{\tilde{\mathbf{H}}_{k}} = \text{Span}(\tilde{\mathbf{H}}_{k})$. Given the bases of $\mathcal{V}_{\mathbf{H}_{k}}$ and $\mathcal{V}_{\tilde{\mathbf{H}}_{k}}$, the principal angle is associated with the largest coefficient of correlation [cf. (\ref{eq:coefficient_correlation})] between the bases of both subspaces.
	\item The \textit{chordal distance} is extensively used in limited feedback systems for codebook design \cite{Barg2002,Love2003}, user grouping and scheduling \cite{Chae2008,Yi2011,Ko2012,YiXu2014}. This metric can be computed either from the principal angles between subspaces or from the projection matrices of $\mathcal{V}_{\mathbf{H}_{k}}$ and $\mathcal{V}_{\tilde{\mathbf{H}}_{k}}$ in heterogeneous MU-MIMO scenarios, i.e., $N_{i} \neq N_{j}$ $\forall i, j \in \mathcal{K}$ with $i \neq j$. 
	\item The \textit{subspace collinearity} \cite{Czink2009,Yi2011} quantifies how similar the subspaces spanned by two channel matrices are, following the rationale behind (\ref{eq:coefficient_correlation}). Given $\mathbf{H}_{i}, \mathbf{H}_{j} \in \mathbb{C}^{N \times M}$ the subspace collinearity of the matrices compares the singular values and the spatial alignment of their associated singular vectors. 
	\item Other metrics measuring the distance between subspaces are the \textit{weighted likelihood similarity measure}, the \textit{subspace projection measure}, and the \textit{Fubini-Study similarity metric}. These metrics have been recently propounded in \cite{YiXu2014} and used for user grouping based on statistical CSI, i.e., $\mathbf{\Sigma}_{k} \ \forall k$.	
\end{enumerate}

{\color{black}{		
\subsection{Discussion}
\label{section:metrics-spatial-compatibility:discussion}

It is worth mentioning that neither the principal angles, nor the chordal distance can fully measure spatial compatibility in heterogeneous MU-MIMO scenarios. This is due to the fact that these metrics take into account the smallest dimension between two subspaces, potentially neglecting useful spatial correlation information \cite{Yi2011}. Moreover, metrics that only evaluate the spatial separation between hyperplanes or the eigenvalue dispersion of $\mathbf{H}(\mathcal{K})$ neglect the degradation of the MIMO channel magnitude due to the interference subspace. Such metrics fail at maximizing the capacity since they do not evaluate or approximate the effective channel gain \cite{Ko2012}. As the number of active users grows, the set of users that maximizes a given  spatial compatibility metric may diverge from the set that maximizes the capacity, either in their elements, cardinalities, or both \cite{Yi2011,Castaneda2014b}.

Authors in \cite{Wang2010a} pointed out that user grouping should be a function of the spatial correlation between $\mathbf{H}_{k}$ and $\tilde{\mathbf{H}}_{k}$, and also consider the inner correlation of each multi-antenna user, i.e., the magnitudes of the eigenvectors of $\mathbf{H}_{k}$ $\forall k$.
According to \cite{Vu2007,Jorswieck2006,Godana2013a}, the precoding performance is primarily affected by the correlation between transmit antennas, whereas receive antenna correlation has marginal or no impact on the precoding design. Nonetheless, the effects of receive antenna correlation has not been fully studied in the user selection literature and conclusions are usually drawn based on specific correlation models. 
The knowledge of statistical CSI, i.e., $\mathbb{E}[\mathbf{H}_{k}^{H} \mathbf{H}_{k}]$, may be assumed in scenarios with practical constraints such as limited feedback rates [cf. Section~\ref{section:precoding:partial-csit}].  If the transmitter has knowledge of statistical CSI, the dominant eigen-directions of the channel covariance matrix ($\mathbf{\Sigma}_{k}$, $\forall k$) can be used as metrics to identify compatible users, e.g., \cite{Kountouris2006b,Hammarwall2008,Gao2009,Li2016,YiXu2014,Adhikary2013a,Nam2014}. Spatial clustering based on $\mathbf{\Sigma}_{k}$, $\forall k$, has been recently proposed to identify users with similar channel statistics in massive MIMO settings, see \cite{YiXu2014,Nam2014,Zheng2015}. 

}}

\section{System Optimization Criteria}
\label{section:performance-uf}

\begin{table*}[t!] \scriptsize
	\renewcommand{\arraystretch}{1.3}	
	\caption{Summary of System Optimization Criteria for Scheduling in MISO ($N=1$) and MIMO ($N>1$) configurations with full ($B=\infty$) and partial ($B<\infty$) CSIT. QoS refers to SINR, BER, or Queue Stability requirements}
	\label{tab:optimization-criteria}%
	\centering		
	\begin{tabularx}{\textwidth}{l *{5}{>{\centering\arraybackslash}X} }  
		\toprule		
		& \textbf{Sum Rate} & \textbf{Fairness} & \textbf{WSR}   & \textbf{QoS} & \textbf{Round Robin} \\
		\midrule
		\textbf{MISO}, $B=\infty$ & \cite{Dimic2005, Huang2013, Tu2003, Wang2008a, Yoo2006, Jiang2006, Dai2009, Lee2014, Adhikary2013, Huang2012b, Razi2010, Mao2012, Bjornson2014, Shi2012, Maciel2010, Driouch2012, Yoo2005a, Christopoulos2013, Xia2009, Lau2005a, Cottatellucci2006, Wang2015, Yang2011, Boccardi2006, Zhang2007b, Dartmann2013, Castaneda2015, Jang2011, Park2014,  Ku2015,YiXu2014,Bogale2015,Lee2014c} & \cite{Moon2013, Yoo2006, Maciel2010, Driouch2012, Christopoulos2013, Lau2005a, Lee2014, Dartmann2013, Seifi2011,YiXu2014,Liu2015a} & \cite{Jagannathan2007, Hammarwall2007, Huh2012,Hammarwall2008,Liu2014a} & \cite{Jagannathan2007, Maciel2010, Chung2010, Stridh2006a, Matskani2008, Koutsopoulos2008, Destounis2015, Swannack2004, Tsai2008, Shirani-Mehr2011, Song2008,  Chung2010, Koutsopoulos2008,Christopoulos2013,Kountouris2006b} & \cite{Yoo2006, Cottatellucci2006, Lee2014, Dartmann2013, Jang2011,Liu2015a} \\
		\midrule
		\textbf{MISO}, $B<\infty$ & \cite{Jindal2006, Wagner2008, Nam2014, Lee2014, Adhikary2013, Adhikary2013a, Maciel2010, Xia2009, Zorba2008, Yang2011, Zakhour2007, Tang2010, Huang2009a, Vicario2008, Yoo2007, Kountouris2007, Kountouris2008, Dai2008, Kountouris2006, Zhang2011, Ravindran2012, Kountouris2006a, Wang2007a, Choi2007c, Min2013, Sohn2010, Kountouris2005a, Huang2007, Sohn2012, Khoshnevis2013,Xu2009a} & \cite{Moon2013, Adhikary2013a, Wagner2008, Zorba2008, Lee2014, Huang2012a, Shirani-Mehr2010, Kountouris2008a, Kountouris2005, Xu2010, Khoshnevis2013,Xu2009a} & \cite{Conte2010, Shirani-Mehr2010, Kobayashi2007} & \cite{Huang2012a, Shirani-Mehr2010, Simon2011,Tang2010,Lau2009,Kountouris2006b} & \cite{Lee2014, Zhang2011,Simon2011} \\
		\midrule
		\textbf{MIMO}, $B=\infty$ & \cite{Bayesteh2008, Shen2006, Wang2010, Bayesteh2008, Ko2012, Elliott2009, Chan2007, Tran2010, Wang2010a, Nam2014a, Zhang2009, Park2010, Lossow2013,Fuchs2007, Tejera2006, Wang2006, Chen2008, Sigdel2009a, Yi2011, Sun2010, Tran2012, Elliott2012, Hei2009, Lim2009, Cheng2014} & \cite{Elliott2009, Nam2014a, Torabzadeh2010, Cui2011a, Yu2013,Fuchs2007, Sigdel2009a, Cheng2014, Aniba2007} & \cite{Tran2012} & \cite{Souihli2010, Moretti2013,Zhang2005a, Aniba2007,Chen2007a} & \cite{Wang2010, Bayesteh2008, Chan2007,Fuchs2007} \\
		\midrule
		\textbf{MIMO}, $B<\infty$ & \cite{Sharif2005, Chae2008, Trivellato2008, Jindal2008, Nam2014a,Fuchs2007, Wang2006, Rico-Alvarino2014, Wang2012d, Zhang2007a} & \cite{Fan2014, Nam2014a, VanRensburg2009, Hosein2009,Fuchs2007, Rico-Alvarino2014} & \cite{Schellmann2010} & \cite{Chen2013,Zhang2005a,Rico-Alvarino2014} & \cite{Fuchs2007,Jindal2008} \\
		\bottomrule
	\end{tabularx}%
\end{table*}%

\begin{figure}[t!]
	\centering
	\includegraphics[width=0.98\linewidth]{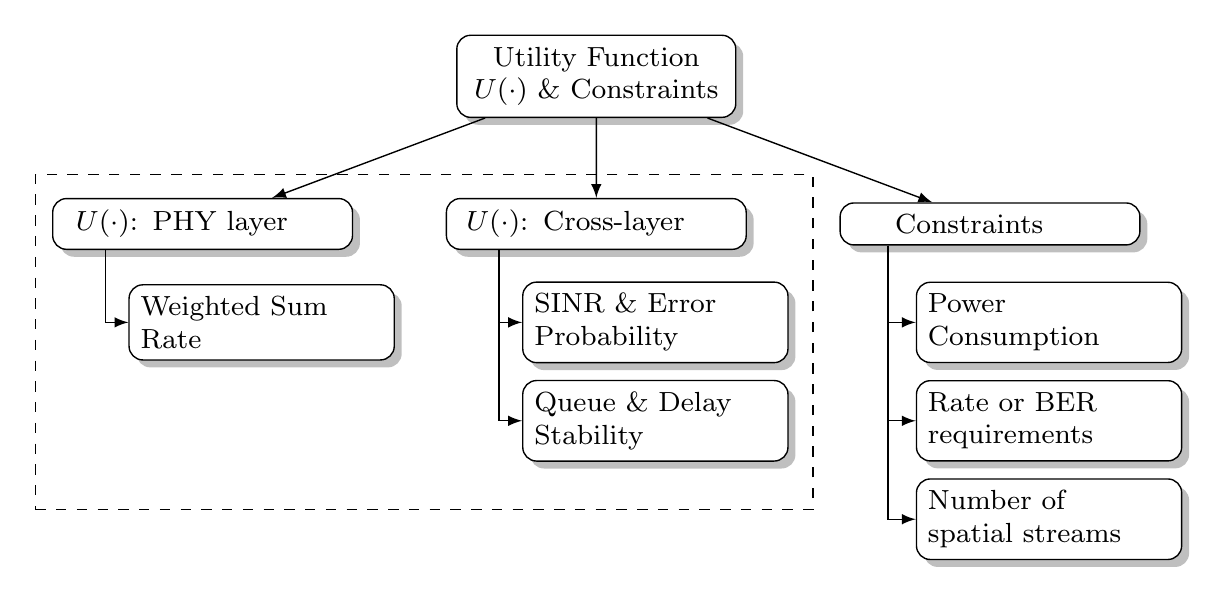}\\
	\caption{Utility Functions and constraints for PHY layer and cross layer based optimization in MU-MIMO systems} 
	\label{fig:utility-functions-and-constraints}
\end{figure}

The optimization criteria determines the optimal resource allocation strategy \cite{Jorswieck2006}, and can be  classified in two groups according to the objective function and constraints \cite{Lau2006}. \textit{i}) \textit{PHY layer} based criteria, where a objective function, $U(\cdot)$, must be optimized and channel information is the only input to the resource allocation algorithms. \textit{ii}) \textit{Cross layer} based criteria, where optimization of $U(\cdot)$, takes into account QoS requirements (defined by upper layers) and channel information, see Fig.~\ref{fig:system-processing-blocks}.
This section presents an overview and classification of the objective functions (criteria), and their associated constraints in the MU-MIMO literature. A summary of the content and general organization of this section are presented in Fig.~\ref{fig:utility-functions-and-constraints} and Table~\ref{tab:optimization-criteria}.
Two relevant concepts in multiuser system optimization are defined below.

{\color{black}{
\begin{definition}\label{defn:resource_feasibility}
	\textit{Resource allocation feasibility}. For given a set of users $\mathcal{K}$, a resource allocation strategy is called feasible if it fulfills all individual and global constraints (e.g. power and QoS), implementing precoding, power control, or a combination of both.
\end{definition}

\begin{definition}\label{defn:feasible_set_of_users}
	\textit{Feasible set of users}. Given the set of all competing users $\bar{\mathcal{K}}$, the subset   $\mathcal{K} \subseteq \bar{\mathcal{K}}$, is called feasible if there exist precoding weights and powers, $\forall k \in \mathcal{K}$, such that $U(\cdot)$ has a solution meeting individual QoS and power constraints.
\end{definition} 

}}

\subsection{Weighted Sum Rate (WSR) Maximization}
\label{section:performance-uf:weighted-sum-rate}

The system optimization is achieved by maximizing the sum of individual utility functions, $U_{k}(\cdot)$, $\forall k \in \mathcal{K}$, subject to a set of power constraints and set cardinality ($|\mathcal{K}|$). 
Global optimization based on concave and differentiable utility functions is desirable, since efficient numerical methods can be applied to guarantee optimality. A common family of functions studied in the literature is defined by the $\alpha$\textit{-fair} utility function \cite{Kelly1997,Mo2000,Hong2012}, which can be used to characterize the objective of the resource allocation strategies.

The $\alpha$\textit{-fair} function is mathematically expressed as \cite{Mo2000}:

\begin{subnumcases}{ U\left( \{{R}_{k}\}_{k=1}^{K} \right) =}
\sum_{k=1}^{K} \log({R}_k), & \text{if $\alpha=1$} \label{eq:alpha-fair-utility-function-1}
\\
\sum_{k=1}^{K} \frac{ ({R}_k)^{1-\alpha}  }{  1-\alpha  }, & \text{otherwise} \label{eq:alpha-fair-utility-function-2} 
\end{subnumcases}
where ${R}_k$ denotes the achievable transmission rate of the $k$-th user. 
Changing the values of $\alpha$ yields different priorities to the users, and can be used to define performance tradeoffs (e.g. throughput-fairness). For instance, $\alpha = 1$, yields maximum fairness, $\alpha = 0$, generates the sum-rate utility, and $\alpha \rightarrow \infty$ defines the minimum-rate function. These functions are, in general, non-concave,\footnote{For instance, non-concave utility functions may arise in the context of real-time applications with hard QoS constraints \cite{Utschick2012}.} and several methods have been developed to optimize them subject to power and QoS constraints, see \cite{Hong2012} for an in-depth review.

A more common formulation of the weighted sum rate in MU-MIMO scenarios is given by the following expression: 

\begin{equation}\label{eq:weighted-sum-rate-function}
U\left( \{{R}_{k}\}_{k=1}^{K} \right) = \sum_{k=1}^{K} \omega_{k} {R}_{k}
\end{equation}
where $\omega_{k}$ is a non-negative time-varying weight, which defines the priority of the $k$-th user. This expression is related to the $\alpha$-fair function, e.g. if $\omega_{k}=1$, then  (\ref{eq:weighted-sum-rate-function}) converges to (\ref{eq:alpha-fair-utility-function-2}) with $\alpha = 0$. The dynamic adaptation of $\omega_{k}$ can provide different levels of fairness (temporal, utilitarian, rate-proportional, fair-queuing, etc.) over the resource allocation at the \textit{user level} \cite{Lau2005}.

Another approach to provide fairness is defining $U\left( \{{R}_{k}\}_{k=1}^{K} \right)$ as a function of the rates' statistics. For instance, one might use the cumulative distribution function (CDF) of ${R}_{k}$ $\forall k$, which can be empirically estimated at the transmitter \cite{Park2005,Kountouris2008a}. 
Fairness at the \textit{system level} is another optimization criterion, measured over the average rates, transmit powers, time-slots, or any other allocated resource.  Different metrics to quantify fairness can be found in the literature of resource allocation \cite{HuaizhouShi2014}. For a single resource, one can use the Gini index \cite{Bartolome2006},  Jain's index \cite{Jain1984}, or other fairness measures in \cite{Lan2010} and reference therein. Generalized fairness metrics over multiple resources allocated simultaneously were defined in \cite{Joe-Wong2013,HuaizhouShi2014} and references therein.
The weights in (\ref{eq:weighted-sum-rate-function}) might be determined by upper layer requirements, such as buffer sizes, traffic/service priority, packet loads, or any other QoS-related metric. The individual weights affect the scheduling decisions, and can be used to improve precoding design \cite{Park2012a}.

A special case of (\ref{eq:weighted-sum-rate-function}) is given by setting $\omega_{k}=1$ $\forall k$, which represents the sum rate. This criterion defines the maximum amount of error-free information successfully received by a set of co-scheduled users, regardless of fairness \cite{Goldsmith2005}. 
The resource allocation strategies are usually assessed in terms of sum rate, since it quantifies the effectiveness of an algorithm and simplifies scheduling rules. However, maximizing the sum rate in scenarios where the users have heterogeneous SNRs might be very inefficient, since users with poor channel conditions would experience starvation \cite{Yoo2006,Jagannathan2007}. A classical solution to such a problem is assigning the weights $\omega_{k} \ \forall k$, according to a fairness criterion.
The works listed under the \textit{fairness} category in Table~\ref{tab:optimization-criteria}, optimize the long-term proportional fairness, extending the original idea from \cite{Chaponniere2002,Viswanath2002} to MU-MIMO scenarios.

Round robin is a channel-unaware scheduling method, that allocates the same amount of time to all competing users. It is the simplest form of fair resource-access, but neglects MUDiv, the amount of occupied resources, and QoS requirements \cite{Capozzi2013,Lee2014b}. Equal opportunity for resource sharing may not mean equal distribution of resources, which results in wastage, redundant allocation, or resource starvation \cite{HuaizhouShi2014}. Nonetheless, round robin might be useful in ultra dense heterogeneous networks, where transitions from non-LoS to strong LoS channels alter the performance of conventional resource allocation strategies \cite{Lopez-Perez2015}. The works listed under the round robin category in Table~\ref{tab:optimization-criteria}, use the method as a benchmark to assess MUDiv gains in MU-MIMO systems.
It is worth mentioning that in the reviewed literature, the system performance is assessed over a large number of channel realizations. For example, using the average sum rate per channel use, also called spectral efficiency, measured in bit/s/Hz. Other common performance metrics are \cite{Feng2013}: the \textit{throughput} or average rates achieved for packet transmission with practical modulations and realistic coding schemes; and  the \textit{goodput}, which is the total average rate successfully delivered to the scheduled users, including layer 2 overheads and packet retransmission due to physical errors \cite{Capozzi2013}.

\subsection{WSR with QoS Optimization}
\label{section:performance-uf:wsr-with-qos}

The WSR optimization problem can incorporate different types of QoS requirements. The QoS defines  a network- or user-based objective [cf. Definition~\ref{defn:quality-of-service}], and the weights $\omega_{k} \ \forall k$, establish priorities according to the service provided to each user.
Taking into account QoS requirements reduces the flexibility of the resource allocation strategies. For instance, in the case of opportunistic transmission, achieving a target QoS reduces the value of CSIT, i.e., the channel conditions have to be weighted by requirements coming from the upper layers \cite{Gesbert2007a}. System optimization considering QoS has a broader scope and once information from upper layers is considered, a more complex cross-layer optimization problem arises.

The QoS is defined according to the system model, which might include individual traffic characteristics or SINR requirements. Satisfying QoS requirements can be realized at different time granularities: each user must achieve a prescribed performance metric, e.g. instantaneous SINR or data rate \cite{Schubert2006}; or a network performance metric must reach a stable behavior over time, e.g. attaining finite buffer sizes \cite{Kobayashi2007}. Below, we provide a classification of the optimization criteria subject to QoS constraints in MU-MIMO scenarios.

\subsubsection{Target SINR and Error Rates}
\label{section:performance-uf:qos:target-rates}

The QoS can be guaranteed as long as individual SINR requirements are met, which also satisfies target error and peak rates. The QoS can be defined by a monotonic and bijective function of the SINR: one example is the bit error rate (BER), given a particular modulation and coding scheme (MCS) \cite{Tsai2008,Chung2010}; another example is the Shannon capacity \cite{Schubert2006}. 
Different tradeoffs between global and individual performance are achieved under SINR constraints: maximizing the WSR \cite{Weeraddana2012}, max-min weighted SINR \cite{Schubert2004,Bartolome2006,Cai2011,Tan2012,Stanczak2009}, sum power minimization \cite{Schubert2004,Schubert2006,Zhang2007b,Hong2012}, and other hybrid formulations, see \cite{Hong2012} and references therein. 

It is worth noticing that in systems with SINR requirements, it is usually assumed that for each channel realization, a set of  compatible users $\mathcal{K}$ has been previously selected. The system optimization is realized by power allocation and precoder design \cite{Weeraddana2012}. This means that $\mathcal{K}$ must be a feasible set of users [cf. Definition~\ref{defn:feasible_set_of_users}], and user scheduling is required only if the constraints associated with $\mathcal{K}$ are infeasible \cite{Stanczak2009,Hong2012}. In that case, it is necessary to relax the initial conditions taking one of the following actions: 

\begin{enumerate}
	\item Reducing the number of selected users applying \textit{admission control} \cite{Ahmed2005,Stridh2006a,Bartolome2006} or \textit{user removal} schemes \cite{Koutsopoulos2008,Mahdavi2010, Maciel2010,Castaneda2013}. These procedures are implemented in systems where the achievable SINRs take any positive real value, e.g. \cite{Schubert2004,Caire2006,Zhang2007b,Cai2011,Tan2012}. The removal process can be thought as a form of scheduling, used to generate a feasible set of users [cf. Definition~\ref{defn:feasible_set_of_users}]. By identifying users with unfeasible constraints, one can drop them temporarily (according to priorities), or re-schedule them over orthogonal resources (e.g. other carriers or time slots) \cite{Wang2007}.
	\item By relaxing the individual QoS requirements $\forall k \in \mathcal{K}$, one can achieve resource allocation feasibility [cf. Definition~\ref{defn:resource_feasibility}]. For instance, consider homogeneous target rates, then all scheduled users can be balanced to a common SINR \cite{Koutsopoulos2008}. For heterogeneous QoS, one can iterate over different SINR levels per user (e.g. provided a set of MCSs), so that link adaptation can take place, see \cite{Weeraddana2012,Castaneda2013,Destounis2015,Bartolome2006}.  
\end{enumerate}

Cross-layer algorithms subject to rate constraints can be designed so that $\mathcal{K}$ and its associated resources are jointly optimized. Analytical results in \cite{Lau2009} show that under mild conditions, there exists, with probability one, a feasible set of users $\mathcal{K}$ that can be found though low-complexity algorithms [cf. Section~\ref{section:algorithms:aggregated-utility}]. 

The transmission errors (bit, frame, symbol, packet) are usually due to noisy channels and inaccurate CSIT. For the former factor, the errors occur due to the effect of non-ideal channel coding and finite blocklength channel coding. The error rates can be reduced by implementing stronger channel codes and longer blocklengths.\footnote{In practice, for reasonable block length (e.g. 8 kbits) and strong coding (e.g. LDPC), the Shannon capacity can be approached to within 0.05 dB for a target FER of 10$^{−3}$ \cite{Lau2005a,Tse2005,Lau2009}.} The error rates also occur due to CSI quantization (limited feedback), estimation errors, delays (channel outage), and distortion during the feedback \cite{Shirani-Mehr2010,Zhang2011,Kobayashi2011}. Therefore, IUI  cannot be fully suppressed by precoding techniques [cf. Definition~\ref{defn:interference-limited-system}], which degrades the achievable SINRs and BER figures. Nonetheless, multi-antenna receivers can implement efficient processing techniques to increase their SINRs and combat quantization errors, e.g., quantization-based combining (QBC) \cite{Jindal2008}. Under partial CSIT conditions, any practical rate adaptation scheme achieves performance between the worst-case outage rate and the ideal-case, where the achievable rate equals the mutual information \cite{Shirani-Mehr2010}.

\subsubsection{Queue Stability in MU-MIMO Scenarios}
\label{section:performance-uf:qos:queue-delay-stability}

A network-based optimization requires cross-layer algorithms taking as inputs the queue state information (QSI), CSIT, and their inter-dependence \cite{Swannack2004}. There exist a rich literature on optimization subject to queue or packet delay constraints, specially for systems with orthogonal resource allocation (e.g. one user per time-slot or sub-carrier),  see \cite{Wang2007,Capozzi2013,Asadi2013} for an in-depth overview.  
According to the QSI relevance, MU-MIMO systems fall in two categories  \cite{Lau2006,Kobayashi2007}:

\begin{enumerate}
	\item \textit{Systems with infinite backlogs}: 
	In this scenario, the transmitter is assumed to always have data to send and infinite buffer capacity. In other words, the transmitter fully knows the intended data for each user. The objective of the resource allocation strategies is, in general, to maximize the WSR [cf. Section~\ref{section:performance-uf:weighted-sum-rate}]. The weight $\omega_{k}$ of the $k$-th user is used to reach a desired throughput-fairness tradeoff, instead of expressing the urgency of data flows. Thus, it can be assumed that the queues are balanced and users cannot be discriminated based on them \cite{Swannack2004}. The service provided is delay-insensitive, and the traffic is managed so that the scheduled users attain non-zero rates \cite{Souihli2010}. MU-MIMO systems without QSI information fall into this category.
	\item \textit{Systems with bursty traffic and limited buffer size}: 
	In these systems, packet data models with stochastic traffic arrivals are considered. The performance assessment is analyzed from the delaying and queuing perspectives. The WSR maximization attempts to balance between opportunistic channel access and urgency of data flows. The resource allocation algorithm must guarantee finite average buffer occupancy, i.e., \textit{stability of the queues lengths} for all users \cite{Caire2006,Chen2013}. It is worth mentioning that by Little's theorem \cite{Lau2006,Wang2007,Chen2013}, achieving stability in the average queues is equivalent to minimize the average packet delay in the steady state.\footnote{For systems with queue or delay requirements, a channel/QoS-aware scheduling algorithm must be supported by admission control mechanisms in order to guarantee feasibility \cite{Capozzi2013}.} The system model might consider different sources of delay, e.g., buffer congestion or destination unavailability (outage delay), which define the type of scheduling policy to be implemented \cite{Souihli2010}. The choice of the system model and problem formulation depend on the desired tradeoff between tractability and accuracy in each particular scenario \cite{Lau2006}.
\end{enumerate}

In MU-MIMO systems where queue stability is optimized, the scheduling algorithm has to guarantee that the average queue lengths of all users are bounded. Simultaneously, the available CSIT should be opportunistically exploited for throughput maximization \cite{Caire2006}. 
This goal can be achieved by establishing queue-based WSR as the optimization criterion. The weights $\omega_{k}$ in (\ref{eq:weighted-sum-rate-function}) can be defined according to the system requirements: considering the packet queue length or packet departure rates at each scheduling interval \cite{Swannack2004,Shirani-Mehr2010,Huang2012a,Chen2013,Destounis2015}; considering fairness with heterogeneous traffic rates \cite{Torabzadeh2010,Tan2012}; or considering service-oriented requirements, e.g., BER, delay tolerance, and packet dropping ratio for real- or non-real-time services \cite{Chung2010,Souihli2010,Tsai2008}. 
The queue lengths not only define the priority or urgency of the traffic, but they can also define the encoding order, $\pi(\cdot)$ in (\ref{eq:general-combinatorial-problem:cost-function}), see \cite{Swannack2004,Caire2006,Destounis2015}. In general, queue lengths are not symmetric or balanced, which means that the WSR optimization is dominated by the QSI, rather than by the CSIT \cite{Swannack2004,Huang2012a}.

The characteristics of the MU-MIMO system model and the set of constraints define the resource allocation strategy that achieves stability. Several factors may affect the performance under QoS constraints:  CSIT accuracy \cite{Huang2012a,Chen2013}, CSIT statistics \cite{Shirani-Mehr2010}, and the number of resources used for channel estimation \cite{Destounis2015}; the spatial compatibility between co-scheduled user and the SNR regime \cite{Swannack2004}; the user priority based on the type of service \cite{Tsai2008,Chung2010}; or the number of simultaneous spatial streams \cite{Souihli2010}.
The overall optimization can also consider delays due to packet losses and retransmissions. This is supported by ACK/NAK (handshake) signaling exchange in the upper layers. Such a protocol is used for automatic repeat request (ARQ), to convey an error-free logical channel to the application layers \cite{Love2008}. Works in \cite{Shirani-Mehr2010,Shirani-Mehr2011} and references therein, discuss algorithms to solve the WSR with queue constraints taking into account ARQ protocols. We refer the reader to \cite{Lau2006,Georgiadis2006} for a comprehensive texts on cross-layer design under queue and delay constraints.

{\color{black}{

\subsection{Discussion and Future Directions}
\label{section:performance-uf:discussion}

The sum-rate maximization is the main criterion to assess conventional cellular system, specially in scenarios with scarce and expensive radio resources, e.g., crowded sub-6 GHz bands. Conventional cellular planning usually considers few high-power transmitters that provide high spectrum efficiency, at expense of other performance metrics, such as energy efficiency (EE) \cite{Tombaz2011}. 5G mobile communications will include dense deployments, operating at higher frequencies, i.e., mmWave, and with very heterogeneous radio resources per base station \cite{Andrews2014,Baldemair2015,Nguyen2015a}.
The criteria presented in Sections~\ref{section:performance-uf:weighted-sum-rate} and \ref{section:performance-uf:wsr-with-qos} are used to define single objective functions subject to a set of constraints. However, 5G networks will require the simultaneous optimization of multiple criteria: peak data rates, traffic and user load across the network, fairness, quality of service and experience, EE, etc. These multiple objectives are usually coupled in a conflicting manner, such that optimization of one objective degrades the other objectives.

One approach to find a tradeoff between objectives is by deriving an explicit expression that can be used as a single objective function. For instance, EE has been jointly optimized with peak rates, WSR, or load balance in heterogeneous networks, see \cite{Li2015,Yang2015a,Liu2016,Nguyen2015a} and references therein. Fundamental tradeoffs among EE and delay, sum rate, bandwidth and deployment cost have been derived in \cite{Chen2011d}. Balancing  fairness and spectral efficiency is a well known problem strived for wireless communications \cite{HuaizhouShi2014}, and several works have formulated its joint optimization as a single objective function, e.g., \cite{HoTingCheng2008,Sediq2013,BinSediq2014}. To illustrate several conflicting objectives, consider that allocating resources to users with strong channels can satisfy QoS requirements and improve EE. However, it might also incur in unfair resource distribution across users and unbalanced load among BSs. Now assume that all users have equally good channels, and transmitting low traffic loads to all users increases the coverage area, but this might be very energy inefficient. Unconstrained EE maximization may result in operating points with low spectral efficiency per user \cite{Bjornson2016a}.

Another approach for the joint optimization of multiple objectives, is to sample the solution space and chose the operative point that satisfies a predefined tradeoff. A mathematical framework to address multi-objective optimization problems for wireless communications has been proposed in \cite{Bjornson2013,Bjornson2014a}. The solution space of such kind of mathematical problems, generally, does not have a unique point that can optimally satisfy all objectives. The main challenges are to characterize and understand efficient operating points within the solution space, so that the objectives are balanced.
Authors in \cite{Bjornson2014a} provide an example of this approach by jointly optimizing individual peak rates, average area rates, and EE for a massive MIMO setting. Numerical results illustrate the conflicting nature of the objectives: the average area rates increases with the number of served users; the individual peak rates can be increased when the power is split among few users; and high EE is attained if the rate per user is small.

Techniques such as multi-objective optimization \cite{Bjornson2014a} and metaheuristic optimization [cf. Section~\ref{section:algorithms:bio-inspired}], sample the solution space to find close-to-optimal solutions. However, the former is constructed from a mathematical framework, which characterizes the solution space, in particular, the desirable operative points.
We foresee that multi-objective optimization techniques will play a primal role towards 5G, as the overall network optimization become more complex. Those techniques can provide guidelines for network design, simplify parametrization, and assess compounded optimization criteria for heterogeneous networks. The network designer must establish resource allocation strategies to conciliate conflicted objectives, keeping in mind that the ultimate goals are user experience, satisfaction, and operators' interest \cite{Sharifian2016}. 
}}

\section{MU-MIMO Scheduling}
\label{section:algorithms}

\begin{table*}[t] \scriptsize
	\renewcommand{\arraystretch}{1.3}	
	\caption{User Scheduling Methods for MISO ($N=1$) and MIMO ($N>1$) configurations, and full ($B=\infty$) and partial ($B<\infty$) CSIT}
	\label{tab:algorithms}%
	\centering		
	\begin{tabularx}{\textwidth}{ *{6}{>{\centering\arraybackslash}X} }  
		\toprule	
		\textbf{Method} &   \textbf{Utility-based}    &   \textbf{CSI-Mapping}    & \textbf{Metaheuristic}\newline \textbf{(stochastic)}  & \textbf{Classic}\newline \textbf{Optimization}& \textbf{Exhaustive}\newline \textbf{Search} \\
		\midrule
		\textbf{MISO}, $B=\infty$ & \cite{Dimic2005, Moon2013, Huang2013, Wang2008a, Huang2012b, Maciel2010, Driouch2012, Lau2005a, Koutsopoulos2008, Kobayashi2007,Liu2014a,YiXu2014,Bogale2015} & \cite{Tu2003, Wang2008a, Yoo2006, Jiang2006, Dai2009, Lee2014, Adhikary2013, Adhikary2013a, Razi2010, Mao2012, Shi2012, Maciel2010, Chung2010, Driouch2012, Yoo2005a, Christopoulos2013, Wang2015, Yang2011, Zhang2007b, Swannack2004, Shirani-Mehr2011, Castaneda2015, Jang2011,Kountouris2006b,Lee2014c,Liu2015a} & \cite{Bjornson2014, Stridh2006a, Lau2005a, Cottatellucci2006, Wang2015, Zhang2007b, Park2014} & \cite{Maciel2010, Stridh2006a, Matskani2008, Ku2015, Song2008} & \cite{Lau2005a, Destounis2015, Swannack2004} \\
		\midrule
		\textbf{MISO}, $B<\infty$ & \cite{Xia2009, Conte2010, Tang2010, Vicario2008, Kountouris2007, Kountouris2006, Shirani-Mehr2010, Zhang2011, Ravindran2012,Hammarwall2008} & \cite{Nam2014, Lee2014, Adhikary2013, Adhikary2013a, Xia2009, Zorba2008, Yang2011, Conte2010, Zakhour2007, Huang2009a, Lau2009, Huang2012a, Yoo2007, Kountouris2007, Kountouris2008, Dai2008, Ravindran2012, Kountouris2006a, Wang2007a, Choi2007c, Min2013, Sohn2010, Kountouris2008a, Kountouris2005a,Kountouris2006b, Kountouris2005, Huang2007, Xu2010, Sohn2012, Khoshnevis2013, Simon2011, Wagner2008,Xu2009a,Hammarwall2008} & \cite{Nam2014, Zhang2011} & -   & \cite{Zakhour2007} \\
		\midrule
		\textbf{MIMO}, $B=\infty$ & \cite{Shen2006, Ko2012, Chan2007, Tran2010, Torabzadeh2010, Zhang2009, Cui2011a, Yu2013, Park2010, Lossow2013,Chen2007a, Fuchs2007, Zhang2005a, Chen2008, Tran2012, Lim2009, Cheng2014} & \cite{Bayesteh2008, Shen2006, Wang2010, Ko2012, Tran2010, Wang2010a, Nam2014a, Souihli2010,  Jagannathan2007, Chen2007a, Fuchs2007, Tejera2006, Zhang2005a, Wang2006, Sigdel2009a, Yi2011, Sun2010, Tran2012, Lim2009, Cheng2014, Aniba2007} & \cite{Elliott2009,Elliott2012, Hei2009} & \cite{Chan2007, Moretti2013} & \cite{Chan2007, Gesbert2007a} \\
		\midrule
		\textbf{MIMO}, $B<\infty$ & \cite{Chae2008, Jindal2008,Rico-Alvarino2014, Fan2014, Wang2012d} & \cite{Chae2008, Chen2013, Trivellato2008, Nam2014a, VanRensburg2009, Hosein2009, Schellmann2010, Sharif2005,Zhang2007a} & -  & \cite{Fan2014} & - \\
		\bottomrule
	\end{tabularx}%
\end{table*}%

Resource allocation can be performed over orthogonal radio resources \cite{Capozzi2013,Asadi2013},  where the allocation decisions are made using individual estimates of  BER, SINR, or rate. Scheduling the best user per resource is performed by low-complexity algorithms that do not depend on full CSIT. For instance, a sorting-based algorithm at the transmitter can schedule users based on the CQI. In contrast, the scheduling algorithms for MU-MIMO systems attempt to allocate resources in a non-orthogonal fashion. The users cannot compute their achievable BER or rates since those metrics depend on the CSIT, powers, and precoders assigned to other co-scheduled users.

To illustrate the scheduling in problem (\ref{eq:general-combinatorial-problem:cost-function}), consider a MU-MIMO system with a single transmitter equipped with $M$ antennas, and $K$ competing users, each equipped with $N$ antennas. Assuming that linear precoding is implemented [cf. Section~\ref{section:precoding:linear}], the maximum number of co-scheduled users per resource is given by $K_{\max} =\lfloor M/N \rfloor$. 
Define $\hat{\mathcal{K}}^{(ss)} = \{ \mathcal{K} \subseteq \{1,2, \ldots, K\}: 1 \leq |\mathcal{K}| \leq K_{\max} \}$ as the \textit{search space} set containing all possible user groups with limited cardinality.\footnote{$\hat{\mathcal{K}}^{(ss)}$ contains the sets of cardinality one, i.e., SU-MIMO configuration. Efficient scheduling rules must dynamically switch between SU- and MU-MIMO configurations  \cite{Chen2008,Ho2009,Sigdel2009a,Schellmann2010,Liu2012,Fan2014}.} The optimal set, $\mathcal{K}^{\star}$, that solves (\ref{eq:general-combinatorial-problem:cost-function}) lies in a search space of size $|\hat{\mathcal{K}}^{(ss)}| = \sum_{m=1}^{K_{\max}} {K \choose m}$. The set $\mathcal{K}^{\star}$ can be found by brute-force exhaustive search over $\hat{\mathcal{K}}^{(ss)}$, which is computationally prohibitive if $K \gg K_{\max}$. 
The search space, and in turn the complexity of the problem, varies according to the system model and constraints:  the values of $M$ and $N$, the number of streams per user, individual QoS or power constraints, the set of MCSs, etc.

\begin{figure}[t!]
	\centering
	\includegraphics[width=0.98\linewidth]{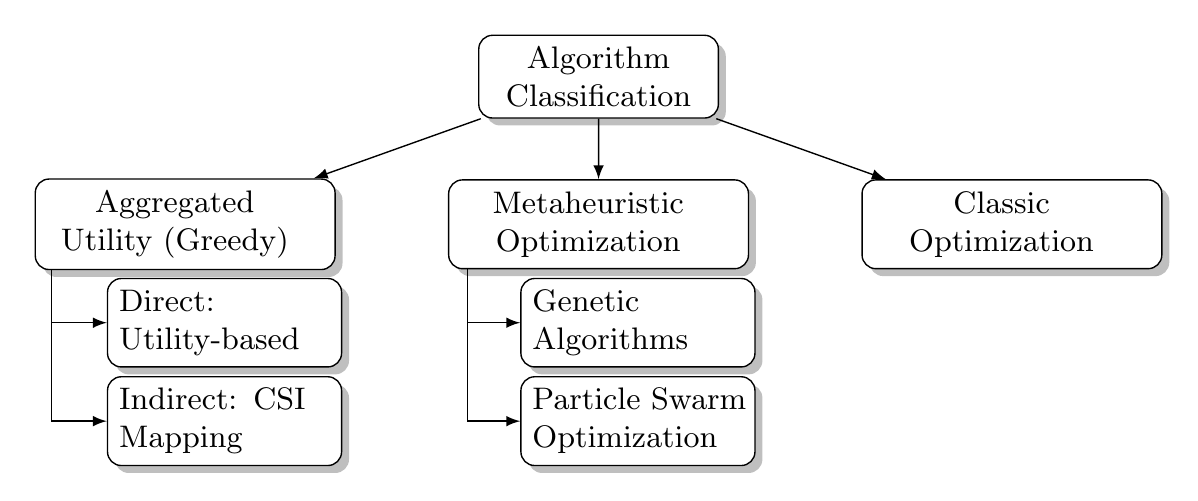}\\
	\caption{Classification of the User Scheduling Algorithms for MU-MIMO}
	\label{fig:algorithm_calssification}
\end{figure}

By plugging the utility function $U(\cdot)$, [cf. Section~\ref{section:performance-uf}], into (\ref{eq:general-combinatorial-problem:cost-function}), we obtain a mathematical formulation of the MU-MIMO resource allocation problem. 
The optimal scheduling rule that solves (\ref{eq:general-combinatorial-problem:cost-function}) in a feasible and efficient way is still an open problem. A number of algorithms have been proposed in the literature to circumvent the high complexity of finding $\mathcal{K}^{\star}$, and its associated resource allocation. 
A sub-discipline of optimization theory, known as heuristic search \cite{Williamson2010}, provides solutions (approximation algorithms) to a category of discrete problems. Such problems cannot be solved any other way or whose solutions take a very long time to be computed (e.g. NP-hard problems). User scheduling falls in such a problem category, and the common approach in the literature is to design suboptimal heuristic algorithms that  balance complexity and performance. The optimal solution, $\mathcal{K}^{\star}$, is usually computed for benchmarking in simple MU-MIMO scenarios. 

In this section, we provide a classification of the most common scheduling algorithms in the literature of MU-MIMO. Table~\ref{tab:algorithms} and Fig.~\ref{fig:algorithm_calssification} show the scheduling algorithms classification, the scenarios in which they are applied, and the methodology they follow. Most algorithms operate at the single resource level, i.e., considering optimization over one sub-carrier or time-slot. Extensions to scenarios with multiple carriers or codes are straightforward, e.g., \cite{Chung2010, Driouch2012, Tsai2008, Park2014,Conte2010, Khoshnevis2013,Elliott2009, Chan2007, Yu2013,Schellmann2010, Lossow2013,Fuchs2007, Tejera2006, Zhang2005a, Cheng2014}.

\subsection{Aggregated Utility Based Selection}
\label{section:algorithms:aggregated-utility}

Let us reformulate (\ref{eq:general-combinatorial-problem:cost-function}) as a multiuser WSR maximization problem as follows

\begin{equation} \label{eq:inner-outer-problem}
	\underbrace{ \underset{ \mathcal{K} \subseteq \bar{\mathcal{K}}: |\mathcal{K}| \leq c_p }{\max} \ \
	\underbrace{ \underset{ \mathbf{W}, \mathbf{P}, \pi }{\max} \ \  \sum_{k \in \mathcal{K}} \omega_{k} \log_{2} \left( 1+ \text{SINR}_{k} \right)  }_\text{Inner problem: $R(\mathcal{K})$} }_\text{Outer problem}
\end{equation}
where $\mathbf{W}$ and $\mathbf{P}$ summarize the precoding weights and the powers assigned to the users in $\mathcal{K}$, and $\pi$ defines an encoding order. 
The formulation in (\ref{eq:inner-outer-problem}) illustrates the fact that the resource allocation problem can be decomposed into different subproblems, which are related to different network layers. Furthermore, under specific system settings and constraints the subproblems can be decoupled. 

The weights $\omega_{k}$, $\forall k \in \mathcal{K}$, can be defined according to the service or application delivered to the users, [cf. Section~\ref{section:performance-uf:wsr-with-qos}]. The inner problem in (\ref{eq:inner-outer-problem}) attempts to maximize the WSR for a given set $\mathcal{K}$. Finding a solution requires joint admission control, precoding design, and power control, e.g. \cite{Stridh2006a, Matskani2008, Liu2011, Hong2012}. A solution to the inner problem exists when the multiuser channel associated with $\mathcal{K}$, has specific spatial dimension defined by the precoding scheme. The channel dimensions must be guaranteed by the solution of the outer problem. If linear precoding schemes are implemented, [cf. Section~\ref{section:precoding:linear}], the original problem can be relaxed: the cardinality of $\mathcal{K}$ is upper bounded by $K_{\max}$; the encoding order $\pi$ can be omitted since it does not affect the precoding performance;  and power allocation can be computed via convex optimization or by heuristic algorithms [cf. Section~\ref{section:power-allocation}]. 
Finding a solution to the combinatorial outer problem implies that the scheduler accomplishes several goals: it exploits MUDiv to maximize multiplexing gains; and it finds the multiuser channel with the best set of spatially compatible users [cf. Definition~\ref{defn:compatible_set_of_users}].
Problem (\ref{eq:inner-outer-problem}) might include constraints over several discrete parameter sets, such as $\mathcal{K}$, the encoding orders, MCSs, spatial streams per user, etc. The complexity of the combinatorial problem grows exponentially with each discrete set size \cite{Koutsopoulos2008}.

Due to the hardness of problem (\ref{eq:inner-outer-problem}), one common approach is to find suboptimal solutions  through heuristic \textit{greedy opportunistic algorithms}. Given a fixed precoding structure and CSIT, the set $\mathcal{K}$ is constructed by a sequence of decisions. Every newly selected user finds a locally maximum for an objective function. The optimized utility $U(\cdot)$ can be defined by the WSR (direct approach) or by a metric of spatial compatibility (indirect approach). The greedy scheduling rule requires low computational complexity and is easy to implement. However, it does not guarantee neither performance nor convergence to the optimal solution \cite{Williamson2010}.

\subsubsection{Direct Approach}
\label{section:algorithms:aggregated-utility:objective-function}

The seminal works \cite{Dimic2005,Shen2006} proposed a simple construction of $\mathcal{K}$, by iteratively selecting users as long as the aggregated objective function improves its value. 
In \cite{Dimic2005}, a greedy user selection (GUS) based on sum-rate maximization was proposed. First, the user with the highest rate is selected, and the following selected user will maximize the total sum-rate. This method was improved in \cite{Wang2008a},  using sequential water-filling to eliminate users with zero transmit power after selection and power allocation. The algorithms in \cite{Dimic2005,Wang2008a} might be computational demanding because calculation of the precoders and power allocation is required each iteration. Furthermore, in the case of imperfect CSI, the above algorithms fail to determine the optimal number of selected users. 
A generic structure of this kind of user selection is presented in Algorithm~\ref{alg:algorithms:direct-approach}. In this approach, the transmitter must have channel information so that every iteration the precoders $\mathbf{W}_k$, $\forall k \in \mathcal{K}$, are recalculated to solve the inner problem in (\ref{eq:inner-outer-problem}). If linear precoder schemes are implemented, the maximum number of iterations is at most $K_{\max}$. 
Since the transmitter knows the precoders and the transmit powers, the achievable rates, SINR, or BERs are known each iteration, allowing the scheduler to identify when it is worth adding more users to $\mathcal{K}$.

This approach is particularly effective if there are no constraints on the cardinality of $\mathcal{K}$, or if the network is sparse, i.e., $|\batchmode\mathcal{K}|\approx M$ \cite{Gesbert2007a}. For instance, if ZF-based precoding is used, the multiplexing gain is not maximized at the low SNR regime, i.e., the total number of scheduled users must be strictly less than $K_{\max}$. In the high SNR regime, multiplexing gains can be maximized if full CSIT is available \cite{Caire2003}, otherwise the system may become interference limited [cf. Definition~\ref{defn:interference-limited-system}].
Several works have modified the structure of Algorithm~\ref{alg:algorithms:direct-approach} to reduce complexity and improve performance, see \cite{Lau2005a,Jiang2006,Hammarwall2007,Chen2008,Wang2008a,Lim2009,Tran2010,Yi2011,Sigdel2009a,Tran2012,Adhikary2013a}. The works \cite{Dimic2005,Hammarwall2007,Wang2008a,Huang2013} showed that the solution space can be shrunk at each iteration, reducing the computational complexity. Based on the characteristics of the powers that solve the inner problem in (\ref{eq:inner-outer-problem}), the scheduler defines a reduced set of candidate users, $\mathcal{K}^{temp}$, for the next iteration.

Nonetheless, such methods also exhibit the flaw of not being able to identify the optimal cardinality, $|\mathcal{K}|$. As such, there are redundant users in the selected user subset, i.e., users that can be deleted from the selected user subset to yield a performance increase. This is an inherent flaw of any greedy incremental algorithm, due to the non-iterative cumulative user selection procedure. In \cite{Huang2012b,Huang2013}, it is proposed to add the \textit{delete} and \textit{swap} operations as means to tackle the redundant user issue. However, this approach increases the complexity, even as compared to GUS, as it involves matrix inversions, projections, and an iterative procedure. Moreover, its performance is sensitive to CSI inaccuracies due to the increased error in the projection operation using imperfect CSI, and it does not necessarily find the best user subset in the imperfect CSI case.

\begin{algorithm}[t!] \small
	\begin{algorithmic}[1]
		\STATE Set $\mathcal{K}^{temp} = \bar{\mathcal{K}}$, select the user $k$ with the strongest channel, update $\mathcal{K} = \{k\}$, calculate precoders and powers, compute the achievable rate $R(\mathcal{K})$ in (\ref{eq:inner-outer-problem}).
		\STATE Find the user $i \in \mathcal{K}^{temp} \setminus \mathcal{K}$ that maximizes $R(\mathcal{K} + \{i\})$.
		\STATE	If $|\mathcal{K} + \{i\}| \leq K_{\max}$ and $R(\mathcal{K} + \{i\}) > R(\mathcal{K})$, then update $\mathcal{K} = \mathcal{K} + \{i\}$, modify $\mathcal{K}^{temp}$, and go back to step 2.\\ Otherwise go to step 4.	
		\STATE Compute final precoders, powers, and WSR for $\mathcal{K}$.
	\end{algorithmic}
	\caption{Utility-Based Scheduling (Direct)}
	\label{alg:algorithms:direct-approach}
\end{algorithm}

\subsubsection{Indirect Approach}
\label{section:algorithms:aggregated-utility:csi-mapping}

In this approach, the inner and outer problems in (\ref{eq:inner-outer-problem}) are decoupled. To illustrate the structure of the resource allocation strategy, let us reformulate (\ref{eq:inner-outer-problem}) as two problems that must be solved sequentially:

\begin{subnumcases}{}  
	\mathcal{K}^{map} =  \underset{ \mathcal{K} \subseteq \bar{\mathcal{K}}: |\mathcal{K}| \leq K_{\max} }{ \arg \max} \ f(\mathbf{H}(\mathcal{K})) &   \label{eq:csi-mapping-problem-1} 
	\\
	\underset{ \mathbf{W}, \mathbf{P}, \pi }{\max} \ \  \sum_{k \in \mathcal{K}^{map}} \omega_{k} \log_{2} \left( 1+ \text{SINR}_{k} \right) & \label{eq:csi-mapping-problem-2}
\end{subnumcases}

Equation (\ref{eq:csi-mapping-problem-1}) describes a combinatorial user grouping problem, where $f(\mathbf{H}(\mathcal{K}))$ is a metric of spatial compatibility, [cf. Section~\ref{section:metrics-spatial-compatibility}]. The user grouping is an NP-C problem, whose solution is found via ExS \cite{Maciel2010}. Observe that finding a solution to problem (\ref{eq:csi-mapping-problem-1}) does not require the computation of the precoders and powers. Instead, it depends on CSIT, algebraic operations defined by a given mapping function $f(\mathbf{H}(\mathcal{K}))$, and an iterative procedure. It is worth noting that (\ref{eq:csi-mapping-problem-1}) may include the weights $\omega_{k}$ to cope with QoS or fairness requirements, see \cite{Yoo2006}. 
In contrast, provided the set $\mathcal{K}^{map}$, problem (\ref{eq:csi-mapping-problem-2}) is tractable and can be solved efficiently via convex optimization \cite{Lau2005a,Caire2003,Godara2002Ch18,Schubert2004}.

Early works on MU-MIMO systems (e.g. \cite{Caire2003,Tu2003,Swannack2004}) pointed out that under certain conditions of the weights, $\omega_{k}$, and SNR regime, the objective function of the inner problem in (\ref{eq:inner-outer-problem}), is dominated by the geometry of the multiuser channel $\mathbf{H}(\mathcal{K})$.
The rational behind the reformulation in (\ref{eq:csi-mapping-problem-1})-(\ref{eq:csi-mapping-problem-2}), is that scheduling a set of spatially compatible users [cf. Definition~\ref{defn:compatible_set_of_users}], is crucial to suppress IUI. Refining the power allocation and precoding weights in (\ref{eq:csi-mapping-problem-2}) requires less computational effort than solving the combinatorial problem (\ref{eq:csi-mapping-problem-1}).
A generic structure of this scheduling algorithms is presented in Algorithm~\ref{alg:algorithms:indirect-approach}.

The complexity of finding a solution to problem (\ref{eq:csi-mapping-problem-1}) can be simplified under certain conditions. Assuming that ZF-based precoding schemes are used, results in \cite{Caire2003,Lau2006} establish that the optimal admitted set of users achieves full multiplexing gains, i.e., $|\mathcal{K}|=K_{\max}$, if the system operates in the high SNR regime. Therefore, the cardinality constraint in (\ref{eq:csi-mapping-problem-1}) is given by $|\mathcal{K}| = K_{\max}$, which shrinks the search space size. A common approach in the literature is to assume that $K_{\max}$ users can be co-scheduled when solving problem (\ref{eq:csi-mapping-problem-1}), and refining the set $\mathcal{K}$ when solving problem (\ref{eq:csi-mapping-problem-2}).

The majority of the works that have proposed algorithms to solve problem (\ref{eq:csi-mapping-problem-1}), e.g. \cite{Tu2003, Jiang2006, Dai2009, Tejera2006,Yoo2006, Nam2014, Bayesteh2008,Mao2012,Maciel2010}, use ZF precoding for the second problem (\ref{eq:csi-mapping-problem-2}). Consequently, the NSP is the metric of spatial compatibility that maximizes the sum-rate for the set of selected users $\mathcal{K}$, [cf. Section~\ref{section:metrics-spatial-compatibility:projection-spaces}].
A considerable amount of research has been focused on reducing the complexity of the NSP computation, e.g., by reusing previous NSP calculations at each iteration, implementing efficient multiuser channel decompositions, or computing approximations of the NSP. The methods highly depend on $M$, $N$, the CSIT accuracy, and the system model.

Different works have proposed a search space reduction per iteration, recalculating and reducing the number of competing users so that spatial compatibility is preserved. A common approach is to preselect a group of candidate users, $\mathcal{K}^{temp}$ in Algorithm~\ref{alg:algorithms:indirect-approach}, based on spatial clustering [cf. Section~\ref{section:metrics-spatial-compatibility:spatial-clustering}]. The candidate users at the next iteration will exhibit channel directions fulfilling an $\epsilon$-orthogonality criterion. This approach, coined as semi-orthogonal user selection (SUS), was originally proposed by Yoo and Goldsmith \cite{Yoo2006,Yoo2005a}. Several variations of this technique can be found in the literature, see \cite{Wang2008a,Lau2009,Sun2010,Razi2010,Sohn2010,Mao2012,Driouch2012,Tran2012,Min2013,Lee2014c}.
Once that $\mathcal{K}$ has been constructed, it may be necessary to modify it, so that problem (\ref{eq:csi-mapping-problem-2}) yields the maximum WSR. A common approach to refine $\mathcal{K}$, is to apply user removal techniques [cf. Section~\ref{section:performance-uf:wsr-with-qos}]. This might be necessary if: the selected channels lacks spatial compatibility; the system operates in extreme SNR regimes; or if the constraints related to $\mathcal{K}$ turn problem (\ref{eq:csi-mapping-problem-2}) infeasible \cite{Lim2009,Maciel2010}. If linear precoding and water-filling are used to solve (\ref{eq:csi-mapping-problem-2}), the set $\mathcal{K}$ can be refined by dropping users that do not achieve a target rate or whose allocated power is zero.

Authors in \cite{Fuchs2007} pointed out that regardless of the number of deployed antennas, it is more efficient to schedule users so that $|\mathcal{K}| < K_{\max}$. This means that the scheduler must seek a tradeoff between maximizing spatial multiplexing gains, and optimizing the WSR for a small set of spatially compatible users. To solve this cardinality problem, some algorithms, e.g. \cite{Fuchs2007,Maciel2010,Ko2012,Cheng2014}, maximize the WSR by comparing the solution of (\ref{eq:csi-mapping-problem-2}) for multiple sets that solve (\ref{eq:csi-mapping-problem-1}). This is performed by sequentially constructing groups of spatially compatible users with different cardinalities, and then selecting the best group based on the achievable WSR.

For the general heterogeneous MU-MIMO scenario, $M\geq N_k > 1$, $\forall k$, the function $f(\mathbf{H}(\mathcal{K}))$ in problem (\ref{eq:csi-mapping-problem-1}) can be defined by a metric of subspace compatibility [cf. Section~\ref{section:metrics-spatial-compatibility:other-metrics}]. The user grouping procedure using such a metric can operate in a greedy fashion, as in Algorithm~\ref{alg:algorithms:indirect-approach}, but every scheduler design depends on the system model.
It is worth pointing out that in heterogeneous systems, maximizing spatial compatibility is not enough to guarantee WSR maximization. Multidimensional metrics measure the potentially achievable spatial separability but not the effective channel gain per spatial dimension, [cf. Definition~\ref{defn:effective_channel_gains}]. Efficient user selection algorithms in scenarios with heterogeneous users should not only consider spatial compatibility, but also an estimation of the achievable WSR, SINRs, or effective channel gains \cite{Yi2011}.
Moreover, dynamic allocation of the number of data streams per user must be included in the scheduler, to take into account CSIT accuracy and fulfill QoS constraints. Users with small signal spaces experience less IUI, and results in \cite{Bjornson2013a} show that under certain conditions, transforming a MIMO channel into an equivalent MISO channel can improve spatial separability at the expense of multiplexing gain.

The reliability of scheduling based on metrics of spatial compatibility is highly sensitive to CSIT accuracy. Most of the works in the literature (e.g. \cite{Bayesteh2008, Shen2006, Wang2010, Ko2012, Tran2010}), perform scheduling by evaluating a form of spatial correlation between MIMO channels of different user. However, the correlation between antennas at each receiver is usually not taken into account. Such a correlation may affect the achievable SINR, depending on the signal processing at the receiver, e.g., receive ZF processing \cite{Wang2010}, or receive combining \cite{Jindal2008,Bjornson2013a}.

The inner correlation of the channel $\mathbf{H}_{k}$ of the $k$-th user can be measured by its rank or by the magnitude (dispersion) of its eigenvalues \cite{Tejera2006}. Transmitting multiple spatial streams per user cannot be reliable, specially for high inner correlation. A common approach to simplify the scheduling designs is by limiting the number of streams per user, e.g., using only the strongest eigenvector per selected user  \cite{Bayesteh2008,Bjornson2013a, Wang2010}. For each channel $\mathbf{H}_{k}$, there exists a dominant spatial direction that can be selected for transmission without creating severe IUI, or performance degradation after precoding. In this way, problem (\ref{eq:csi-mapping-problem-1}) can be solved based on antenna or eigen-direction selection, e.g., \cite{Chen2007a,Chen2008,Bayesteh2008,Lim2009,Sun2010,Ko2012,Wang2010}.

\begin{algorithm}[t!] \small
	\begin{algorithmic}[1]
		\STATE Set $\mathcal{K}^{temp} = \bar{\mathcal{K}}$, select the user $k$ with the strongest channel, update $\mathcal{K} = \{k\}$.
		\STATE Find the user $i \in \mathcal{K}^{temp} \setminus \mathcal{K}$ that maximizes $f(\mathbf{H}(\mathcal{K}+\{i\}))$.
		\STATE	If $|\mathcal{K} + \{i\}| \leq K_{\max}$, then update $\mathcal{K} = \mathcal{K} + \{i\}$, preselect users in $\mathcal{K}^{temp}$, and go back to step 2.\\ Otherwise go to step 4.	
		\STATE Perform operations to refine $\mathcal{K}$.
		\STATE Compute precoders and powers to solve (\ref{eq:csi-mapping-problem-2}) for $\mathcal{K}$.
	\end{algorithmic}
	\caption{CSI-Mapping-Based Scheduling (Indirect)}
	\label{alg:algorithms:indirect-approach}
\end{algorithm}

\subsection{Metaheuristic Algorithms}
\label{section:algorithms:bio-inspired}

Several works in the literature address problem (\ref{eq:inner-outer-problem}) using stochastic optimization methods, which find close to optimal solutions to complex and non-convex mixed problems. These non-traditional or metaheuristic methods are an alternative to classical programming techniques \cite{Rao2009,Williamson2010}. The mathematical proof of convergence to the optimal solution cannot be demonstrated, and analytical results or performance analysis cannot be derived. However, these methods may find a close to optimal solution with high probability, without the need of computing derivatives or satisfying convexity requirements (as in standard optimization techniques).    
The principle behind metaheuristic methods is the following: a set of feasible solutions is iteratively combined and modified until some convergence criterion is met. The term stochastic comes from the fact that initial conditions are generally chosen randomly. For each iteration, the internal parameters change according to dynamic probability functions or even randomly. 
The algorithms designed under this approach are based on certain characteristics and behavior of natural phenomena, e.g., swarm of insects, genetics, biological or molecular systems \cite{Rao2009,Zhang2014b}.  
In the context of MU-MIMO systems, two techniques have been used to solve the user scheduling problem: genetic algorithms (GA) and particle swarm optimization (PSO) algorithms.

GA are bio-inspired or evolutionary algorithms, that can solve multiple objective optimization problems with many mixed continuous-discrete variables, and poorly behaved non-convex solution spaces. Unlikely standard optimization methods that rely on a single starting point (e.g. algorithms sensitive to initial conditions \cite{Hammarwall2007}), GA start with a set of points (population). These points increase the argument domain of the optimized function and avoid local optimal problems due to the evaluation of the solution space \cite{Rao2009}.
GA encode potential solutions to the optimization problem in data structures called chromosomes. Due to the binary nature of the variables $\xi_{k}$ in problem (\ref{eq:general-combinatorial-problem:cost-function}), the chromosomes can be represented as binary strings containing valid configurations of the variables $\xi_{k}$, [cf. (\ref{eq:general-combinatorial-problem:set-cardinality})]. A set of chromosomes is referred to as a population, and at each iteration, the chromosomes combine, exchange critical information, mutate, and eventually evolve toward the optimal solution \cite{Lau2005a}.
Practical application of GA to solve the WSR maximization problem (\ref{eq:inner-outer-problem}) have been proposed for different precoding schemes: ZFBF \cite{Lau2005a,Lau2006}, ZFDP \cite{Elliott2012}, and DPC \cite{Elliott2009}. 
Recall that given a set $\mathcal{K}$, the performance of DPC is sensitive to a given user order $\pi$ selected out of $|\mathcal{K}|!$ valid orders. The works \cite{Tu2003,Tejera2006} have proposed heuristic algorithms to define $\pi$. Since the chromosome structure contains different optimization variables, it can include not only the set of selected users but can also the order in which the users are encoded for ZFDP \cite{Elliott2012} and DPC \cite{Elliott2009}.

PSO are behavior-inspired algorithms that mimic the distributed behavior of social organisms (particles). The particles use their own intelligence (based on an individual utility function) and the swarm intelligence (global utility function), to discover good directions that can be followed and verified by the swarm. 
PSO initially defines the swarm as a set of particles randomly sampled from the search space. Each particle has two associated parameters that define how fast it can move toward the optimal solution: position and velocity. In this approach, each particle wanders around in the search space to collect information about the achievable utility function at different locations. This means that the particle looks for the configuration that steers toward the global maximum or the average direction of the swarm. The particles exchange information regarding the best directions, and adjust their positions and velocities accordingly until the swarm converge to the same solution \cite{Rao2009}.  
Authors in \cite{Hei2009} designed PSO algorithms to solve problem (\ref{eq:inner-outer-problem}) using BD precoding. Such an optimization uses the utility function, $U(\cdot)$, of the direct and indirect approaches described in Section~\ref{section:algorithms:aggregated-utility}. The PSO approach can define the swarm direction using the function $R(\mathcal{K})$ in the inner problem (\ref{eq:inner-outer-problem}), or the mapping function $f(\mathbf{H}(\mathcal{K}))$ in (\ref{eq:csi-mapping-problem-1}). This illustrates the fact that the computational complexity of metaheuristic algorithms depends on the optimized objective function.

Algorithm~\ref{alg:algorithms:metaheuristic} shows a sketch of the steps used to solve problem (\ref{eq:inner-outer-problem}) following the GA or PSO approaches. The constants $K_{smp}$ and $K_{best}$ are arbitrarily defined to limit the overall computational complexity and to speed up convergence. The convergence criterion can be defined by the maximum number of iterations or a performance threshold. The details of the operations required to generate new chromosomes or particles, [Step 3 in Algorithm~\ref{alg:algorithms:metaheuristic}], can be found in the reference aforementioned. The robustness of GA and PSO lies in the fact that the best solutions are systematically combined, and the algorithms have reasonable immunity from getting stuck in local minimums. 

Metaheuristic algorithms are efficient techniques for problems where the desired solution does not need to be computed in short time scales.  Although bio-inspired algorithms have been used to optimize resource allocation in wireless networks, e.g. \cite{Rondeau2009,Sampaio2013,Lee2014b,Zhang2014b}, they seem to have limited application in practical MU-MIMO systems. This is because the channel conditions, rate demands, and even the number of competing users may change very rapidly over time. Metaheuristic algorithms require centralized processing and the convergence time remains a critical issue \cite{Lee2014b}. They might suffer from a high computational cost because the objective function is evaluated for each member of the population (swarm). Nevertheless, these methods can provide performance benchmarks to assess other suboptimal, yet practical, resource allocation algorithms.

\begin{algorithm}[t!] \small
	\begin{algorithmic}[1]
		\STATE For the $i=1$ iteration, extract $K_{smp}$ elements (chromosomes or particles) from the search space  $\hat{\mathcal{K}}^{(smp)}(i) =  \{\mathcal{K}_{i,1}, \ldots, \mathcal{K}_{i,K_{smp}}\} \subset \hat{\mathcal{K}}^{(ss)}$. 
		\STATE For the $i$-th iteration, evaluate the objective function for each element in $\hat{\mathcal{K}}^{(smp)}(i)$, select the $K_{best}$ elements with the largest objective function, and discard the other bad elements.
		\STATE Perform operations over the remaining elements to generate an improved population/swarm, $\hat{\mathcal{K}}^{(smp)}(i+1)$, for iteration $i+1$. 
		\STATE If convergence criterion is met, then compute precoders and powers for the solution set.\\ Otherwise go to step 2.
	\end{algorithmic}
	\caption{Metaheuristic Optimization}
	\label{alg:algorithms:metaheuristic}
\end{algorithm}

\subsection{Classical Optimization}
\label{section:algorithms:classical-optimization}

Classical optimization techniques \cite{Rao2009,Boyd2004},  have been used for several resource allocation problems in MU-MIMO systems, e.g., WSR maximization, sum power minimization, max-min SINR, and other objective functions, see \cite{Bjornson2013,Hong2012,Utschick2012,Stanczak2009,Athanasiou2015,Bethanabhotla2016} and references therein. Numerous works have included user scheduling for the optimization of the WSR \cite{Chan2007}, sum power minimization \cite{Matskani2008,Ho2009,Moretti2013}, or some metrics of spatial compatibility \cite{Maciel2010,Castaneda2014}. A large number of optimization problems in MU-MIMO systems are non-convex or non-polynomial time solvable, depending on the system models and constraints \cite{Hong2012}. Due to the high complexity of the resource allocation problem, the conventional optimization solutions must relax the original problem, approximate non-convex constraints in an iterative fashion, or change the domain of the optimization, sacrificing optimality for the sake of tractability.  Nevertheless, by employing a fixed structure for the precoders, the original problem can be simplified, and efficient resource allocation can be performed.

Authors in \cite{Chan2007} presented different approaches to solve (\ref{eq:general-combinatorial-problem:cost-function}) in a OFDMA MU-MIMO scenario. The mathematical formulation renders the combinatorial problem into a convex problem. By analyzing the entire search space, the proposed algorithms perform channel assignment and signal space selection, i.e., the algorithm defines the scheduled users per carrier and the number of spatial streams per user. Such a reformulation of (\ref{eq:general-combinatorial-problem:cost-function}) yields algorithms that decouple the channel assignment in the OFDMA system, i.e., the global optimization is attained by solving problem (\ref{eq:inner-outer-problem}) per sub-carrier. Furthermore, the objective function of the inner problem in (\ref{eq:inner-outer-problem}) is defined in terms of rates and powers, which optimizes the total capacity and power consumption.  
For classical programming techniques, the set of constraints might turn the problem infeasible. The resource allocation algorithms proposed in \cite{Chan2007} identify the assignment sets for which time sharing is the best resource allocation strategy. As discussed in Section~\ref{section:performance-uf:wsr-with-qos}, admission control or user removal techniques can be implemented as a form of scheduling to guarantee feasibility.

Authors in \cite{Moretti2013} have reformulated the user scheduling problem in multi-carrier scenarios as an extension of problems (\ref{eq:csi-mapping-problem-1}) and (\ref{eq:csi-mapping-problem-2}). A close to optimal solution can be found by a sequence of convex and linear programming optimizations. To reduce complexity of the combinatorial problem (\ref{eq:csi-mapping-problem-1}), the mapping function $f(\mathbf{H}(\mathcal{K}))$ is defined as the magnitude of the MIMO channels, and a close to optimal scheduling sequence is designed based on that metric. According to \cite{Ho2009,Moretti2013}, the generalization of (\ref{eq:csi-mapping-problem-2}) for OFDMA systems yields a non-convex problem, but quasi-optimal solutions can be attained by applying the classical dual decomposition method. 
The works \cite{Moretti2013,Maciel2010} have modeled the user scheduling problem in OFDMA systems as a channel assignment problem, which can be solved by efficient algorithms that run in strong  polynomial time \cite{Burkard2009}.

A novel reformulation of the WSR maximization problem, (\ref{eq:inner-outer-problem}), as a semi-definite program  in \cite{Ku2015}, sheds some light on non-explicit convex properties of the joint precoding design, power allocation, and users selection problem. The reformulation requires semi-define relaxation, and  the solution is found by combining convex optimization and sub-gradient projection methods. 
Another approach in \cite{Song2008} extends the max-min fair rate allocation problem in \cite{Schubert2004,Tan2012,Cai2011,Schubert2006} so that joint optimization of precoding, power allocation, and user selection was attained. The problem was formulated as the difference of convex functions and solved by the branch and bound (BB) technique, which is a method used to solve mixed-integer programming problems \cite{Rao2009}. 
The approach in \cite{Song2008} yields an interference limited systems [cf. Definition~\ref{defn:interference-limited-system}], i.e., it is allowed that $|\mathcal{K}|\geq M$, which highly increases the search space and computational complexity. Notice that even if the search space is constrained so that $|\mathcal{K}|\leq M$, there are $2^K -1$ possible user schedules in a MISO configuration that will be enumerated by the BB technique.

As discussed in Section~\ref{section:performance-uf}, admission control is a form of user selection that may be required to guarantee feasibility. In \cite{Matskani2008}, the joint sum power minimization and maximization  of $|\mathcal{K}|$ was formulated as a second-order cone program (SOCP). The set $\mathcal{K}$ is refined iteratively by dropping the user with the largest gap to its target SINR,\footnote{See \cite{Mahdavi2010} for optimal and suboptimal user removal algorithms for QoS and power constrained systems.} (i.e. the most infeasible user), and optimal precoders and powers are found by the SOCP optimization.  
Although the approaches in \cite{Song2008,Matskani2008,Ku2015} provide suboptimal solutions due to the relaxation of non-convex problems, they show that joint optimization of the three variable sets $\mathcal{K}$, $\mathbf{P}$, and $\mathbf{W}$ in (\ref{eq:inner-outer-problem}), is feasible through convex optimization.

Other classical optimization methods have been used in the single-carrier MISO scenario with ZF precoding. 
Quadratic optimization was used in \cite{Maciel2010} to solve problem (\ref{eq:csi-mapping-problem-1}), by optimizing a heuristic function $f(\mathbf{H}(\mathcal{K}))$ that linearly combines channel magnitudes and spatial correlation of the MIMO channels.
In \cite{Castaneda2014}, a heuristic objective function that approximates the NSP has been proposed, and the set $\mathcal{K}$ was found via integer programming. 
The authors in \cite{Chen2013} jointly addressed scheduling and WSR maximization subject to QoS constraints using geometric programming. However, similar to the approach in \cite{Chan2007}, the optimization in \cite{Chen2013} iterates over all solutions of the search space, which limits its application for practical scenarios.

Other works model the resource allocation problem (\ref{eq:csi-mapping-problem-1}), as mathematical problems that have been extensively studied \cite{Korte2006,Burkard2009}: the minimum set cover problem from graph theory (e.g., \cite{Yoo2005a,Driouch2012}); and the sum assignment problem (e.g., \cite{Aniba2007,Castaneda2013b,Dartmann2013,Wang2011b}). The modeling used in these approaches requires the calculation of weights or costs values associated with every possible subset in the solution space, i.e., $\{ \forall \mathcal{K}  \subseteq \bar{\mathcal{K}} : |\mathcal{K}| \leq c_p \}$, where $c_p$ depends on the system parameters. The solutions of such general problems can be found by heuristic algorithms.
Due to the exponentially increasing complexity on $M$, $N$, and $K$, the algorithms based on classical optimization are suitable for offline implementation, benchmarking, and to assess other suboptimal heuristic resource allocation strategies for relatively small values of $K$ \cite{Utschick2012,Hong2012}.

\subsection{Scheduling in Multi-Cell Scenarios}
\label{section:algorithms:multi-cell-selection-classification}

In Section~\ref{section:system-set-ups:multi-cell-scenarios}, we have described the types of transmission schemes in multi-cellular scenarios. Different levels of coordination are required to improve and maintain the performance of users geographically spread within the coverage area. For users located at the cell edge, the simultaneous transmission from different clustered BSs can boost their throughputs (e.g. CoMP \cite{Marsch2011}), whereas users close to their BS do not require interaction with adjacent BSs.
In full signal-level coordinated systems (e.g., Network MIMO or JT), the global user data and CSI knowledge enables a CU to perform CS using the approaches described in Section~\ref{section:algorithms:aggregated-utility} and Section~\ref{section:algorithms:selection-partial-csit}. 
Assuming partial coordination, the CS/CBF can offer flexibility, scalability, and viability, since centralized or distributed algorithms can be implemented with limited signaling overhead and using local CSIT.
The minimum level of cooperation between transmitters can be attained through orthogonal resource allocation. For instance, in LTE systems a technique known as \textit{dynamic point blanking},  prevents transmission at a certain time-frequency resource to reduce interference over such a resource used at a neighboring transmission point \cite{Dahlman2013,Lopez-Perez2015}.

The optimization of CS and CBF can be done based on the methodologies described in Section~\ref{section:algorithms:aggregated-utility}, i.e., the two problems can be addressed either independently or jointly. If CS and CBF are decoupled, CS can be tackled by processing shared information related to statistical CSI or performance metrics extracted from local CSI [cf. Section~\ref{section:metrics-spatial-compatibility}]. The CBF optimization can be solved by optimal or heuristic techniques for precoding design and power allocation, see \cite{Bjornson2013} for a comprehensive review.  
A joint optimization of CS and CBF requires iterative calculation of the precoders and selected users, since both optimization spaces are coupled \cite{Li2014,Ku2015}. Efficient scheduling rules must determine the set of users that maximizes the performance, and generates less interference to neighboring cells.   
The following classification presents some existing methods to implement CS and CBF in multi-cell networks \cite{Pateromichelakis2013,Li2014}.

\subsubsection{Joint CS-CBF}
\label{section:algorithms:multi-cell-selection-classification:join-cs-cbf}

In full coordinated multi-cellular systems, a cluster of BSs defines a super-cell or virtual distributed antenna system with per-BS power constraints. Full coordination allows different transmission strategies: \textit{i}) all clustered BSs send data to all the users in $\mathcal{K}$; \textit{ii}) some BSs jointly serve a subset of $\mathcal{K}$; \textit{iii}) each BS serves a set of users associated to it; or \textit{iv}) each BS performs single transmission (ST) and serves only one user per scheduling interval, which is known as the interference channel model (IFC), see \cite{Cover2006,Jorswieck2008,Larsson2008,Bjornson2013}. The fist and second strategies are exclusive of fully coordinated systems since payload data is shared among BSs. The other strategies are also implemented for distributed CS/CBF. Recent works have extended these transmission schemes to heterogeneous networks, e.g., \cite{Lossow2013,Park2014,Ku2015}. 

Accounting for global CSIT provides flexibility to the CS/CBF design, and several ICI cancellation techniques can be applied. Using classical optimization, the authors in \cite{Ku2015} solved the joint CS/CBF optimization, formulated as an extension of problem (\ref{eq:inner-outer-problem}) for heterogeneous networks.  
The authors in \cite{Zhang2009,Huh2012} addressed the CS by iteratively evaluating the precoders and powers, following the methodology described in Section~\ref{section:algorithms:aggregated-utility:objective-function}. 
The approach in \cite{Park2010} tackled the CBF problem using interference alignment \cite{Jafar2011}, whereas the CS was sequentially solved in a greedy fashion. A dynamic transmission switching between ST and JT was presented in \cite{Park2014}, where a centralized CS exploited individual rate statistics (CDF scheduling). 
In \cite{Marsch2011}, problem (\ref{eq:csi-mapping-problem-1}) was solved using metrics of spatial compatibility (NSP-based user selection), and a centralized resource allocation solved problem (\ref{eq:csi-mapping-problem-2}), [cf. Section~\ref{section:algorithms:aggregated-utility:csi-mapping}]. 
Practical CS schemes were proposed in \cite{Lossow2013} for heterogeneous cellular systems, where macro and small cells coordinate scheduling decisions using an the approach described in Section~\ref{section:algorithms:aggregated-utility:objective-function}. The CBF was optimized using linear precoding and dynamic SU/MU-MIMO switching \cite{Schellmann2010}.

\subsubsection{CS with cyclic CBF}
\label{section:algorithms:multi-cell-selection-classification:cs-cyclic-cbf}

Given a pool of precoding weights, each BS picks a different subset for transmission every scheduling interval, and switches them periodically, in a temporal beam-reuse fashion \cite{VanRensburg2009,Hosein2009}. This method requires minimum coordination and uses information regarding the allocated beams per BS and the interference environment. This allows practical scheduling and mitigates the flashlight effect (changes of the active beams at neighboring transmitters). Another approach in \cite{Dartmann2013}, performs user scheduling by solving a series of assignment problems heuristically, and the precoding weights are optimized every iteration.

\subsubsection{Sequential CS-CBF}
\label{section:algorithms:multi-cell-selection-classification:sequential-cs-cbf}

The clustered BSs jointly select users and assign resources in a sequential fashion. The first BS selects its users and broadcast its decision, then the second BS selects its user based on the decision made by the first BS and so on, see \cite{Cui2011a,Wang2011b,Hossain2011,Dartmann2013,Moon2013}. The user selection per BS can be done using the approaches presented in Section~\ref{section:algorithms:aggregated-utility}, and the scheduling order of the cells can be assigned according to interference levels or using the round robin approach. 
The CS can be performed based on interference constraints as in \cite{Seifi2011}. The BSs schedule their users and  perform resource allocation sequentially, so that the ICI generated to previously selected users is below a threshold. If the interference cannot be mitigated as desired, the interfering BSs remain silent during the scheduling interval, which is a low-complexity dynamic clustering strategy.

\subsubsection{Decoupled CS-CBF}
\label{section:algorithms:multi-cell-selection-classification:decoupled-cs-cbf}
	
A distributed optimization policy interconnects clustered BSs through a CU or master BS, limiting the message exchange to local CSI and scheduling control signaling \cite{Bjornson2013}. 
Problem (\ref{eq:inner-outer-problem}) has three optimization variable sets, namely, $\mathcal{K}$, $\mathbf{W}$, and $\mathbf{P}$, and their joint optimization is difficult to solve optimally and distributively. Still, authors in \cite{Chiang2008} proposed an iterative approach that decouples and optimizes some optimization variables, while the others remain fixed. 
Yu \textit{et.al} in \cite{Yu2013} have followed such an approach and developed a strategy to tackle (\ref{eq:inner-outer-problem}) in a semi-distributed manner, demanding limited communication between BSs and the CU: fix the first two set of variables and optimize the third set, then fix the second and third sets and optimize the first one, and so on until convergence. The BSs jointly update parameters associated with all variable sets in an semi-distributed fashion. By fixing the set of users and precoder weights, the power allocation can be performed based on the information shared between BSs, see \cite{Stridh2006a,Chiang2008,Stanczak2009}. By fixing the precoders and powers, the user scheduling can be performed using the approaches described in Sections~\ref{section:algorithms:aggregated-utility}. By fixing the set of users and power allocation, the precoder weights can be computed using distributed algorithms, see \cite{Bjornson2013}. 	
A semi-distributed CS was proposed in \cite{Castaneda2015}, where the users are selected using an approximation of the NSP [cf. Section~\ref{section:metrics-spatial-compatibility:projection-spaces}], and the CBF is computed with local CSI. A similar approach was proposed in \cite{Jang2011}, where the BSs dedicate spatial DoF to cancel ICI for some selected users at neighboring cells. This approach performs semi-distributed spatial clustering for CS [cf. Section~\ref{section:metrics-spatial-compatibility:spatial-clustering}], and distributed linear precoding for CBF.

{\color{black}{

\begin{figure*}[t!]
	\centering
	\includegraphics[width=0.98\linewidth]{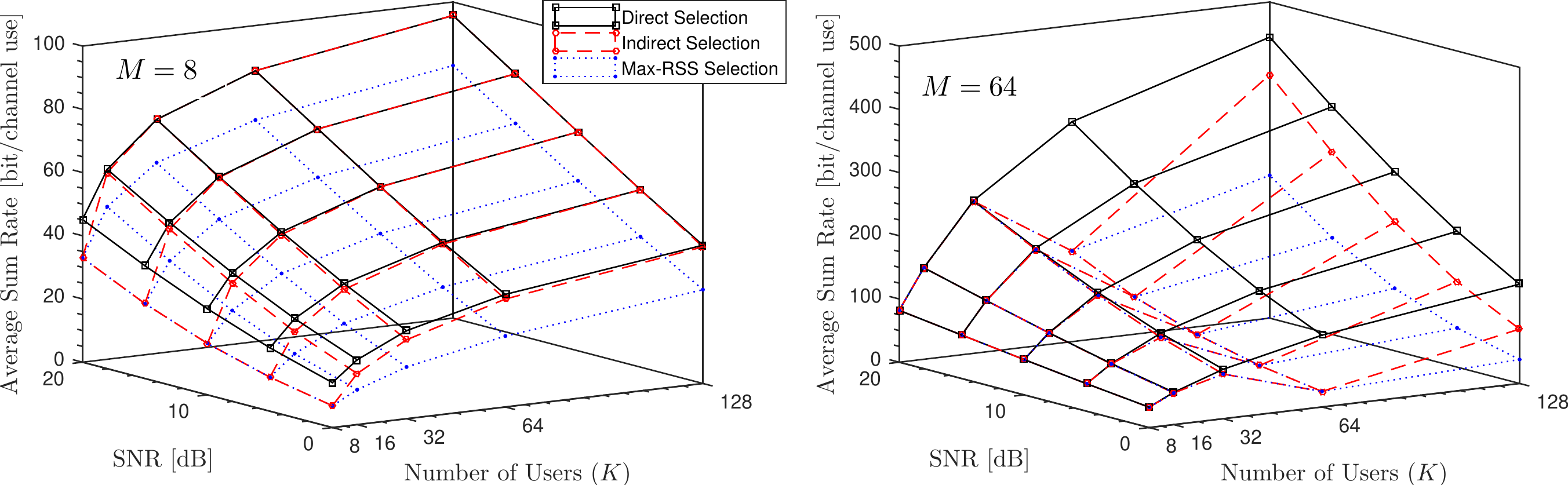}\\
	\caption{Average sum rate vs SNR vs the number of users ($K$), for $M=8$ (left plot) and $M=64$ (right plot).} 
	\label{fig:k-vs-snr}
\end{figure*} 

\subsection{Scheduling for Massive MIMO}
\label{section:algorithms:aggregated-utility:massive-mimo}

Massive or large-scale MIMO has been widely envisaged as a key transmission technology for next generation of  wireless communication, 5G \cite{Andrews2014,Zheng2015}. In massive MIMO systems, the BS is equipped with a large number of antennas (few hundreds) and serve multiple users (normally few tens). The excess amount of antennas enables focusing the transmission and reception of signal energy into smaller regions of space. Joint optimization over the spatial and multi-user domains can provide significant gains in throughput and EE  \cite{Marzetta2010, Yang2013b, Rusek2013, Zheng2015}.

In conventional MU-MIMO systems, the number of co-scheduled users $K$ is usually larger than the number of BS antennas $M$. In contrast, in massive MIMO systems the  transmit antennas outnumber the active users ($M\gg K$), which reduces the signal processing complexity and achieves large peak rates. However, the promising performance improvements come at the expense of hardware complexity. Due to cost and power consumption, practical transceiver architectures have a different number of RF chains ($M_{RF}$) compared to the number of antennas $M$, [cf. Section~\ref{section:precoding:massive-mimo}]. If every antenna has its own RF chain, i.e. $M_{RF} = M$, digital beamforming, [cf. Section~\ref{section:precoding:linear}], can allocate the whole spectrum to each active user \cite{Bjornson2016}. In practice, the total number of simultaneous users is constrained by the number of RF chains at the base stations \cite{Bogale2015,Shokri-Ghadikolaei2015,Bogale2016}. When the number of antennas is larger than the number of RF chains and users, i.e., $M\gg K \geq M_{RF}$, the system performance can be boosted from the diverse path losses and shadow fading conditions of different users.

The channel hardening effect in massive MIMO removes frequency selectivity, and avoids complex scheduling and power control designs \cite{Bjornson2016}. As $M \rightarrow \infty$, the MIMO channels become spatially uncorrelated, user separability (in the sense of Definition~\ref{defn:compatible_set_of_users}) plays a minor role in the scheduling decisions and the performance optimization relies on the channel magnitudes, see \cite{Zhang2007b,Bjornson2009a,Liu2015a,Hong2015}. If sum-rate is the ultimate metric to be optimized, a highly efficient scheduler only needs to assign resources to the users with the largest channel magnitudes.

Most of the scheduling designs for massive MIMO are based on the greedy approaches presented in Section~\ref{section:algorithms:aggregated-utility}, e.g. \cite{Huh2012,Adhikary2013,Nam2014,YiXu2014,Lee2014c,Bogale2015,Bogale2016a}. In these works is common to assume clusters of users sharing similar slow fading characteristics. Approaches such as the \textit{Joint Spatial Division and Multiplexing} (JSDM) \cite{Adhikary2013,Nam2014} first partition cell users into groups with distinguishable linear subspace spanned by the dominant eigenvectors of the group’s channel covariance matrix.  The transmit beamforming design is performed in two stages: a pre-beamforming that separates groups by filtering the dominant eigenvectors of each group’s channel covariance matrix, followed by precoding for separating the users within a group based on the effective channel.

The user clustering in \cite{Adhikary2013,Nam2014} can be implemented using  spatial subspace compatibility metrics [cf. Section~\ref{section:metrics-spatial-compatibility:other-metrics}]. It is worth noticing that such a grouping  process might be necessary to eliminate pilot contamination, by allocating identical pilot sequences to groups of users with similar channel characteristics \cite{Elijah2016}. The overlap between user clusters and their sizes are features dictated by two factors, the number of RF chains, $M_{RF}$, and the number of geographically co-located uses \cite{Shokri-Ghadikolaei2015}. The principle of the two-stage beamforming technique used by JSDM can be applied to multi-cell systems, where each  cluster of users is associated to one BS, and the outer precoders are used to cancel ICI \cite{Liu2014a}. Note that widely used scheduling algorithms in small-scale MU-MIMO, e.g. SUS \cite{Yoo2006}, result in extremely high complexity (of the order of $O(M^3K))$ in the large antenna regime \cite{Lee2014c,Hong2015}. In contrast, low-complexity scheduling algorithms may attain acceptable performance with full digital beamforming in certain scenarios \cite{Hong2015,Bai2014,Liu2015a}. However, further research must be done to design efficient scheduling rules for heterogeneous networks using hybrid precoding [cf. Section~\ref{section:precoding:massive-mimo}].

A process closely related to scheduling is user association, whose mathematical formulation resembles problem (\ref{eq:general-combinatorial-problem:cost-function}). Matching a user to a particular transmitter, so that the association optimizes an objective function [cf. Section~\ref{section:performance-uf}], is a complex combinatorial assignment problem \cite{Liu2016}. Ongoing research on this topic have covered cellular (e.g. \cite{Bethanabhotla2016}) and WLAN (e.g. \cite{Athanasiou2015}) systems, but literature on heterogeneous networks is still limited \cite{Liu2016}. 
Heuristic and simple association rules employed in current cellular networks, e.g. the max-RSS (received signal strength) or the biased-received-power based criteria, cannot include all system constraints neither fully exploit massive MIMO. In heterogeneous networks,  one must consider  per-base-station load and power constraints, per-user QoS constraints, and different number of antennas per transmitter. Construction of efficient rules for user association for future heterogeneous 5G networks is an emerging research topic, which must consider channel conditions, load balancing, and EE \cite{Liu2016}.   
Recent results in \cite{Li2015} show that user association to a single BS is optimal most of the times, which can simplify the association rules and algorithms. However, in dense scenarios where the coverage range per transmitter is limited (e.g., due to power constraints), multiple user-base station associations can reduce handover and waste of resources in the backhaul network \cite{Shokri-Ghadikolaei2015,Bogale2016}.

\begin{figure}[t!]
	\centering
	\includegraphics[width=.98\linewidth]{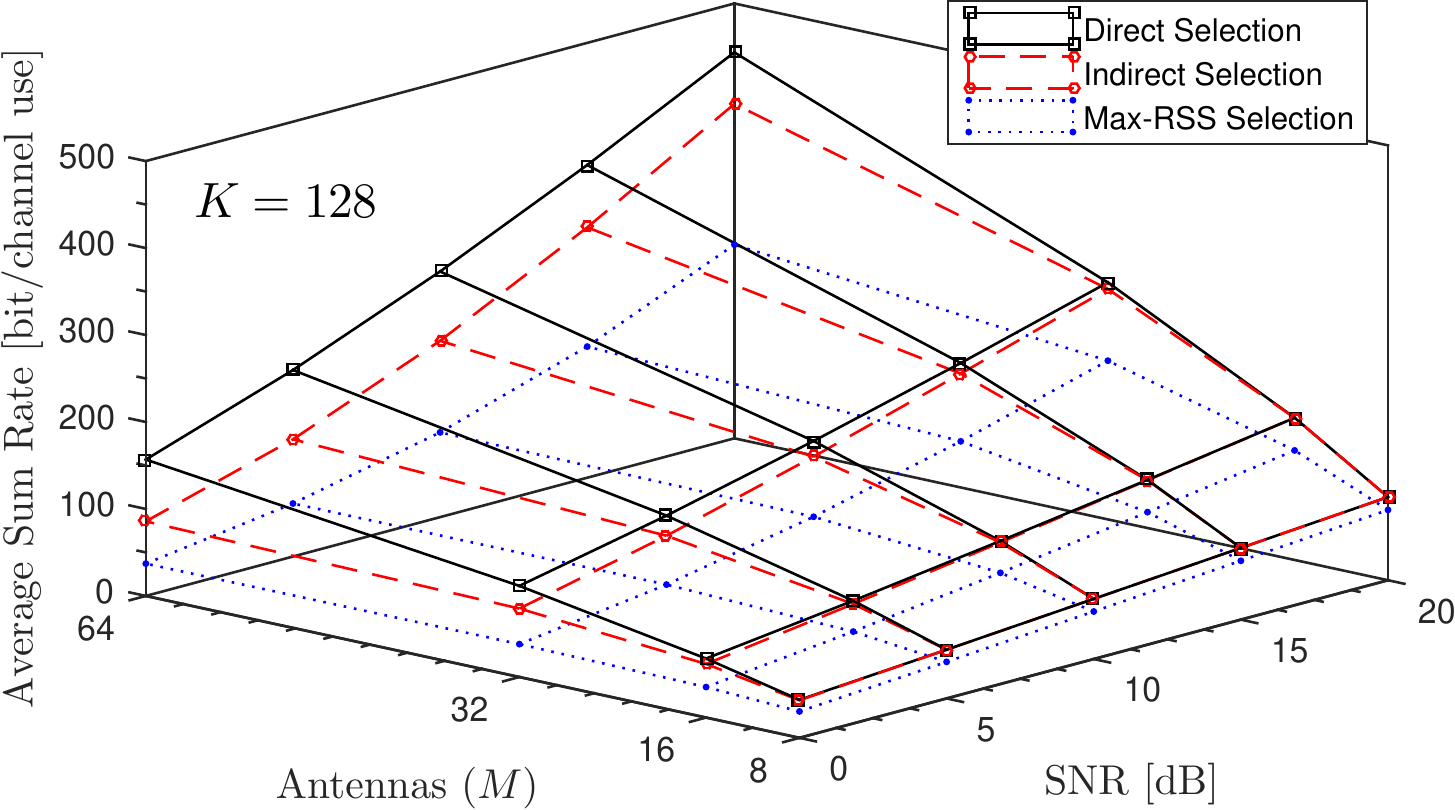}\\
	\caption{Average sum rate vs number of antennas ($M$) vs SNR, and fixed number of active users, $K=128$.} 
	\label{fig:m-vs-snr}
\end{figure}

}}

{\color{black}{

\subsection{Discussion and Future Directions}
\label{section:algorithms:discussion}

To analyze the performance of the scheduling schemes described in Section~\ref{section:algorithms:aggregated-utility}, we consider uniformly distributed users deployed within a single cell with radius 250 m. The users have heterogeneous channel conditions, the path loss exponent is 3.5, and the log-normal shadow-fading has 8 dB in standard deviation.  For the direct selection [cf. Algorithm~\ref{alg:algorithms:direct-approach}], the utility function is the sum rate, in the indirect selection [cf. Algorithm~\ref{alg:algorithms:indirect-approach}], the mapping function is the NSP, and the Max-RSS selects the users with larger channel gains. It is assumed full CSIT at the BS, the precoding scheme is full digital ZFBF (i.e. $M_{RF}=M$), and water-filling is used for power allocation.

The average achievable sum rate is shown in Fig.~\ref{fig:k-vs-snr}, for a MISO configuration with $N=1$, $M \in \{8, 64\}$, a cell edge SNR in the range $[0,20]$ dB, and several number of users, $K \in [8, 128]$.
For $M=8$ (left-plot), we have a conventional MU-MIMO scenario with $K \geq M$, $|\mathcal{K}| \leq M$, and the performance gap between the direct and indirect selection vanishes as $K$ increases. The Max-RSS selection partially exploits the MUDiv by only considering the channel magnitudes, which results in a considerable performance gap as $K$ grows.
For $M=64$ (right plot), we can divide the users ($K$) axis in two regions. For $K\leq 32$, the system operates in the massive MIMO region, where there are enough spatial resources to allocate one stream per active user. Notice that as $K$ approaches $M$, e.g. $K=32$, the direct selection overcomes the other methods specially in the low SNR regime. For $K=64$, the indirect and Max-RSS methods attempt to allocate resources to $M$ users, whereas the direct selection optimizes its objective function regardless the attainable multiplexing gain. This result shows that even if $M$ is large, the performance metric is optimized when $|\mathcal{K}| < M$, see \cite{Bjornson2016}. It is worth noticing that for $M=64$ and $K=128$, the performance gap between the direct and indirect selection decreases, but the latter approach might require less computational processing. This illustrates the fact that each selection method provides different performance-complexity tradeoffs, depending on the parameters $M$, $N$, $K$, and the SNR regime. 
Fig.~\ref{fig:m-vs-snr} compares the performance of the selection methods with fixed $K=128$, and different values of $M$. The performance gap between the direct and indirect selection is small for $M\leq 32$, where the MUDiv is rich. The Max-RSS selection might be an efficient alternative, if $M$ is large and the number of co-scheduled users is small, i.e., $M \gg |\mathcal{K}|$.

Scheduling in massive MIMO systems has not attracted the same attention as in conventional MU-MIMO systems because the excess number of antennas, additional DoF in the scheduling procedure, and the channel hardening effect result in marginal scheduling gains. However, in real-world massive MIMO systems with imperfect CSI, errors in the channel estimation and calibration, there will be some remaining interference among users, especially using linear precoding. The system will inevitably become interference-limited and the sum-rate may saturate (ceiling effect) at high SNR if the CSI and estimation accuracies do not improve accordingly. In this case, it is crucial to select the right number of users to serve, independently of spatial compatibility and channel correlation metrics. Conventional user selection algorithms [cf. Section~\ref{section:algorithms:aggregated-utility}], may fail to serve the optimal number of users, even if the selected users have good spatial characteristics. 
In other words, in massive MIMO systems, it is more important to identify the optimal number of users to serve rather than a set of users with certain channel characteristics \cite{Bjornson2016}. Even random or max-RSS user selection will perform well if they select the right amount of users.

The overhead due to channel estimation is proportional to the number of transmit antennas $M$. Thus, massive MIMO is more adequate for TDD operation, and its implementation in FDD mode is still an open problem \cite{YiXu2014,Bjornson2016}. Although uplink and downlink scheduling are independent tasks with different traffic loads and access techniques \cite{Abu-Ali2013}, the CSI necessary to optimize the downlink transmission is estimated via uplink pilots (channel reciprocity).\footnote{Experimental work for distributed MIMO \cite{Hamed2016}, presented a channel reciprocity protocol that can achieve the performance of explicit channel feedback for mobile users, which is critical for massive MIMO settings.} Thus, further research is necessary to understand the relationship between uplink and downlink scheduling and their associated resource allocation for TDD-based massive MIMO. 

}}

\section{MU-MIMO Scheduling with partial CSIT}
\label{section:algorithms:selection-partial-csit}

In MU-MIMO systems with partial CSIT, resource allocation algorithms can multiplex up to $M$ users accounting for the feedback of one scalar (CQI) and one index (CDI) \cite{Sharif2005}, [cf. Section~\ref{section:introduction:needof-csit}]. The feedback information load is proportional to the number of deployed users and antennas, but scheduling usually requires rough quantization resolution in the CQI to differentiate between high and low rate users \cite{Gesbert2004}. The CDI plays a more relevant role to achieve spatial multiplexing, thus, it requires a higher quantization granularity.
User scheduling requires two main steps: \textit{i}) the transmitter sends pilots for CSI acquisition, the users quantize their channels and feed back the CQI and CDI; \textit{ii}) subsets of users and beams are selected for data transmission based on a particular performance metric \cite{Wagner2008}.
As discussed in Section~\ref{section:precoding:partial-csit}, there are two approaches to acquire channel information using codebooks, i.e., quantizing either the channel, or the precoder that better fits the channel. In the following, we classified different scheduling approaches, based on the type of quantization.

\subsection{Scheduling using quantized channels}
\label{section:algorithms:selection-partial-csit:qunatized-csi}

For the sake of exposition, assume a MISO scenario ($N=1$), linear precoding, and equal power allocation ($\rho=P_k = P/M$,  $\forall k$). The $\text{CQI}_{k}$ of the $k$-th user can be given by its SINR defined as \cite{Jindal2006}:

\begin{equation}\label{eq:pcsit_sinr_user_k}
\text{SINR}_{k} = \frac{  \rho \| \mathbf{H}_{k} \|^{2} | \hat{\mathbf{c}}_{k} \mathbf{W}_{k} |^{2} }
{ 1 + \rho \| \mathbf{H}_{k} \|^{2} \sum_{j\neq k} |  \hat{\mathbf{c}}_{k}\mathbf{W}_{j} |^{2}   }, 
\end{equation}
where $\{\mathbf{W}_{i}\}_{i=1}^{M}$, are the precoding vectors extracted from the CDIs (e.g. using the ZFBF).  $\hat{\mathbf{c}}_{k}$ is the actual quantized unit-norm channel or $\text{CDI}_{k}$ given by
\begin{equation}\label{eq:pcsit_channel_quantization_best_beam_user_k}
\hat{\mathbf{c}}_{k} = \underset{\mathbf{c}_{b} \in \mathcal{C}}{ \arg \max} \ \  \cos (\measuredangle(\mathbf{H}_{k},\mathbf{c}_{b})),
\end{equation}
where $\mathcal{C}=\{ \mathbf{c}_{1}, \ldots, \mathbf{c}_{b}, \ldots, \mathbf{c}_{2^B} \}$ is a predefined codebook, [cf. Section~\ref{section:precoding:partial-csit}]. The $k$-th user determines its CDI using (\ref{eq:pcsit_channel_quantization_best_beam_user_k}), according to a minimum distance criterion, see \cite{Mukkavilli2003} for MISO or \cite{Love2003,Chae2008} for MIMO setting. The solution of (\ref{eq:pcsit_channel_quantization_best_beam_user_k}) has the following geometrical interpretation: the user $k$ selects the most co-linear or correlated codeword to its channel, which is equivalent to find the cone containing $\mathbf{H}_{k}$, as illustrated in Fig.~\ref{fig:hyperslab_cones}. 
Notice that the $\text{SINR}_{k}$ takes into account channel magnitudes, quantization errors of the CDI, and the spatial compatibility of channel $\mathbf{H}_{k}$ regarding $\mathcal{C}$. However, to compute the precoders $\{\mathbf{W}_{i}\}_{i=1}^{M}$, it is necessary to know a priory $\mathcal{K}$, and the associated CDIs $\{\hat{\mathbf{c}}_{i}\}_{i=1}^{|\mathcal{K}|}$, which results in a chicken-and-egg problem. Therefore, the exact value of (\ref{eq:pcsit_sinr_user_k}) is unknown at the transmitter or receiver sides. Several works have proposed different ways to approximate (\ref{eq:pcsit_sinr_user_k}), mainly by its upper/lower bounds or expected value, cf. \cite{Kountouris2007,Yoo2007,Trivellato2008,Conte2010,Xu2010,Ravindran2012,Sohn2012,Wang2013a,Huang2009a}. If the transmitter has statistical CSI knowledge, the formulation of (\ref{eq:pcsit_sinr_user_k}) must take into account the covariance matrices, $\mathbf{\Sigma}_{k}$, $\forall k$, to have a more accurate estimate of the interference powers \cite{Hammarwall2008a}.

The CQI can also be given by the channel magnitude, i.e., $\|\mathbf{H}_{k}\|^2$, but this measure ignores the valuable information contained in the codebook (spatial compatibility), and neglects quantization errors. The number of competing users, the feedback load, and scheduling complexity can be reduced by setting thresholds to the CQIs. This means that only users with strong channels will be considered for scheduling \cite{Conte2010,Min2013}. Multiplexing and MUDiv gains are realized by carefully tuning the threshold according to statistics (e.g. \cite{Xu2010}) or numerical characterization (e.g., \cite{Yoo2007,Sohn2010,Min2013}) of the CQI, for a set of fixed parameters $B$, $M$, $K$, and $\epsilon$, [cf. Section~\ref{section:metrics-spatial-compatibility:spatial-clustering}]. 
In some scenarios where the maximum number of spatial streams per user is at most one, each receiver uses either a weight vector \cite{Trivellato2008} or a combining technique \cite{Jindal2008,Bjornson2013a,Schellmann2010}  to transform its MIMO channel matrix into an \textit{effective} MISO channel vector.

Different approaches to reduce feedback load and  compute  the CQI and CDI can be implemented (e.g. \cite{Wang2013a}). 
In general, the CQI requires less bits than the CDI, but an optimized bit allocation must take into account the SNR regime, the number of competing users $K$ \cite{Xu2010,Yoo2007}, and practical quantization levels (e.g. MCSs) \cite{Vicario2008}, [cf. Section~\ref{section:algorithms:selection-partial-csit:two-stage-feedback}].
Assume that the CQI and CDI are fed back by all competing users, so that the quantized channels can be reconstructed at the transmitter. Resource allocation can be performed using the methods described in  Section~\ref{section:algorithms} or by low-complexity approaches based on antenna selection, e.g. \cite{Khoshnevis2013,Castaneda2015}. 
ZF precoding is the most common technique used in scenarios with quantized channels, hence, scheduling based on metrics of spatial compatibility [cf. Section~\ref{section:metrics-spatial-compatibility}], can be directly applied, e.g., \cite{Yoo2007, Kountouris2007,Wang2007a,Chae2008}. The user selection for the general MIMO scenario can be performed by computing effective MISO channels or by treating each receive antenna as a virtual user \cite{Xu2010}. 
It is worth mentioning that ZF becomes interference-limited in the high SNR regime with fixed $B$. However, if $K$ or $B$ scale with the SNR, the achievable rates can be improved and the quantization errors can be mitigated \cite{Jindal2006,Yoo2007}.

A common approach to combat IUI (generated by quantization errors), is by reducing the multiplexing gain. Finding the optimal number of data streams is an optimization problem that depends on the system parameters, e.g. SNR regime, mobility, $K$, $B$, and $M$, see \cite{Wang2007a,Dai2008,Kountouris2012,Zhang2011}. If $K \approx M$,  it is likely to have a codebook with limited granularity (fixed $B$) yielding irreducible IUI. In such a case, the subset of selected users $\mathcal{K}$, must have cardinality strictly less than $M$, and their CDIs must have high resolution. In that way, the precoding performance can be improved and additional spatial DoF can be used to cancel interference out \cite{Ravindran2012}. The results in \cite{Wang2007a,Dai2008,Zhang2011} show that, the number of selected users highly depends on $B$, whose optimal value is a function of the SNR regime. For  extreme values of the SNR (very low or high), the optimal transmission schemes are TDMA or SU-MIMO, which completely avoid IUI caused by inaccurate CSIT. 
Experimental results in \cite{Jones2015} show that selecting about $\frac{3}{4}$ of the total available beams maximizes performance in different MU-MIMO WLAN configurations.

It is worth noting  that the SNR regime plays an important role in the scheduling rule design. Authors in \cite{Huang2009a} suggested the following guidelines to improve performance: 
\begin{enumerate}
	\item In the high SNR regime with fixed codebook size $2^B$, the system becomes interference limited [cf. Definition~\ref{defn:interference-limited-system}]. The scheduling rules should prioritize users whose fed back channels reduce the quantization error, i.e., the CDI and the available spatial DoF defined the attainable performance. Results in \cite{Yoo2007} show that the error quantization can be mitigated either in the large user regime ($K \rightarrow \infty$), or the high resolution regime ($B \rightarrow \infty$).  
	\item If the variance of the noise and the IUI are comparable, both the CQI and CDI should be taken into account for user selection.
	\item In the low SNR regime,  the system is noise limited and the scheduling rules should prioritize the CQI, since the CDI or error quantization play a negligible role.
\end{enumerate}

These  guidelines for scheduling design can be applied to MU-MIMO scenarios with full CSIT, and they have been extended to multi-cellular cooperative systems in \cite{Castaneda2015}.

\subsection{Scheduling using RBF}
\label{section:algorithms:selection-partial-csit:rbf}

\begin{figure}[t!]
	\centering
	\includegraphics[width=0.98\linewidth]{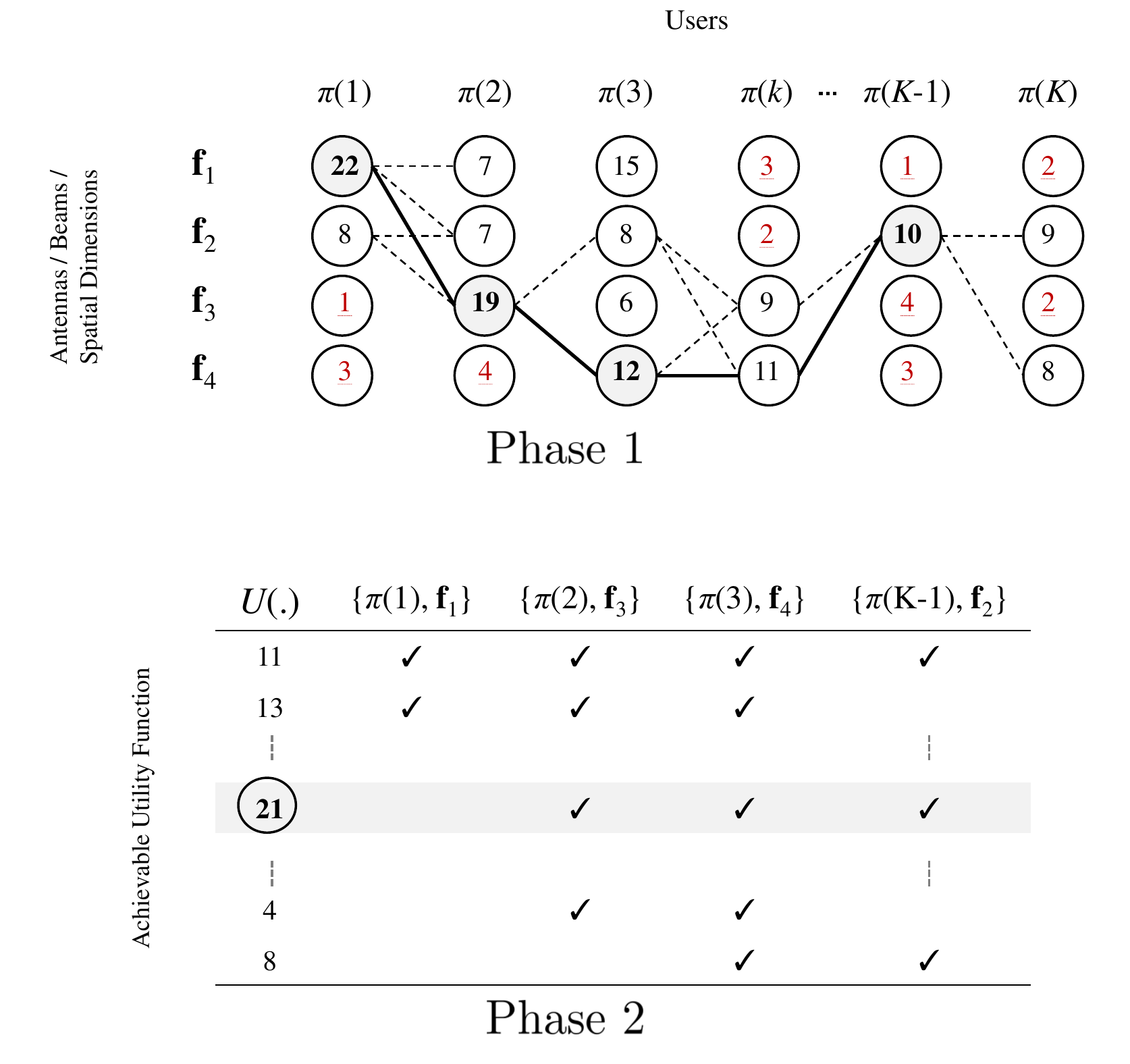}\\
	\caption{Given the user ordering $\pi(\cdot)$, the user selection is done by finding the strongest user per beam/antenna/spatial dimension associated with $\mathcal{F}$. Phase 1 matches a user with a beam based on the CQI.  Phase 2 maximizes the utility function by selecting a subset of users and their associated beams according to some enumeration or search technique.}
	\label{fig:beam_selection}
\end{figure}

Channel information acquisition based on RBF uses a codebook $\mathcal{F} = \{ \mathbf{f}_{1}, \ldots, \mathbf{f}_{b}, \ldots, \mathbf{f}_{2^B} \}$, to define the precoders and provide high flexibility for  scheduling, [cf. Section~\ref{section:precoding:partial-csit}]. In contrast to the methods described in Section~\ref{section:algorithms}, where the precoders are unknown before user selection, the approach using RBF can efficiently compute a performance metric that includes the effective channel gain and interference due to quantization errors. However, the accuracy of the quantized channel information is limited by the basis (RBF) or bases (PU2RC) that comprise the codebook \cite{Sharif2005}.
In general, user scheduling is performed by joint user selection and precoding allocation, and the overall complexity depends on the parameters $M$, $B$, and $K$.  
The set $\mathcal{K}$ that solves problem (\ref{eq:inner-outer-problem}), can be found as illustrated in Phase 1 of Fig.~\ref{fig:beam_selection}, which is based on the algorithm proposed in \cite{Sharif2005} assuming that $K>M$ and $M\geq N$.

$\bullet$  \textit{Training Phase}: The precoding vectors in $\mathcal{F}$ can be generated according to a known distribution and chosen randomly. The transmitter sends pilots on all the spatial beams $\mathbf{f}_{m}, \forall m \in \{1, \ldots, 2^B\}$, so that all users can estimate their channels. The codebook design can be simplified by defining codewords as the orthonormal basis of $\mathbb{R}^{M \times 1}$, as is \cite{Zhang2007a}. For this codebook design, antenna selection is implicitly performed at the receiver side. This means that each receive antenna is treated as an individual user, which can reduce scheduling complexity and feedback load.
Another approach to construct $\mathcal{F}$, is by eigen-codebook design \cite{Kountouris2006,Jindal2008}. Each user computes eigen-decomposition of its covariance matrix $\mathbf{\Sigma}_{k}$, $\forall k$, and feeds back the eigenvector associated to its maximum eigenvalue. This approach requires extra feedback load, but can provide flexibility to the scheduler.
The quantization granularity of $\mathcal{F}$ can be enhanced if the codebook comprises multiple bases, as in PU2RC \cite{Ravindran2012}.

$\bullet$ \textit{CSI Feedback}: The users compute a CQI metric per beam, and report their largest CQI to the transmitter. The \textit{effective SINR} of the $k$-th user in the $m$-th beam can be defined as \cite{Sharif2005}:

\begin{equation}\label{eq:pcsit_rbf_sinr_user_k}
\text{SINR}_{k,m} =  \frac{  \rho  | \mathbf{H}_{k} \mathbf{f}_{m} |^{2} }
{ 1 + \rho \sum_{n\neq m} |  \mathbf{H}_{k}\mathbf{f}_{n} |^{2}   },
\end{equation}
and the index of the best precoder is given by
\begin{equation}\label{eq:pcsit_rbf_best_beam_user_k}
i_{k} =  \underset{m \in \mathcal{B}}{\arg \max} \ \  \text{SINR}_{k,m}
\end{equation}
where $\mathcal{B} \subseteq \{1,\ldots,2^B\}$, is an index subset of active codewords defined by the transmitter so that $|\mathcal{B}| \leq M$. The user designates its CDI$_{k}=i_{k}$ and CQI$_{k}=Q(\text{SINR}_{k,i_{k}})$ with $i_k \in \mathcal{B}$, and $Q(\cdot)$ is a quantization function, see \cite{Zhang2007a,Vicario2008} and discussion in Section~\ref{section:algorithms:selection-partial-csit:two-stage-feedback}. 
For $N>1$, evaluating (\ref{eq:pcsit_rbf_sinr_user_k}) and (\ref{eq:pcsit_rbf_best_beam_user_k}) can be performed in different ways \cite{Schellmann2010}: using receiver combining techniques, e.g. MRC, if one beam is assigned per user; or equalization techniques, e.g. MMSE \cite{Tse2005}, when multiple beams are assigned per user.
The CQI computation is not limited to the achievable SINR or peak rate, e.g., it can be computed from sufficient statistics of the channel conditions and past CQIs \cite{Wagner2008}. For instance, authors in \cite{Simon2011} defined the CQI as an estimation of the minimum power required to achieve a target SINR. In \cite{Kountouris2008a}, the CQI is defined as a function of the current and previous channel conditions, which is an approach similar to the CDF-based scheduling \cite{Nguyen2015}. This probabilistic method uses the channel or rate distributions\footnote{Observe that the nature of the objective function modifies the statistics of the CQI. For instance, if we want to maximize the sum rate, then users that experience better channel conditions are more likely to be selected \cite{Bjornson2009a}.} to schedule the users that are more likely to achieve high performance \cite{Kountouris2008a}.

One can preselect the competing users based on a performance threshold, which reduces feedback and scheduling complexity. Accounting for i.i.d. channels allows to simplify the threshold design, since the channel directions follow a uniform distributions \cite{Huang2007}.
The $k$-th user feeds back information only if its associated CQI is above a predefined threshold, $\gamma_{th}$, a scheme called \textit{selective multiuser diversity}, see \cite{Gesbert2004,Sharif2005,Huang2007,Zhang2007a,Kountouris2008a,Xu2010}. 
An extension of such a method is the multi-threshold selection, where the scheduling rule also takes into account the MCSs supported over each link, e.g. \cite{Vicario2008,Tang2010}. If the system optimization involves queue stability constraints, other thresholds based on QSI can be imposed on top of $\gamma_{th}$, provided that the users have knowledge of their respective queue lengths \cite{Chen2013,Destounis2015}.

Preselection can be performed in the spatial domain [cf. Section~\ref{section:metrics-spatial-compatibility:spatial-clustering}], where the users meet a constrained in the quantization error. The set of competing users can be defined as $\{k \in \bar{\mathcal{K}}: \cos (\measuredangle(\mathbf{H}_{k},\mathbf{f}_{i_{k}})) < \epsilon\}$, for a predefined threshold $\epsilon$, see \cite{Huang2007,Kountouris2008a}.
In the large user regime ($K \rightarrow \infty$) with fixed $M$ and $N$, the MUDiv\footnote{Numerical analysis of capacity versus the number of users $K$, for both full and partial CSIT, e.g. \cite{Yoo2006,Shen2006,Yi2011,Nam2014a,Ko2012,Lee2014,Tran2010, Choi2007c,Kobayashi2007,Park2010,Huang2009a}, show diminishing returns of MUDiv over i.i.d. Rayleigh fading channels, i.e., the capacity gain  flattens as $K\rightarrow \infty$, see  \cite{Viswanath2002,Hassibi2007}.} scales as $\log(\log(KN))$ \cite{Sharif2007}, which suggests that preselection based on channel statistics, long-term throughput, or even randomly (e.g., \cite{Huh2012}), can reduce the feedback load and achieve MUDiv gains.

$\bullet$ \textit{User Selection}: The transmitter must find the set of users that maximizes the performance metric, e.g. WSR. There are several scheduling approaches, e.g., treating each receive antenna as a single user,  assigning at most one beam per user, or allocating multiple beams per user. 
The simplest selection is assigning the $m$-th beam to user $k_m$, which is defined as
\begin{equation}\label{eq:pcsit_rbf_best_user_per_beam}
k_{m} =  \underset{k \in \bar{\mathcal{K}} \ : \ \text{CDI}_{k} = m }{\arg \max} \ \  \text{CQI}_{k} 
\end{equation}
If the quantization granularity of the function $Q(\cdot)$ is low, it may happen that (\ref{eq:pcsit_rbf_best_user_per_beam}) accepts more than one solution, in which case $k_{m}$ can be chosen randomly among the best user for beam $m$ \cite{Zhang2007a}. 
If the elements of $\mathbf{H}_{k}$ are i.i.d., each selected user cannot achieve maximum SINR for more than two beams on one antenna provided that $K > M$. This is a valuable design guideline that allows to reduce scheduling complexity \cite{Sharif2005,Chen2013}. Analytical results in \cite{Hassibi2007} show that $M$ must be scaled proportional to $\log(K)$, so that user starvation is avoided.
Once that the set $\mathcal{K} = \{k_{m}\}_{m=1}^{M}$ has been found, power allocation and link adaptation can be performed.

Owing to the fact that the quantization granularity is fixed and bounded ($B < \infty$), the precoders $\mathbf{f}_{m} \in \mathcal{F}$ cannot fully separate users in the spatial domain. Hence, IUI is unavoidable and particularly harmful in high SNR. Moreover, in the sparse user regime, $K \approx M$, RBF cannot benefit from MUDiv and performs poorly. Several user scheduling algorithms have been proposed in \cite{Lee2016} to handle sparsity and limited MUDiv over mmWave channels.
The optimum number of beams that maximizes the WSR is a function of the number of competing users and the SNR regime \cite{Wagner2008,Kim2005a,KountourisThesis}. Analytical results establish that for RBF with fixed $K$, and extreme values of the SNR regime, the optimal number of active beams is one\footnote{These results are equivalent for the case where channel are quantized in Section~\ref{section:algorithms:selection-partial-csit:qunatized-csi}, see e.g., \cite{Wang2007a,Dai2008,Zhang2011}.} \cite{Wagner2008,KountourisThesis}.
Depending on the channel conditions and the system constraints, user scheduling might yield a dynamic switching between TDMA (time-sharing) and MU-MIMO transmission. This can take place if IUI is very high, or if the individual rate or power constraints are violated \cite{Kountouris2005a,Kountouris2008}. 
Therefore, efficient operation in an arbitrary SNR requires a reduced number of scheduled users and active beams, i.e., $|\mathcal{B}| < M$. Analytical results in \cite{Wagner2008} show that $|\mathcal{K}| = M$ can be attained if $K \rightarrow \infty$ for a fixed $B$. 
In general, the optimal set of selected users is such that $|\mathcal{K}|<M$,  \cite{Vicario2008}. 
Finding the best subset of users and beams is a combinatorial problem whose optimal solution is found by ExS \cite{Wagner2008}. One has to enumerate all user combinations, compute their associated performance metrics $U(\cdot)$, and select the subset with maximum $U(\cdot)$, as illustrated in Phase 2 of Fig.~\ref{fig:beam_selection}. One alternative is to formulate an integer programming optimization, as in \cite{Kim2005a}, or to perform greedy selection, see \cite{Vicario2008} and  Section~\ref{section:algorithms:aggregated-utility}.

Several extension to the approach presented in Fig.~\ref{fig:beam_selection} have been developed: improving codebook resolution, e.g. \cite{Huang2007,Choi2007c}; dynamically updating multi-bases codebooks \cite{Kountouris2005}; computing the CQI by different functions, e.g. \cite{Kountouris2008a}; combining RBF with deterministic precoding schemes, e.g. \cite{Zorba2008}; dynamically adjusting the number of active beams, e.g. \cite{Wagner2008,Kountouris2008}, etc. 
To enhanced the quantization granularity, authors in \cite{Choi2007c} generate several sets of $2^B$ beams, and the users report the best CQI and CDI per set.  
The user selection in \cite{Zorba2008} combines RBF and SZF for the MISO scenario: given an initial random precoding $\mathbf{f}_{1}$, its associated user $k_1$ solves $\max_{k \in \bar{\mathcal{K}}} \cos (\measuredangle(\mathbf{H}_{k},\mathbf{f}_{1}))$, and feeds back a finer CDI that is used to generate the precoder $\mathbf{f}_{2}$ for the second user, $k_2$, and so on. In this way, the $k$-th selected user will only receive interference from the previous $k-1$ selected users. This algorithm fits particularly well in heterogeneous systems, where users can be clustered according to their locations or channel statistics, e.g., \cite{Ho2009,Nam2014,Moretti2013}.

In systems with QoS constraints [cf. Section~\ref{section:performance-uf:wsr-with-qos}], the opportunistic nature of the algorithm in Fig.~\ref{fig:beam_selection} changes and plays a secondary role, which has a direct impact in the MUDiv gains. This is because the competing users are preselected based on QSI, since scheduling exclusively based on CQIs incurs in delay penalties \cite{Hassibi2007}. In \cite{Chen2013}, a subset of user is preselected using QSI, and only those users feed back their CSI. The final selection is done as shown in Fig.~\ref{fig:beam_selection}, where the CQIs are computed as the effective SINRs weighted by their respective QSIs. Notice that the parameters of the deployment may affect the latency experienced by the users. For instance,  large values of $B$ might result in more scheduled users since IUI can be suppressed more efficiently.

\subsection{Two-stage Feedback Scheduling}
\label{section:algorithms:selection-partial-csit:two-stage-feedback}

\begin{figure}[t!]
	\centering
	\includegraphics[width=0.8\linewidth]{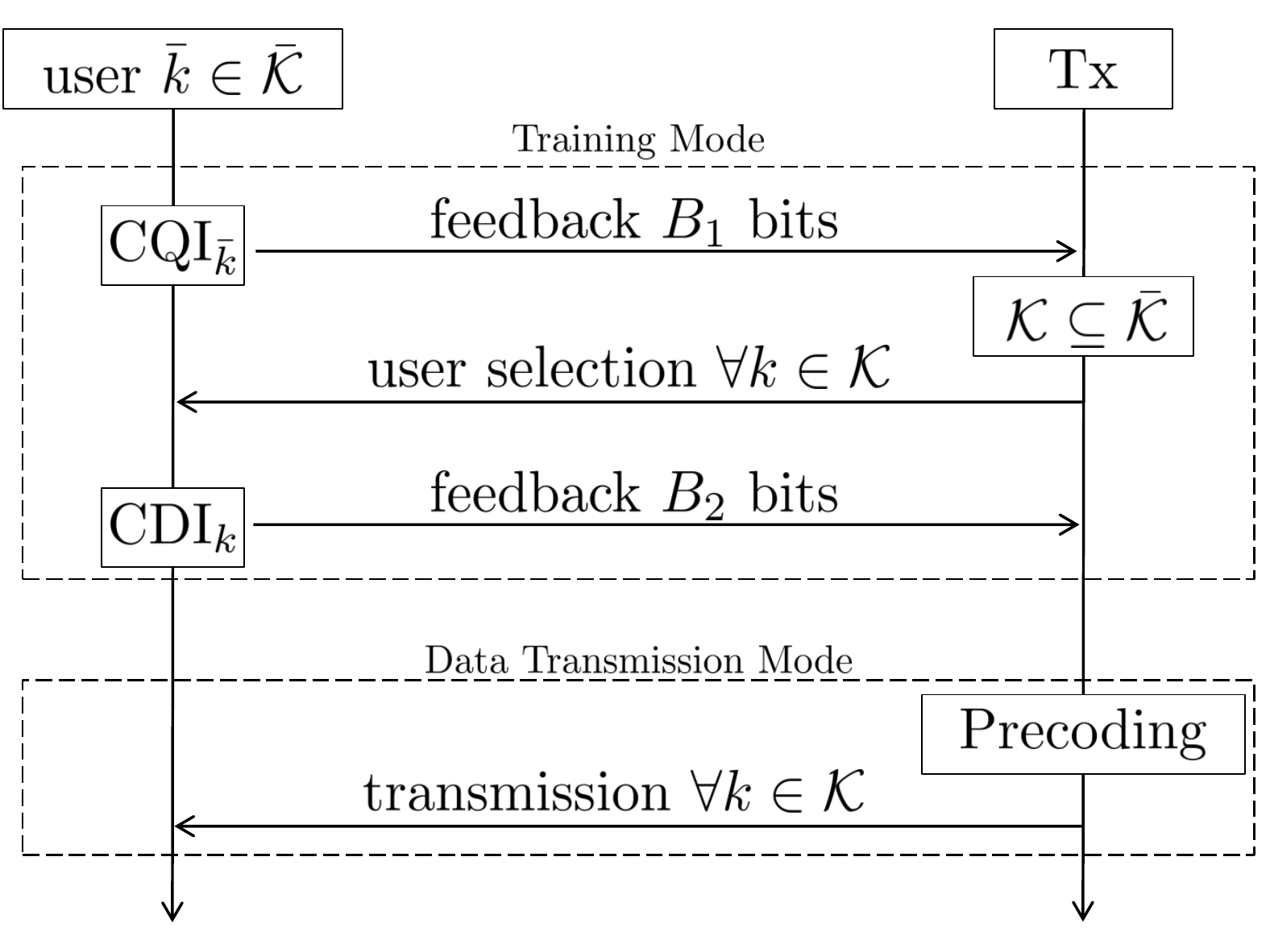}\\
	\caption{Two-stage feedback MU-MIMO scheduling}
	\label{fig:two-phase-feedback}
\end{figure}

Several works in the literature (cf.  \cite{Kountouris2006a,Gesbert2007a,Zakhour2007,Wagner2008,Kountouris2008,Kountouris2008a,Xu2010,Sohn2012,Khoshnevis2013}), have studied the user selection and precoding design problems in terms of the fed back information required to implement them. The total amount of feedback bits per user, $B_{t}$, is partitioned such that $B_{t}=B_1 + B_2$, where each part is used to solve a problem, as illustrated in Fig.~\ref{fig:two-phase-feedback}. 
Notice that the beam allocation problem in Fig.~\ref{fig:beam_selection} and the bit allocation problem Fig.~\ref{fig:two-phase-feedback}, bear some resemblance to the methodology described in Section~\ref{section:algorithms:aggregated-utility:csi-mapping}. The user selection problem (\ref{eq:csi-mapping-problem-1}), is solved using some properties of the CSIT of all competing users, and the precoding design and power allocation problem (\ref{eq:csi-mapping-problem-2}) is solved only for the set of users previously selected or grouped.

The communication phases between the transmitter and the receivers in Fig.~\ref{fig:two-phase-feedback}, are part of a training or pilot mode \cite{Wagner2008,Kobayashi2011}, which is applied to MU-MIMO system with partial CSIT. 
Taking into account the type of implemented quantization, some remarks are in order \cite{Sohn2012,Khoshnevis2013}:
\begin{enumerate}
	\item for high-rate feedback systems, a large value of $B_{t}$ for channel quantization might be more efficient. The accurate CDI can be used for precoding design and to enhance of the multiplexing gain at high SNR \cite{Hassibi2007}.
	\item for low-rate feedback systems with small values of $B_{t}$, the RBF or PU2RC schemes, [cf. Section~\ref{section:algorithms:selection-partial-csit:rbf}], are more suitable since they can meet rate constraints and simplify the scheduling.
\end{enumerate}

In scenarios with limited feedback rates, $B_t < \infty$, it is desirable to: \textit{i}) achieve significant MUDiv using $B_1$ bits for the CQI; \textit{ii}) design codebooks using $B_2$ bits of resolution for the CDI, to efficiently mitigate quantization errors, IUI, and achieve multiplexing gains; and \textit{iii}) reduce the scheduling complexity and limit the number of users that feedback $B_2$ bits.  
The optimal partition of $B_{t}$ depends on system parameters ($N$, $M$, and $K$), SNR regime, and type of precoding scheme \cite{Kountouris2006a,Zakhour2007}. Moreover, practical systems not only limit the amount of bits per user,\footnote{In LTE-Advanced, one physical uplink control channel (PUCCH) is assigned per user \cite{Dahlman2013}. The control information such as CQI, CDI, channel rank, scheduling request, and hybrid-ARQ ACK/NAK are conveyed though the PUCCH, which has few bits dedicated to CSI reporting (format 2).} $B_{t}$, but also the total feedback rate, $KB_{t}$, e.g., if the users share the feedback link, see \cite{Khoshnevis2013,Huang2007,Xu2010,Sohn2012}. 
Observe that the multiuser scheduling in Fig.~\ref{fig:two-phase-feedback} requires CSI feedback and  signaling of the scheduling decisions to the selected users. These feedback and control loop introduce a non-negligible overhead, $KB_{1}$, and latency in the system that must be carefully considered in the scheduling rule design \cite{Gesbert2007a}.

In the first feedback phase in Fig.~\ref{fig:two-phase-feedback}, the competing users send a coarse quantized version of their achievable performance metric. The $\text{CQI}_{k} \in \{q_{1},\ldots,q_{2^{B_1}}\}$ is quantized using $B_1$ bits, where $q_{i}$ is the $i$-th quantization level. The set of users in $\mathcal{K}$, can be chosen in a ranking-based per-beam selection, as in Phase 1 of Fig.~\ref{fig:beam_selection}, or per-antenna selection as in \cite{Zhang2007a,Khoshnevis2013}. 
A number of  works  have shown that depending on the system parameters, choosing $B_1 \in \{1,2,3,4\}$ can be enough to attain the MUDiv gain of the unquantized CQI, see \cite{Xu2010,Zhang2007a,Ravindran2012}. Results in \cite{Khoshnevis2013} and references therein, show that efficient precoding design is attained by assigning roughly $1/M$ out of $B_{t}$ bits to the CQI, and allocating the remaining bits to the CDI. 
If the system operates in the large user regime ($K \rightarrow \infty$), small values of $B_1$ are enough to achieve asymptotically optimal system performance, see \cite{Xu2010} and references therein. 
MUDiv is defined by users with SINR above a threshold, and constructing $\mathcal{K}$ depends on CDIs, since spatially compatible users will be grouped with high probability \cite{Chen2013,Sharif2005}. In the high resolution regime ($B_t \rightarrow \infty$), an efficient partition meets $B_1 > B_2$, and the fraction of bits dedicated for CDI scales proportionally to $\log(B_t)/B_t$, see  \cite{Khoshnevis2013}.

If $K \approx M$, larger values of $B_1$ are required to circumvent the lack of MUDiv and properly identify strong users. However, in the high SNR regime is more beneficial to have $B_1 < B_2$, since quantization errors in the CDI are the main performance-limiting factor \cite{Zhang2011,Ravindran2012}.
MUDiv clearly affects the optimal partition of $B$. Numerical results in \cite{Kobayashi2011} show the effects of $K$ over the feedback rates, and the interdependence between MUDiv and $B_2$.
Accounting for link adaptation, $B_1$ must be chosen so that the CQI fully exploits the granularity of the available MCSs, either for channel quantization \cite{Trivellato2008} or RBF \cite{Vicario2008}. 
The value of $B_1$ depends on the type of metric that defines the CQI. For example, the SINR and the achievable rate are real numbers, whereas other metrics might already be defined as integer numbers, see \cite{Kountouris2008a}.

In the second feedback phase in Fig.~\ref{fig:two-phase-feedback}, each user in $\mathcal{K}$ uses $B_2$ bits to report its CDI. This mean knowledge of refined channel information at the transmitter, e.g., more accurate spatial directions or effective channel gains.
If the system is constrained in the total feedback bandwidth, $|\mathcal{K}|B_2$, heterogeneous and dynamically allocated bits can be used to quantize the CDI. By assigning different number of bits per user, i.e., $B_{2(k)}$, $\forall k \in \mathcal{K}$, efficiency at the CDI feedback phase can be achieved \cite{Sohn2012}. The dynamic assignment of $B_{2(k)}$ is more effective in scenarios where the users have heterogeneous average long-term channel gains. The bit allocation rules might consider the following guidelines \cite{Xu2009,Sohn2012,Khoshnevis2013}:

\begin{enumerate}
	\item for high SNR users, large values of $B_{2(k)}$ reduce quantization errors and mitigate IUI. This is particularly important in scenarios where the CDI is used to compute the precoder weights (e.g., MMSE \cite{Kountouris2005a} or ZFBF \cite{Xu2010,Khoshnevis2013}), [cf. Section~\ref{section:algorithms:selection-partial-csit:qunatized-csi}].  
	\item for low SNR users, the performance is noise limited, i.e., the noise variance is larger than the interference, and the value of $B_{2(k)}$ is relatively less important to optimize performance.  
	\item dynamic assignment of $B_{2(k)}$ can be extended to sequential transmissions with error correction mechanisms (e.g. ARQ \cite{Xu2009}). Such an assignment can be used to further refine $\mathcal{K}$, by temporarily dropping users retransmitting the same packet multiple times.
	\item in multiple-transmitter scenarios, the level of coordination and transmission scheme [cf. Section~\ref{section:algorithms:multi-cell-selection-classification}], define the methodology to assign the value of $B_{2(k)}$ at each transmitter \cite{Khoshnevis2013}. 
\end{enumerate}

Results in \cite{Sohn2012} suggest that grouping users based on their locations or long-term channel gains (e.g.,  \cite{Ho2009,Nam2014,Moretti2013}), yields a more efficient assignment of the feedback bandwidth and simplify the user grouping. 
Another parameter to be considered when defining $B_1$ and $B_2$, is the time used for hand-shaking between feedback phases. The total time used for training and data transmission must be kept below the coherence time of the channel, so that the impact of delayed/outdated CSI is minimized \cite{Kountouris2008,Sohn2012}. Otherwise, increasing $B_2$ would not be enough to achieve MIMO gains \cite{Zhang2011}. 
Analytical results in \cite{Ravindran2012} show that for multi-basis codebooks  (e.g., PU2RC), large values of $B_2$ usually do not provide considerable capacity gains, and the overall performance is highly sensitive to MUDiv. A design rule to bear in mind is that the accuracy of the CSIT is more valuable than MUDiv in practical MU-MIMO scenarios \cite{Ravindran2012,Xu2010}.

\subsection{Implementation in Cellular Scenarios}
\label{section:algorithms:multi-cell-selection-classification:discussion}

The scheduling algorithms for MU-MIMO in LTE-Advanced rely on quantized CSI: \textit{i}) the RI/PMI metrics contain information regarding the SU-MIMO channel (due to backward compatibility\footnote{The limitations due to PMI feedback in SU-MIMO mode can be overcome for certain scenarios by means of efficient estimations of the CDI/CQI. One can take into account spatial compatibility between the MIMO channels and the codebooks, i.e., estimating the effective channel gains \cite{Wang2013a}.}) projected onto the subspace defined by the codebook; \textit{ii}) the CQI metric  indicates the energy of the projection, and it might include ICI and noise \cite{Liu2012}. Since the aggregated multiuser channel can be constructed by concatenating the users PMI metrics, and the CQIs are known at the transmitter, the scheduling algorithms described in the subsections above can be directly used in cellular scenarios.

In general, the scheduling algorithms in LTE-Advanced are proprietary and implementation-specific and there is no standardized procedure to define them \cite{Liu2012}. It is worth noting that the amount of available CSIT defines the transmission mode (SU- or MU-MIMO), and the scheduling decisions. Practical scheduling rules with switching mode can be defined by evaluating the achievable performance of SU-MIMO and MU-MIMO (with multi-rank transmission) modes, and simply choosing the one with better performance \cite{Tang2010,Wang2012d,Fan2014}. However, robust switching between these modes still requires research to guarantee adaptability to changes in channel and traffic conditions, as well as viable computational complexity and scalability \cite{Li2014}.

Authors in \cite{Wang2012d}, highlight the fact that for known precoding matrices at the transmitter, it is possible to generate look-up-tables (LUTs) that contain information regarding spatial compatibility and potential interference. Resource allocation based on LUTs could be used to significantly reduce the complexity of the scheduling algorithms. There are several factors that must be considered for CS/CBF design in cellular systems \cite{Zhang2009,Huh2012,Marzetta2010,Marsch2011,Gesbert2010,Bjornson2013}: dynamically determining whether or not coordination is required, switching between JT and CS/CBF, optimizing cluster formation and sizing, and scalability issues. The resource allocation strategies must also include constraints on backhaul bandwidth, CSI acquisition, latency, QoS, mobility, and synchronization.

\subsection{Implementation in WLAN Scenarios}
\label{section:algorithms:wlan-selection}

In the IEEE 802.11ac standard, MU-MIMO transmission consists of two tasks \cite{Aboul-Magd2013}: \textit{i}) to identify the users that belong to the transmission set, which is implementation-specific and might depend on priority weights or traffic related parameters [cf. Section~\ref{section:performance-uf}]; \textit{ii}) the assignment of a Group ID for each set of co-scheduled users, and the downlink transmission signaling. For a given time instance, the AP constructs a Group ID assignment table. The number of rows is defined by the available number of ID groups, and the columns are given by the number of associated users. The condition to have a proper Group ID, is to have different users assigned to each one of the available stream positions. 
Selecting the proper Group ID and its associated scheduled users is equivalent to a coloring problem over a two dimensional array, whose optimal solution can be found by ExS. A heuristic algorithm was proposed in \cite{Aboul-Magd2013}, where the positions of the users in a particular Group ID are determined according to the probability of occurrence. 
The Group ID selection and success allocation probability can be improved by modeling the discrete searching process as a linear sum assignment problem, for which there exist reliable polynomial-time algorithms \cite{Burkard2009}.

For the general MU-MIMO scenario, a selected user can receive more than one data stream, and all of them must use the same MCS \cite{Arubanetworks2014}. The authors in \cite{Cheng2014} proposed a user selection algorithm inspired by \cite{Fuchs2007}, originally designed for cellular scenarios. The scheme applied BD with geometric-mean decomposition to guarantee equal MCS allocation over all spatial streams assigned to a single user. That work highlights the need of careful stream allocation per selected user, where the best number of streams is not necessarily equal to rank of the channel.
Another approach to guarantee maximum sum rate with joint link adaptation (MCS selection), BD, and user/stream selection was proposed in \cite{Rico-Alvarino2014}. The authors extended the scheduling algorithm in \cite{Shen2006}, originally designed for cellular scenarios, to WLANs with partial CSIT. The link adaptation is performed by a machine learning classifier that provides robustness to CSI inaccuracy. Numerical results suggest that estimation of the IUI is required to allocate MCS more efficiently. 
%
{\color{black}{The next generation of 802.11 standard will define the conditions and methods for user grouping/association  \cite{Liao2016}. A critical open problem is to build scheduling algorithms to balance overhead, fairness and individual priorities.}}

\subsection{Scheduling Complexity}
\label{section:algorithms:complexity}

The scheduling complexity can be classified in two types \cite{Lee2014b}: implementation complexity and computational complexity. The former, refers to the amount of signaling overhead and information exchanged among different network entities. The latter refers to the processing time required to execute a certain algorithm at the transmitter or CU. 
The implementation complexity is assumed to be reduced whenever the channel information is quantized [cf. Section~\ref{section:precoding:partial-csit}]. This type of complexity can be estimated as a function of the system parameters $K$, $B$, $N$, $M$, and the number of clustered transmitters. Nevertheless, to the best of our knowledge, there is no published work or reference framework that analyze the complexity in a fundamental and general way, at least for the MU-MIMO systems considered here. Most research in the field of limited CSI feedback has focused on the following issues related to complexity \cite{Gesbert2007a,Marsch2011}: \textit{i}) reducing the number of parameters to be quantized, e.g., bits to quantize the CQI and CDI; \textit{ii}) reducing the required quantized feedback resolution; and \textit{iii}) constraining the signaling and message exchange within a cluster, and minimizing the overall coordination.

In the MU-MIMO literature, most works focus on the computational complexity required to select users, to extract the precoders, and to perform power allocation. There are several metrics to estimate the computational complexity: \textit{i}) number of real or complex operations. \textit{ii}) number of flops, where a flop is defined as a floating point operation \cite{Shen2006}, e.g., a real addition, multiplication, or division. \textit{iii}) using the big-Oh notation, ${O}(\cdot)$, which is proportional to the term that dominates the number of elementary operations needed to execute an algorithm \cite{Moretti2013}. \textit{iv}) other metrics, such as the number of vector/matrix operations, the number of iterations needed for convergence, or results based on simulations and execution time.
Therefore, there is no unified approach to characterize the computational complexity for scheduling algorithms, but the aforementioned metrics, summarized in Table~\ref{tab:complexity}, provide a coarse assessment of the computational load order. 

We hasten to say that another sort of complexity analysis can be performed for MU-MIMO systems, in which precoders and powers are jointly optimized for a fixed set of users $\mathcal{K}$, [cf. Section~\ref{section:performance-uf:wsr-with-qos}]. The resource allocation problem only involves $\mathbf{W}$, $\mathbf{P}$, and is formulated as the optimization of a certain system-level utility function, $U(\cdot)$, [cf. Section~\ref{section:performance-uf:weighted-sum-rate}], subject to resource budget constraints, see \cite{Hong2012} and references therein. In these cases, the set of scheduled users is not optimized, and the complexity is normally defined as a function of $M$, $N$, and the characteristics of $U(\cdot)$, e.g., convexity or non-convexity.

\begin{table}[t!]
	\renewcommand{\arraystretch}{1.3}	
	\caption{Summary of Computational Complexity Metrics for MU-MIMO Scheduling Algorithms}
	\label{tab:complexity}%
	\centering
	\begin{tabularx}{0.48\textwidth}{l X}
		\toprule
		\bfseries	Metric & \bfseries	References \\
		\midrule	
		Flop-based & \cite{Shen2006, Chen2008, Tran2010,   Dai2009, Mao2012, Ko2012, Tran2012, Nam2014a,  Elliott2012, Hei2009} \\
		\midrule
		Real/complex Operations & \cite{Huang2013, Dimic2005, Huang2012b, Chen2007a, Wang2010, Conte2010, Wang2008a, Maciel2010, Mao2012, Shi2012}  \\
		\midrule
		${O}(\cdot)$ & \cite{Shen2006, Moon2013, Lim2009, Dartmann2013,  Xia2009, Sohn2012, Aniba2007,   Chan2007, Moretti2013,Wagner2008,Liu2014a,YiXu2014,Liu2015a}  \\
		\midrule
		Other Metrics & \cite{Cheng2014, Vicario2008, Tsai2008, Park2010,  Yoo2006, Razi2010, Wang2010, Fuchs2007, Bayesteh2008, Chung2010, Driouch2012, Yoo2005a, Sigdel2009a, Yi2011, Xia2009, Castaneda2015,  Stridh2006a, Matskani2008, Moretti2013, Elliott2009, Lau2005a,Liu2014a}  \\
		\bottomrule	
	\end{tabularx}%
\end{table}%

{\color{black}{
		
\subsection{Discussion}
\label{section:algorithms:selection-partial-csit:discussion}

The two approaches in Section~\ref{section:algorithms:selection-partial-csit:qunatized-csi} and \ref{section:algorithms:selection-partial-csit:rbf} are used in scenarios where the transmitters have different capabilities and constraints. The former approach is applied when the transmitter can compute linear precoding using the quantized channels [cf. Section~\ref{section:precoding:linear}]. The latter approach is used when the precoders are predefined and cannot be modified based on the instantaneous CSI. The bit allocation for the CQI and CDI, described in Section~\ref{section:algorithms:selection-partial-csit:two-stage-feedback}, can be used to analyze both approaches, channel quantization and RBF.  
Practical resource allocation algorithms work with partial CSI, whose computational complexity depends on the codebook resolution, the number of deployed antennas and active users. On the one hand, high codebook resolution improves peak rates and reduces IUI, but it is a bottleneck for the uplink in FDD mode. On the other hand, dynamic channel-based codebook designs speed up the resource allocation, reduce user-pairing complexity, and mitigate interference \cite{Li2014}. 
The CQI and CDI reports can be optimized to enhance centralize or distribute resource allocation, and they are also functions of the operative wide band and feedback periodicity \cite{Ku2014}. The main issue related to CSI feedback is to find a good trade-off between signaling overhead and accurate channel quality estimation. Enhanced scheduling algorithms for multiuser systems will depend on the type of CSI available at the BSs (e.g. channel quantization or RBF), the deployment configuration, and more complex decision-making rules that can maximize the overall throughput \cite{Capozzi2013}.

In practical massive MIMO scenarios, hybrid precoding will be used for MU-MIMO communication, whose performance will depend on the number of RF chains at the transmitter and the accuracy of the fed back CSI. Moreover, the type of transceiver architecture and precoding scheme must be defined according to the objective function (e.g. sum-rate or EE) and the user sparsity \cite{Gao2016}.  Further research is needed to understand the joint optimization of bit and stream allocation using hybrid precoding, taking into account hardware resolution and channel estimation accuracy \cite{Kutty2016}.

}}

\section{Power Allocation}
\label{section:power-allocation}

The optimal power allocation $\mathbf{P}$ in (\ref{eq:inner-outer-problem}) depends on the CSIT availability, the type of precoding scheme implemented at the transmitter, the individual user priorities, and the objective function. Accounting for full CSIT in single-transmitter scenarios, the optimal power allocation is usually known in analytical closed-form for several performance metrics, e.g., BER, SINR, WSR, or fairness \cite{Bartolome2006,Vu2007}. 
Assuming that $\mathcal{K}$ is fixed, and linear precoding is implemented at the transmitter, a network-centric power control algorithm finds the optimal $\mathbf{P}$ for WSR maximization through convex optimization, i.e., using the water-filling principle, see \cite{Palomar2005,Cover2006,Boyd2004}. The linear precoding decouples the signals into orthogonal spatial directions mitigating the IUI, and water-filling allocates powers according to the effective channel gains [cf. Definition~\ref{defn:effective_channel_gains}]. 

In scenarios with a fixed feasible set of users $\mathcal{K}$, [cf. Definition~\ref{defn:feasible_set_of_users}], and assuming that $\mathbf{W}$ is coupled with $\mathbf{P}$, the power allocation may involve, for example, a WSR maximization with rate control, sum power minimization with individual SINR constraints, or a max-min SINR problem, see \cite{Schubert2004,Tan2012,Koutsopoulos2008,Bjornson2013}. For this kind of problems, power control algorithms are designed based on optimization theory (see \cite{Hong2012,Bjornson2013}), and the Perron-Frobenius theory\footnote{The Perron-Frobenius theory is a fundamental tool to solve congestion control problems and optimize interference limited systems [cf. Definition~\ref{defn:interference-limited-system}], such as wireless sensor or multi-hop networks \cite{Stanczak2009}.} for non-negative matrices (see \cite{Stanczak2009,Chiang2008}). 
In contrast to WSR maximization, where water-filling assigns more power to stronger channels, other objective functions require to allocate powers in a different way. For instance, balancing the power over all users so that weak users are assigned more power to improve their error rates  \cite{Cheng2014}, assigning powers according to target SINRs \cite{Chung2010,Tsai2008,Simon2011,Stridh2006a,Matskani2008,Ku2015}, or considering queue stability constraints \cite{Huang2012a}. Some works model $\mathbf{W}$ so that the power is absorbed into it, and optimizing the directions and magnitudes of the precoder weights implicitly performs power control, e.g., \cite{Dartmann2013,Zhang2007b}.   
The interested reader is referred to \cite{Chiang2008} for a comprehensive survey on theory and algorithms for joint optimization of $\mathbf{W}$ and $\mathbf{P}$.  

Equal power allocation (EPA) is a suboptimal strategy used to simplify the evaluation of the utility function $U(\cdot)$, especially when the transmitter does not have full CSI knowledge \cite{Jindal2006}. In certain scenarios, EPA allows a more tractable system performance analysis or derivation of statistics based on $K$, $M$, $N$, or $B$, which generally yields closed-form expressions. Assuming full CSIT, if the WSR maximization problem [cf. Section~\ref{section:performance-uf:weighted-sum-rate}] is optimized using ZF-based or BD-based precoding schemes, the EPA (directly proportional to the individual weights $\omega_{k}$, $\forall k \in \mathcal{K}$), asymptotically achieves the performance of the optimal water-filling power allocation at high SNR \cite{Lee2007}. 	
The authors in \cite{Song2008} discussed the secondary role of power allocation when optimizing problem (\ref{eq:inner-outer-problem}). The paper suggested that more efforts should to be taken to find $\mathcal{K}$ and the corresponding $\mathbf{W}$, which may justify the adoption of EPA under specific SNR conditions.

In systems with partial CSIT, power allocation requires numerical methods that depend on the optimized performance metric \cite{Vu2007}. Consider a practical scenario where $B$ is finite and fixed, and $\mathcal{K}$ has been found by one of the methods described in Section~\ref{section:algorithms:selection-partial-csit}. In such a scenario, the system is limited by interference [cf. Definition~\ref{defn:interference-limited-system}], at the moderate and high SNR regimes, since the interference components in the denominator of (\ref{eq:pcsit_sinr_user_k}) and (\ref{eq:pcsit_rbf_sinr_user_k}) are non-zero. Therefore, one of the main objectives of power control is to perform interference management. However, to compute the optimal $\mathbf{P}$ for WSR maximization, the transmitter must have knowledge of all interference components for all users. This results in non-convex optimization problems, for which efficient algorithms (depending on the operating SNR regime) can provide suboptimal solutions \cite{Kountouris2008}. Moreover, global knowledge of the interference components at the transmitter may not be attainable in limited feedback systems (e.g. only CQIs are known), and adopting EPA is a common practice in the reviewed MU-MIMO literature.

\begin{table*}[t] \scriptsize
	\renewcommand{\arraystretch}{1.3}	
	\caption{Scheduling Approaches and their associated Power Allocation Methods}
	\label{tab:power-allocation}%
	\centering		
	\begin{tabularx}{\textwidth}{l *{3}{>{\centering\arraybackslash}X} }  
		\toprule
		\textbf{Method} & \textbf{EPA} & \textbf{Water-filling} & \textbf{Adaptive} \\
		\midrule
		\textbf{Utility-based} & \cite{Moon2013, Chen2007a, Chen2008, Xia2009, Rico-Alvarino2014, Fan2014, Wang2015, Tang2010, Vicario2008, Chae2008, Kountouris2007, Kountouris2006, Shirani-Mehr2010, Zhang2011, Ravindran2012, Jindal2008, Torabzadeh2010, Cui2011a, Seifi2011, Park2010, Wang2012d, Lossow2013,Hammarwall2008,YiXu2014,Bogale2015} & \cite{Dimic2005, Huang2013, Wang2008a, Shen2006, Huang2012b, Chen2008, Ko2012, Tran2012, Chan2007, Lim2009, Tran2010, Boccardi2006, Huh2012,Bogale2015} & \cite{Zhang2005a, Cheng2014, Koutsopoulos2008, Kobayashi2007, Tsai2008, Hammarwall2007, Dartmann2013, Zhang2009, Yu2013,Liu2014a} \\
		\midrule
		\textbf{CSI-Mapping} & \cite{Kountouris2006b,Nam2014, Jiang2006, Bayesteh2008, Adhikary2013, Adhikary2013a, Chen2007a, Wang2010, Fuchs2007, Bayesteh2008, Wang2006, Xia2009, Lim2009, Tran2010, Zorba2008, Cheng2014, Yang2011, Zakhour2007, Huang2009a, Chae2008, Yoo2007, Kountouris2007, Kountouris2008, Chen2013, Zhang2007a, Kountouris2006a, Wang2007a, Trivellato2008, Choi2007c, Min2013, Sohn2010, Nam2014a, Kountouris2008a, Kountouris2005, Huang2007, Xu2010, Sohn2012, Khoshnevis2013, VanRensburg2009, Hosein2009, Castaneda2015, Jang2011, Schellmann2010, Aniba2007, Christopoulos2013,Wagner2008,Xu2009a,Liu2015a} & \cite{Tu2003, Yoo2006, Shen2006, Dai2009, Lee2014, Wang2010, Mao2012, Shi2012, Fuchs2007, Tejera2006, Maciel2010, Driouch2012, Yoo2005a, Wang2006, Sigdel2009a, Yi2011, Sun2010, Lau2009, Kountouris2008, Wang2010a, Souihli2010, Swannack2004, Shirani-Mehr2011, Kountouris2005a,Lee2014c} & \cite{Jagannathan2007, Razi2010, Zhang2005a, Chung2010, Christopoulos2013, Huang2012a, Kountouris2008, Zhang2007b, Simon2011} \\
		\midrule
		\textbf{Metaheuristic (stochastic)} & \cite{Wang2015, Zhang2011, Park2014} & \cite{Elliott2012, Elliott2009, Lau2005a, Hei2009} & \cite{Bjornson2014, Cottatellucci2006} \\
		\midrule
		\textbf{Classic Optimization} & \cite{Maciel2010} & \cite{Chan2007, Moretti2013} & \cite{Stridh2006a, Matskani2008, Ku2015, Song2008} \\
		\midrule
		\textbf{Exhaustive Search} & \cite{Zakhour2007, Destounis2015} & \cite{Chan2007,Swannack2004} & - \\
		\bottomrule
	\end{tabularx}%
\end{table*}%

In multi-transmitter scenarios, the level of coordination, the availability of user data and CSIT, and the precoding scheme determine the best power allocation policy. In full coordinated scenarios, all transmitters belong to the same infrastructure, and a CU determines the power allocation across the cluster. Assuming global CSIT, a fixed set $\mathcal{K}$, and linear precoding schemes, the power allocation that optimizes a global performance metric subject to per-transmitter power constraints can be realized numerically or through water-filling, see \cite{Zhang2009,Huh2012}. 
For scenarios with partial coordination between transmitters, assuming that $\mathcal{K}$ is globally known, the power allocation depends on the precoding schemes computed from local CSI and individual priorities $\omega_{k}$, $\forall k \in \mathcal{K}$, known by each transmitter, see \cite{Jorswieck2008,Park2012a,Bjornson2010}.

Table~\ref{tab:power-allocation} summarizes the reviewed scheduling methods and their associated power allocation strategies: EPA, water-filling, and adaptive methods. The references listed under the \textit{adaptive} category depend on the particular objective function and system constraints, see Tables \ref{tab:precoding} and \ref{tab:optimization-criteria}. Those references jointly optimize $\mathbf{P}$ and $\mathbf{W}$ via conventional optimization methods \cite{Boyd2004} and iterative algorithms.

{\color{black}{   
\subsection{Summary and Future Directions}
\label{section:power-allocation:ee}

Power allocation is a dynamic process that compensates instantaneous channel variations to maintain or adjust the peak rates \cite{Dahlman2013}. In MU-MIMO scenarios, power control is required to optimize an objective function, mitigate interference, and satisfy system constraints. Several problem formulations deal with convex objectives and linear constraints, for which water-filling can compute the optimal power allocation. Alternatively, EPA simplifies the optimization by transforming $\mathbf{P}$ into a constant vector, which becomes close-to-optimal for ZF at  high SNR. In multi-cell scenarios, power allocation requires exchange of control messages to mitigate ICI. The amount of signaling overhead depends on the type cooperation and coordination between transmitters, and the coordinated signaling (centralized or distributed). Results in \cite{Hamed2016} show that hardware calibration and distributed power allocation are critical to achieve joint transmission in MU-MIMO settings.

Recently, energy consumption has become a central concern in academia and industry \cite{Tombaz2011,Feng2013}. One of the objectives of 5G is to enhance EE by orders of magnitude \cite{Andrews2014}, tanking into account power consumption at the access (transmit and circuitry power) and core (backhaul) networks. EE depends on the particular network planning and objective function [cf. Section~\ref{section:performance-uf}]. For instance, in dense heterogeneous networks the distance between transmitters and receivers determine EE \cite{Feng2013}. Dense deployments can provide high EE, but the number of transmitter per serving area should be carefully planned to avoid ICI and waste of resources due to handovers \cite{Shokri-Ghadikolaei2015}. 
The EE is expressed as a benefit-cost ratio and measured in bits per Joule \cite{Feng2013,Bjornson2016a}. Complementary metrics such as Gflops per Watt (to calculate the typical computational efficiency \cite{Bjornson2014a}) can be used to assess the power consumption per processing block at the transceivers, or to evaluate the EE of a particular algorithm.  
}}

\section{Final Remarks}
\label{section:remarks}

There are many challenges to be solved for resource management in MU-MIMO systems. Identify and standardize the best practices to group and schedule users is fundamental to profit from MUDiv. This is tightly related to affordable-complexity algorithm design to efficiently manage the allocation of multiple users, carriers, time slots, antennas, and transmitters. One of the main requirements to obtain MUDiv and multiplexing gains is knowledge of the CSI. In practical systems, this is challenging because the feedback load rapidly increases with $K$ and $M$. Multiple transmit antennas require additional pilot overhead proportional to $M$, if the users need to learn the complete MIMO channel \cite{Vu2007}. Further research is required to completely understand  performance bounds advance transmission schemes, e.g., dynamic SU-/MU-MIMO switching and multi-/single-stream allocation.
The viability of user scheduling and simultaneous transmission in multi-cell deployments must be further investigated. Several topics must be included in future studies: synchronization issues in MU-MIMO, metrics for user grouping in HetNets, scheduling signaling load, coordinated power control, latency and delay due to scheduling, mobility, the impact of realistic traffic models, backhaul constraints, and semi-distributed control algorithms. The scheduling policies for the next generation of wireless technologies must balance between the high cost of CSI acquisition and the benefits of cooperative transmission.

The current performance of MU-MIMO processing is still limited in 4G cellular communication systems. This is because user terminals do not perform interference estimation, and in all cases only CSI feedback for the SU-MIMO mode (SU-RI, PMI, and CQI) is employed \cite{Li2014}. Most of the reviewed works assume continuous power control, but LTE only supports discrete power control in the downlink, through a user-specific data-to-pilot-power offset parameter \cite{Dahlman2013}. These facts must be considered to properly assess current and emerging NOMA schemes in multiuser scenarios, as well as their limitations, see \cite{Li2014a,Liu2015,Dai2015}. Resource management rule have a fundamental role to play in future generations of wireless networks, where new and evolving technologies such as mmWave and massive MIMO have already been considered \cite{Kim2014,Nam2015,Zheng2015,Kutty2016}. Such rule will require novel precoding designs and an assertive usage of the multiuser dimension to provide substantial spectral efficiency, energy efficiency and user satisfaction.

{\color{black}{

\subsection{Heterogeneous Networks}
Current cellular networks are facing immense challenges to cope with the ever increasing demand for throughput and coverage owing to the growing amount of mobile traffic. Network densification through
deployment of heterogeneous infrastructure, e.g., pico BSs and small cells, is envisioned as a
promising solution to improve area spectral efficiency and network coverage. Nevertheless, heterogeneous networks and small cell deployments will bring significant challenges and new problems in resource allocation and scheduling for MU-MIMO system. First, network densification reduces MUDiv, since few users are associated to the closest BSs \cite{Lopez-Perez2015}. This can reduce the net throughput and may challenge precoding techniques with imperfect CSI that rely on MUDiv to compensate for the imperfections. Furthermore, the selection metrics may need to change in HetNets as scheduling may be used not for boosting the received signal exploiting MUDiv but for serving users with low ICI. Moreover, putting a large number of antennas in small cells is not envisioned mainly due to excess cost and processing capabilities, thus the promising gains of massive MIMO may not be realized in HetNets. The same applies for cooperative and network MU-MIMO schemes (e.g. CoMP) among small cells, which may require additional signaling and communication among small cells. Access network architectures enhanced with distributed antennas \cite{Tolstrup2015} can complement small cells and increase system throughput, but only if large backhaul capacity is available \cite{Ngo2016}. The existence of signaling interfaces (e.g. X2 in LTE) seem to be sufficient for current deployments, but changes may be required for more advanced coordinated MIMO techniques and joint resource allocation. 
Network densification will result in major challenges in indoor coverage and services, for which the channel models used in existing MU-MIMO literature may not be relevant.

\subsection{Massive MIMO and Millimeter Waves}

Massive MIMO and mmWave  technologies have a fundamental role to play in 5G \cite{Boccardi2014,Andrews2014}, but there are many challenges and open problems. More research is needed to enable FDD mode with acceptable overhead and mitigation of pilot contamination, which will speed up the standardization and commercial adoption of massive MIMO technology. 
Operating in wideband channels at mmWaves will require efficient hybrid precoding and fast beam adaptation. Low-resolution and cost effective transceiver architectures must be designed to cope with these goals, which will bring new techniques to calibrate the antenna array, define the most efficient array geometry, estimate CSI, and mitigate hardware impairments \cite{Puglielli2016,Hamed2016}. 
Although massive MIMO and mmWave will be implemented in cellular networks \cite{Shokri-Ghadikolaei2015}, currently, they are jointly applied for indoor short-range \cite{Kutty2016} and outdoor point-to-point applications \cite{Taori2015,Bogale2016}.  Signal processing and medium access techniques for massive MIMO may provide a cost-effective alternative to dense HetNets \cite{Lopez-Perez2015}, and full understanding of how these technologies complement each other is matter of future research \cite{Bjornson2016a}.

Initially, synchronous massive MIMO systems have been implemented with capabilities for joint signal processing \cite{Vieira2014}. However, coordinated per-user allocation and time-synchronous transmission may be hard to achieve in cellular systems \cite{Bjornson2015a}. Therefore, further research is needed to design, prototype, and assess distributed and asynchronous massive MIMO systems, specially in challenging high mobility scenarios. 
Additionally, the data buses and interfaces at the transmitters must be scaled orders of magnitude to support the traffic generated by many concurrent users. Research on intermittent user activity (bursty traffic) has been presented in \cite{Bjornson2015a}, but more work is needed to optimize resource allocation for users with heterogeneous data rate requirements and services. These challenges will demand enhanced physical and media access control layer interactions, in order to support fast user/channel tracking and dynamic resource allocation.

}}

\subsection{Summary of Asymptotic Analysis and Scaling Laws}
\label{section:remarks:analytical-results}

\begin{table*}[t] \scriptsize
	\renewcommand{\arraystretch}{1.3}	
	\caption{Asymptotic Regimes in Multiuser Systems for MISO and MIMO configurations, with full ($B=\infty$) and partial ($B<\infty$) CSIT}
	\label{tab:analytical-results}%
	\centering		
	\begin{tabularx}{\textwidth}{l *{4}{>{\centering\arraybackslash}X} }  
		\toprule
		\textbf{Parameter} & \textbf{MISO}, $B=\infty$ & \textbf{MISO}, $B<\infty$ & \textbf{MIMO}, $B=\infty$ & \textbf{MIMO}, $B<\infty$ \\
		\midrule
		\textbf{Capacity} & \cite{Wang2015, Yang2011, Sohn2010,Lee2007} & \cite{Jindal2006, Yang2011, Huang2009a, Vicario2008, Al-naffouri2009, Zhang2011, Min2013, Sohn2010, Au-Yeung2007, Kountouris2008a, Huang2007} &  \cite{Sharif2007, Hassibi2007, Lee2007} & \cite{Sharif2007, Al-naffouri2009} \\
		\midrule
		\textbf{Queues} & \cite{Chung2010, Lau2006, Destounis2015} & \cite{Huang2012a, Shirani-Mehr2010} & \cite{Hassibi2007} & \cite{Chen2013} \\
		\midrule
		\textbf{High SNR} & \cite{Caire2003,Lee2007} & \cite{Jindal2006, Xia2009, Wagner2008, Conte2010, Tang2010, Huang2009a, Vicario2008, Zhang2011, Yoo2007, Kountouris2008a}  & \cite{Wang2010, Lim2009, Tran2010, Hassibi2007,Lee2007}, \cite{Chen2008, Sun2010, Tran2012} & \cite{Vu2007} \\
		\midrule
		\textbf{Low SNR} & \cite{Caire2003} & \cite{Wagner2008, Huang2009a, Vicario2008, Zhang2011, Kountouris2008a} & \cite{Wang2010, Tran2010} \cite{Sun2010} & \cite{Vu2007} \\
		\midrule
		\textbf{Large} $K$ & \cite{Viswanath2002,Wang2008a, Yoo2006, Jiang2006, Lee2014, Bjornson2014, Yoo2005a, Sohn2010, Huh2012, Jang2011,Lee2014c}, \cite{Jagannathan2007} & \cite{Wagner2008, Nam2014, Xia2009, Tang2010, Huang2009a, Lau2009, Yoo2007, Kountouris2007,  Dai2008, Huang2007, Xu2010} & \cite{Sharif2007, Bayesteh2008, Tran2010, Hassibi2007}, \cite{Sun2010, Tran2012} & \cite{Al-naffouri2009, Trivellato2008}, \cite{Sharif2005}  \\
		\midrule
		\textbf{Large} $M$ & \cite{Huang2012b, Bjornson2014, Huh2012,Ngo2016}  & \cite{Nam2014, Adhikary2013, Adhikary2013a, Lau2009, Dai2008} & \cite{Tran2010}, \cite{ Chen2007a} & \cite{Sharif2005} \\
		\midrule
		\textbf{Bits} $B_1$, $B_2$ & \cite{Zakhour2007} & \cite{Jindal2006, Xia2009, Wagner2008, Huang2012a, Vicario2008, Yoo2007, Dai2008, Zhang2011, Ravindran2012, Kountouris2006a, Min2013, Sohn2010, Au-Yeung2007, Kountouris2008a,  Khoshnevis2013} & -  & \cite{Trivellato2008, Jindal2008}, \cite{Zhang2007a} \\ 
		\bottomrule	
	\end{tabularx}%
\end{table*}%

\begin{table*}[t!]\scriptsize
	\centering
	\caption{Abbreviations}
	\label{table:abbreviations}
	\begin{tabular}{@{}llllllll@{}}
		\toprule
		\textbf{3GPP}     & \begin{tabular}[c]{@{}l@{}}3rd Generation Partnership\\   Project\end{tabular} &  & \textbf{FDD}     & Frequency-division-duplex                      &  & \textbf{PMI}     & Precoding matrix index             \\
		\textbf{ADC}      & Analog-to-digital converter                                                    &  & \textbf{GA}      & Genetic algorithms                             &  & \textbf{QoS}     & Quality of service                 \\
		\textbf{AP}       & Access point                                                                   &  & \textbf{GUS}     & Greedy user selection                          &  & \textbf{QCA}     & Quantization cell approximation    \\
		\textbf{ACK/NACK} & Acknowledgement handshake                                                      &  & \textbf{HetNet}  & Heterogeneous network                          &  & \textbf{QSI}     & Queue state information            \\
		\textbf{AWGN}     & Additive white Gaussian noise                                                  &  & \textbf{ICI}     & Inter-cell interference                        &  & \textbf{RF}      & Radio frequency                    \\
		\textbf{ARQ}      & Automatic repeat request                                                       &  & \textbf{IFC}     & Interference channel                           &  & \textbf{RBF}     & Random beamforming                 \\
		\textbf{BS}       & Base station                                                                   &  & \textbf{IUI}     & Inter-user interference                        &  & \textbf{RI}      & Rank indicator                     \\
		\textbf{BER}      & Bit error rate                                                                 &  & \textbf{JT}      & Joint signal transmission/processing           &  & \textbf{RSS}     & Received signal strength           \\
		\textbf{BD}       & Block diagonalization                                                          &  & \textbf{LoS}     & Line-of-sight                                  &  & \textbf{SUS}     & Semi-orthogonal user selection     \\
		\textbf{BB}       & Branch and bound                                                               &  & \textbf{LTE}     & Long term evolution                            &  & \textbf{SLNR}    & Signal to leakage plus noise ratio \\
		\textbf{BC}       & Broadcast channel                                                              &  & \textbf{MRT}     & Maximum ratio transmission                     &  & \textbf{SNR}     & Signal-to-noise ratio              \\
		\textbf{CU}       & Central processing unit                                                        &  & \textbf{MSE}     & Mean-square-error                              &  & \textbf{ST}      & Single transmission                \\
		\textbf{CDI}      & Channel direction information                                                  &  & \textbf{MMSE}    & Minimum MSE filter                             &  & \textbf{SU-MIMO} & Single user MIMO                   \\
		\textbf{CQI}      & Channel quality information                                                    &  & \textbf{MCS}     & Modulation and coding scheme                   &  & \textbf{SISO}    & Single-input single-output         \\
		\textbf{CSI}      & Channel state information                                                      &  & \textbf{MAC}     & Multiple access channel                        &  & \textbf{SDV}     & Singular value decomposition       \\
		\textbf{CDMA}     & Code division multiple access                                                  &  & \textbf{MIMO}    & Multiple-input multiple-output                 &  & \textbf{SDMA}    & Space-division multiple access     \\
		\textbf{CIZF}     & Constructive interference ZF                                                   &  & \textbf{MISO}    & Multiple-input single-output                   &  & \textbf{SZF}     & Successive ZF                      \\
		\textbf{CBF}      & Coordinated beamforming                                                        &  & \textbf{MUDiv}   & Multiuser diversity                            &  & \textbf{TDMA}    & Time division multiple access      \\
		\textbf{CoMP}     & Coordinated multi-point                                                        &  & \textbf{MU-MIMO} & Multiuser MIMO                                 &  & \textbf{TDD}     & Time-division-duplex               \\
		\textbf{CS}       & Coordinated scheduling                                                         &  & \textbf{NOMA}    & Non-orthogonal multiple access                 &  & \textbf{THP}     & Tomlinson-Harashima precoding      \\
		\textbf{CSIR}     & CSI at the receiver                                                            &  & \textbf{NP}      & Non-deterministic polynomial time problem      &  & \textbf{Wi-Fi}   & Trademark of IEEE 802.11           \\
		\textbf{CSIT}     & CSI at the transmitter                                                         &  & \textbf{NP-C}    & NP complete                                    &  & \textbf{VP}      & Vector perturbation                \\
		\textbf{CDF}      & Cumulative distribution function                                               &  & \textbf{NSP}     & Null space projection                          &  & \textbf{VQ}      & Vector quantization                \\
		\textbf{DoF}      & Degrees-of-freedom                                                             &  & \textbf{OFDMA}   & Orthogonal frequency-division multiple access  &  & \textbf{WSR}     & Weighted sum rate                  \\
		\textbf{DPC}      & Dirty paper coding                                                             &  & \textbf{OFDM}    & Orthogonal frequency-division multiplexing     &  & \textbf{WLAN}    & Wireless local area network        \\
		\textbf{EE}       & Energy efficiency                                                              &  & \textbf{PSO}     & Particle swarm optimization                    &  & \textbf{ZF}      & Zero forcing                       \\
		\textbf{EPA}      & Equal power allocation                                                         &  & \textbf{PU2RC}   & Per unitary basis stream user and rate control &  & \textbf{ZFBF}    & ZF beamforming                     \\
		\textbf{ExS}      & Exhaustive search                                                              &  & \textbf{PHY}     & Physical layer                                 &  & \textbf{ZFDP}    & ZF dirty paper                    \\
		\bottomrule                     
	\end{tabular}
\end{table*}

Several authors have focused their work on asymptotic analysis and scaling laws of performance metrics or utility functions $U(\cdot)$. The analytical results depend on the \textit{system parameters}, $K$, $N$, $M$, $B$, $P$, and $q_k$ $\forall k \in \mathcal{K}$. The information-theoretic results derived in several works provide fundamental limits of achievable values of $U(\cdot)$ from the user and system perspectives. They also shed light on the relevance of each parameter, the conditions where the parameters are interchangeable, the potential and limitations of MU-MIMO systems, and judicious guidelines for the overall system design. For each utility function $U(\cdot)$, and user priorities $\omega_{k}$ $\forall k$, there exist optimal and suboptimal operation points that can be characterized according to the system parameters and their respective regimes. 
The capacity of MU-MIMO systems has been assessed in various asymptotic regimes: high SNR ($P\rightarrow \infty$), low SNR ($P\rightarrow 0$), large number of users ($K\rightarrow \infty$), large number of transmit antennas ($M\rightarrow \infty$), and large codebook resolution ($B \rightarrow \infty$).

Table~\ref{tab:analytical-results} summarizes the system parameters and the antenna configurations, i.e., MISO ($N=1$) or MIMO ($N>1$), and $M \geq N$. Every single reference in the table has its own system model, assumptions, and constraints, studying one parameter and the corresponding effects on the performance $U(\cdot)$, and other fixed parameters. The table is by no means exhaustive; a comprehensive taxonomy of the asymptotic analytical results is out of the scope of this paper. Our aim is to provide a list of organized results from the reviewed paper, so that the interested reader may use each reference for further studies. 
Notice that the first row in Table~\ref{tab:analytical-results} points to references that analyze the capacity as a function of different parameters, while the second row refers to works that provide analytical results of MU-MIMO systems with queue constraints.

\bibliographystyle{IEEEtran}
\scriptsize

\end{document}